# Unveiling Stimulation Secrets of Electrical Excitation of Neural Tissue Using a Circuit Probability Theory

Hao Wang*[1,3,5], Jiahui Wang[1,2,3,5], Xin Yuan Thow[2], Sanghoon Lee[1,2,3,5], Wendy Yen Xian Peh[2], Kian Ann Ng[2], Tianyiyi He[1,3,5], Nitish V. Thakor[2], Chengkuo Lee*[1,2,3,4,5]

[1]Department of Electrical and Computer Engineering, National University of Singapore, Singapore 117583

[2]Singapore Institute for Neurotechnology (SINAPSE), National University of Singapore, Singapore 117456

[3]Center For Intelligent Sensor and MEMS, National University of Singapore, Singapore 117581

[4]NUS Graduate School for Integrative Sciences and Engineering, National University of Singapore, Singapore 117456

[5]Hybrid Integrated Flexible Electronic Systems, National University of Singapore, Singapore 117583

**Abstract**

Electrical excitation of neural tissue has wide applications, but how electrical stimulation interacts with neural tissue remains to be elucidated. Here, we propose a new theory, named the Circuit-Probability theory, to reveal how this physical interaction really happen. We show that many empirical models, including strength-duration relationship and linear-nonlinear-Poisson model, can be theoretically explained, derived, and amended using our theory. Furthermore, this theory can explain the complex nonlinear and resonant phenomena and fit *in vivo* experiment data. In this letter, we validated an entirely new framework to study electrical stimulation on neural tissue, which is to simulate voltage waveforms using a proper circuit first, and then calculate the excitation probability stochastically.

Neuromodulation by electrical stimulation has proven itself as an effective treatment for medical conditions in many therapeutic situations, including deep brain stimulation (e.g. Parkinson's disease) [1], spinal cord stimulation (e.g. chronic pain) and peripheral nerves stimulation (neuroprosthetics) [2-3]. Despite these wide applications, fine details of the mechanism by which electrical stimulation modulates neural response remains elusive [4-5], and a more complete theoretical model accounting for tissue response to various electrical stimulation parameters is still an ongoing pursuit [6-7]. The conventional approach is to study stimulation of individual neurons based on the Hodgkin–Huxley model (HH model) along with simulated electric field (E-field) distribution [8-11]. However, there is a gap in our knowledge describing the microscopic axon structure leading up to the stimulation and response in complex neural (nerve and cortex) and non-neural (muscle) tissues [12]. To address this issue, empirical models and rules have been developed (e.g., Linear-Nonlinear-Poisson cascade model (LNP model) [13], strength-duration curves [14-15], and stimulation waveforms efficiency difference [16-21]). Still, some important phenomena, such as the frequency dependent response of nerve fibers [22-26] and the stochastically distributed gating pattern of the ion channels [27-30], remain unaccounted for.

Here, we propose a new theory, named Circuit-Probability (C-P) theory, to provide a physical framework, which is completely different from the conventional way of using H-H model with E-field modeling. Then, we show that some widely-used empirical models and rules can be intuitively derived from the C-P theory.

How should we analyze tissue response to an external stimulation? To answer this question, we performed a thought experiment, which ultimately led to our new framework of Circuit-Probability. When considering the electrode-tissue interaction, the first question is how the electrode is bridged to the tissue. We know the activation of action potential is induced by the gating of the voltage-dependent ion channels. Then, for electrode-tissue interaction, the key issue is how the electrical input affects the voltage on these ion channels. Considering the cell membrane is a capacitor, which is impermeable to ions, it affects its electrical response in two aspects. Firstly, the voltage change on the capacitor, which is induced by charging and discharging procedures, will generate a different waveform in response to the input waveform. And the charging and discharging procedures are not only affected by the capacitor itself, but also affected by its peripheral circuit. Secondly, the E-field will always be perpendicular to the plate of the capacitor, which is the cell membrane surface, and the direction of the E-field is only determined by the orientation of the capacitor. Apparently, the correct voltage waveform and correct E-field direction can be both obtained with a proper circuit involving the capacitor of cell membrane. This is why we use a circuit to characterize the electric response on the cell membrane.

With a proper circuit, we can model the voltage waveform. Then, from this voltage waveform, how can we know the stimulation strength? In the *in vivo* testing, the number of activated action potentials shows a continuous change with the electric input. However, single channel measurement shows that an individual ion channel does not display a continuous state change in response to electric input. It acts like a digital bit, which only has two states: closed and open. Meanwhile, the gating pattern of a single ion channel also shows a stochastic behavior. Then, how can we build a bridge from the microscopic discrete to macroscopic continuity? The exclusive option is a probabilistic description of the ion channel gating, just as the situation of thermal dynamics and quantum mechanics. Here we assume the ion channel gating is a quantum event and follows the exponential distribution. Then with the voltage waveform simulated using the circuit, it is easy to calculate the probability of activating an action potential with a certain electrical input.

Up to here, we have obtained a basic framework of Circuit-Probability based on pure physical reasoning. The proper circuit configuration can only be fitted using experimental results, which is a posteriori, while the probability equation can obtained by theoretical derivation, which is a priori.

Here we firstly show how to theoretically derive the probability equation.

In the electrical stimulation of a neuron, we assume that electron transition of the protein causes the opening of sodium ion channel, which then generates an action potential (AP). Electron transition is a quantum phenomenon, which is random. Hence, the generation of APs can be described with an exponential distribution for quantum event:

$$f(\lambda, t) = \begin{cases} 1 - e^{-\lambda t}, & t \geq 0 \\ 0, & t < 0 \end{cases} \quad \lambda > 0$$

Here $f(\lambda, t)$ represents the probability of AP to be generated within a time duration of $t$. $\frac{1}{\lambda}$ is the expected time until AP is generated. Expand the exponential distribution to a general calculus form:

$$f(\lambda, t) = 1 - e^{-\int \lambda(t)dt};$$

then the normal exponential distribution is the special form when $\lambda(t)$ is a constant.

Meanwhile, $\frac{1}{\lambda}$ is also a function of the voltage $V$:

$$\frac{1}{\lambda} = g(V)$$

We have three electrophysiological considerations for $g(V)$:

**Consideration 1**: $\frac{1}{\lambda}$ to be infinitely large when the voltage, $V$, is more positive than the threshold voltage, $V_{Threshold}$. In this condition, the AP cannot be generated.

**Consideration 2**: $\frac{1}{\lambda}$ to be inversely proportional with the amplitude of $|V - V_{Threshold}|$, when $V$ is more negative than $V_{Threshold}$.

**Consideration 3**: $\frac{1}{\lambda}$ to approach a minimal level when $|V - V_{Threshold}|$ goes to infinite. So $\frac{1}{\lambda}$ should get saturated at a certain value.

With these three considerations, one possible form of $g(V)$ can be expressed as:

$$\frac{1}{\lambda} = g(V) = \frac{1}{\alpha} \times (e^{\frac{\beta}{|V-V_{Threshold}|^n}} - c)$$

The equation can be re-written as:

$$\lambda = \frac{1}{g(V)} = \alpha \times \frac{1}{e^{\frac{\beta}{(|V-V_{Threshold}|)^n}} - c}$$

Here, $\alpha, \beta, n$ and $c$ are adjustment parameters, where $\alpha > 0$, $\beta > 0$, $n > 1$ and $0 \leq c \leq 1$.

To simplify the equation, here we assume that $n = 1$ and $c = 0$.

Then the complete expression of $\lambda$ is:

$$\lambda = \begin{cases} \alpha \times e^{-\frac{\beta}{|V - V_{Threshold}|}}, & V < V_{Threshold} \\ 0, & V \geq V_{Threshold} \end{cases}$$

Considering the voltage waveform, $V(t)$, is a function of time, $t$, the complete probability calculus equation is:

$$f(\lambda, t) = 1 - e^{-\int \lambda(t)dt} = 1 - e^{-\int \lambda(V(t))dt} = 1 - e^{-\alpha \int e^{-\frac{\beta}{|V(t) - V_{Threshold}|}}dt}, \quad V(t) < V_{Threshold}$$

In this equation, $\alpha, \beta$ and $V_{Threshold}$ are three parameters to be determined by data fitting.

For a specific voltage waveform as shown in Figure 1(a), the voltage waveform can be converted to a $\lambda$ waveform as shown in Figure 1(b). Then the probability calculus can be further simplified as:

$$f = 1 - e^{-\int \lambda(t)dt} = 1 - e^{-S_\lambda}$$

where $S_\lambda$ is the area of the $\lambda$ waveform. A detailed analysis of the probability calculus can be found in **Supplementary S1**.

Then, we build a proper circuit using the results shown in Fig. 2. Its general configuration and analysis can be obtained by reasoning (**Supplementary S2**). Based on the general configuration, its exact configuration is obtained by fitting the experiment data. This is a parallel RLC circuit. The capacitor refers to the cell membrane. The inductor is included to explain the frequency dependent response observed in the experiments. We validated this by applying a single-frequency input (sinusoid wave) to the Common Peroneal (CP) nerve. Sine wave currents (red curve in Fig. 2(b)) were applied upon the CP nerve to activate the Tibialis Anterior (TA) muscle and the resulting force was recorded. The force measured with respect to frequency forms a curve, here named as 'force mapping curve' in this study. With a specific current waveform, a resultant voltage waveform on the capacitor can be calculated, shown as the blue curve in Fig. 2(b) for probability calculation. Similarly, a probability curve with respect to frequency calculated by modeling is defined as a probability mapping curve. The detailed experiment procedure and testing setup can be found in the **Supplement S3**. The force mapping results (force generated by TA muscle) against the pulse width of single pulses (in Hz) of four different current amplitudes curves are shown in Fig. 2(c). The same data plotted with error bar can be found in the **Supplement S6.1**.

The shapes of these four curves are quite different, showing a complicated changing trend with increasing current amplitude. For the curves of 20 µA and 40 µA, a clear resonance effect can be observed. However, 80 µA curve shows an initial decline, before increasing to a resonance frequency. The curve of 200 µA shows a monotonically decreasing trend without the resonance effect. Despite these variations, C-P theory can still reproduce the general shapes of the curves via probability mapping (Fig. 2(d) and Fig. 2(e) shows a more detailed probability mapping). The parameters for the circuit and probability calculus can be found in Table 1-1(d&e). It clearly shows how the force-frequency curve changes from one shape to another shape with increasing current amplitude over a variety of pulse frequencies and accurately predicts the trend, particularly the existence of local minima and maxima. The probability mapping from the C-P theory reproduces the complex changing trends of the testing results, validating the parallel RLC circuit, probability calculus, and most importantly, the existence of an inductor.

The C-P framework and the probability calculus equation is achieved by reasoning, which is a priori, rather than a posteriori. This is very unusual for biological research. Meanwhile, the circuit is still of a preliminary configuration. To validate the correctness of this priori theory, a series of experiments on four types of non-neural and neural tissues using a rat model were conducted: the skeletal muscles (**Supplement S6.3**), the sciatic nerve (**Supplement S6.4**), the cortex (**Supplement S6.5**) and the pelvic nerve (**Supplement S6.6**). All the testing data can be well fitted or explained by the C-P theory: 1. Different current waveforms will generate force mapping curves with different shapes; 2. Force mapping curves generated by arbitrary current waveforms can be fitted by modeling of C-P theory; 3. The resonance frequency widely exists in nervous systems and can be measured with proper stimulation parameters. To help readers understand how various force mapping patterns are generated and affected by parameters, a general demonstration (**Supplementary S5**) and a detailed case analysis (**Supplementary S4**) of how the circuit parameters affect the probability mapping pattern are provided.

Meanwhile, C-P theory can give a unique prediction: the electrical voltage response by electrical stimulation, which is conventionally considered as the stimulus artifact, can be well fitted by the voltage response of the circuit in Fig. 2(a). This voltage response will show the same voltage response as a parallel RLC circuit. The data by experiment and modeling can be found in **Supplement S6.2**.

This C-P theory provides a physical understanding of the electrical nerve stimulation, which is not available in previous theories and models. Thus, most of the phenomenological models and theories can be directly derived or even amended from C-P theory. Here we just show how to derive and correct two well-known phenomenological models in electrical nerve stimulation: strength-duration relationship [14-15] and LNP model (Linear-Nonlinear-Poisson cascade model) [13].

Firstly, we will derive and amend the strength-duration relationship. Previously, it is widely believed that charge is the factor to induce nerve stimulation. In such charge based theory, there is an empirical linear relationship between the threshold charge level and pulse duration, which is called Weiss's strength–duration equation [14] for negative monophasic square current pulses. This equation describes the threshold charge as a function of pulse width as follows:

$$\boldsymbol{Q_{th}(PW) = I_{rh} \times PW + T_{ch} \times I_{rh}}$$

where $\boldsymbol{I_{rh}}$ is the rheobase current, $\boldsymbol{T_{ch}}$ is the chronaxie, and $\boldsymbol{PW}$ is the pulse width. The rheobase current is defined as the threshold current for infinitely long pulses. The chronaxie is defined as the pulse duration required for excitation when the current amplitude is equal to twice the rheobase current. And Lapicque reiterated Weiss's equation for the strength–duration relationship [15], but in terms of the threshold current, and introduced the rheobase current and chronaxie as the constants:

$$I_{th}(PW) = I_{rh}\left(1 + \frac{T_{ch}}{PW}\right)$$

Apparently, these two equations are just mathematical descriptions without explaining how $I_{rh}$ happen and why the curve follows a specific trend.

As follows is the derivation of this relationship with physical definition of $I_{rh}$.

Fig. 3(a) shows a typical voltage waveform by applying negative monophasic square current with difference SPPW (single phase pulse width). For the voltage waveform of each SPPW, the peak voltage is denoted as $V_P$, which is a function of $I$ and $SPPW$ and written as $V_P(I, SPPW)$. Based on C-P theory, nerve excitation can be induced when $V_P(I, SPPW) \geq V_{Threshold}$. Then both the threshold current $I_{th}$ and the threshold charge, $Q_{th} = I_{th} \times SPPW$, are defined as the current and charge required to make the $V_P$ reaches $V_{Threshold}$.

Then the critical condition is:

$$V_P(I_{th}, SPPW) = V_{Threshold}$$

$I_{th}$ and $Q_{th}$ can be written as functions of $SPPW$ and $V_{Threshold}$:

$$I_{th} = f(SPPW, V_{Threshold})$$

$$Q_{th} = I_{th} \times SPPW = f(SPPW, V_{Threshold}) \times SPPW$$

Since $V_P(I_{th}, SPPW)$ cannot be expressed analytically, only numerical solutions of $I_{th}$ and $Q_{th}$, which are calculated with a set of modeling parameter (Table 3 (b,c)) are provided in Fig. 3(b) and Fig. 3(c). In Fig. 3(b), all curves decrease to a constant value, $I_{rh}$. This is because the $V_P$ will saturate at a maximum value, $V_{P_{max}}$, when $SPPW \geq SPPW_{P_{max}}$, as shown in Fig. 3(a).

Meanwhile,

$$Q_{th} = I_{th} \times SPPW = I_{rh} \times SPPW \quad \text{when } SPPW \geq SPPW_{P_{max}}$$

Since $I_{rh}$ is a constant, $Q_{th}$ increases linearly with $SPPW$, when $SPPW \geq SPPW_{P_{max}}$, as shown in Fig. 3(c).

The physical meaning of $I_{rh}$ is the threshold current when $V_{P_{max}} = V_{Threshold}$. Meanwhile, the nonlinear curve of $I_{th}$ versus $SPPW$, existence of $I_{rh}$ and linear curve of $Q_{th}$ versus $SPPW$, can be directly obtained without any additional hypotheses. The exact analytical equation for this relationship is not available. The curves in Fig. 3(b) and Fig. 3(c) are the numerical solution of strength–duration relationship. It corrects the relationship in two aspects:

1. Rather than infinitely approaching to the $I_{rh}$ as the case in Weiss's strength–duration equation, the threshold current curve will be equal to the $I_{rh}$ when $SPPW \geq SPPW_{P_{max}}$.

2. Rather than being a completely straight line, the threshold charge curve is linear only when $SPPW \geq SPPW_{P_{max}}$. When the $SPPW$ is approaching zero, the slope of threshold charge curve will also approach zero, meaning that the threshold charge will converge in a constant value at low $SPPW$.

These two major special differences with the Weiss's equation have already be confirmed by previous research [31-32] and now can be well explained in the C-P theory.

Moreover, it also explains why such relationship can only be applied for negative monophasic square current waveform. Because the voltage waveforms differs with the current waveforms, inducing a more complicated trend without a stable $I_{rh}$, which was observed in other researches [32]. In Fig. 4, representative strength–duration curves of other waveforms including different types of square waves and sine waves are shown. For the curve of sinewave current, the threshold current curve increases at high SPPW range, this phenomenon has been observed in previous research with triangle current waveform [33]. But these curves also vary with different circuit parameters.

Then, we will derive the LNP model. The LNP model is a simplified functional model of neural spike responses [13]. It has been successfully used to describe the response characteristics of neurons in early sensory pathways, especially the visual system. The LNP model is generally implicit when using reverse correlation or the spike-triggered average to characterize neural responses with white-noise stimuli. The number of action potential generated can be described by the Poisson distribution in LNP model.

Actually the Poisson distribution and exponential distribution describe the same stochastic process. If the Poisson distribution provides an appropriate description of the number of the occurrences per interval of time, then the exponential distribution will provide a description of the time interval between occurrences.

The Poisson distribution is as follow:

$$P(x = k; \lambda) = \frac{\lambda^k}{k!} e^{-\lambda}$$

$P(x = k; \lambda)$ is the probability of the $k$ times occurrences of the event in a unit time interval, $\lambda$ is the expected times of occurrence.

The exponential distribution is as follow:

$$P(t; \lambda) = 1 - e^{-\lambda t}$$

$P(t; \lambda)$ is the probability of the occurrences of the event with the time interval $t$, $\frac{1}{\lambda}$ is the expected time interval.

These two distributions share the same $\lambda$. Apparently, in the C-P theory, if the generation of action potential can be described by exponential distribution, it surely can be described by Poisson distribution.

As follow is the derivation of LNP model.

The white noise involved in LNP model can be simplified as a triangle wave series of frequency $f$ and amplitude $V_w$ as shown in Fig. 5(a). Actually any kind of periodical voltage waveform can be used. The triangle wave is used as an example of simple waveform.

Only part of the voltage can exceed the $V_{Threshold}$. As explained in Fig. 1(b), the voltage curve can be converted to a $\lambda$ curve as shown in Fig. 5(b). The area $S_\lambda$ of the $\lambda$ curve within a period $T = 1/f$ can be calculated. Since the $\lambda$ implemented in the C-P theory is not a constant value while $\lambda$ in Poisson distribution can only be a constant value, an equivalent $\lambda_e$ for Poisson distribution can be calculated based on the $S_\lambda$:

$$\lambda_e = \frac{S_\lambda}{T} = S_\lambda \times f$$

which is the blue straight line in Fig. 5(b). Apparently, the $\lambda$ curve and the $\lambda_e$ curve are of the same area, so they will induce the same statistical results.

So the probability calculus equation can be rewritten as:

$$P = 1 - e^{-S_\lambda} = 1 - e^{-\lambda_e t}$$

The corresponding Poisson distribution is:

$$P(x = k; \lambda_e) = \frac{\lambda_e^k}{k!} e^{-\lambda_e}$$

By increasing the noise amplitude $V_w$, $S_\lambda$ will also increase, result in an increasing $\lambda_e$ as shown in Fig. 5(c) and Fig. 5(d). Since $S_\lambda$ is a function of $V_w$, and $\lambda_e$ is a function of $S_\lambda$, $\lambda_e$ is also a function of $V_w$, shown as the nonlinear curve in Fig. 5(e). This explains how a linear increment of $V_w$ induces a non-linear increment of $\lambda_e$ happened in LNP model. Because the expression of $S_\lambda$ is a piecewise function of $V_w$, the exact function $\lambda_e(V_w)$ can only be calculated numerically with a fixed $\alpha, \beta, V_{Threshold}$ and $f$. The analytical expression of $\lambda_e(V_w)$ is not available.

In summary, we propose a new theory, named the Circuit-Probability theory, to unveil the "secret" of electrical nerve stimulation, essentially explain the nonlinear and resonant phenomena observed when nerves are electrically stimulated. In this theory, an inductor is involved in the neural circuit model for the explanation of frequency dependent response. Furthermore, predicted response to varied stimulation strength is calculated stochastically. Two empirical models, strength-duration relationship and LNP model, can be theoretically derived from C-P theory. This theory is shown to explain the complex nonlinear interactions in electrical nerve stimulation and fit *in vivo* experiment data on stimulation-responses of many nerve experiments. As such, the C-P theory should be able to guide novel experiments and more importantly, offer an in-depth physical understanding of the neural tissue. As a promising neural theory, we can even further explore the more accurate circuit configuration and probability equation to better describe the electrical stimulation of neural tissues in the future.

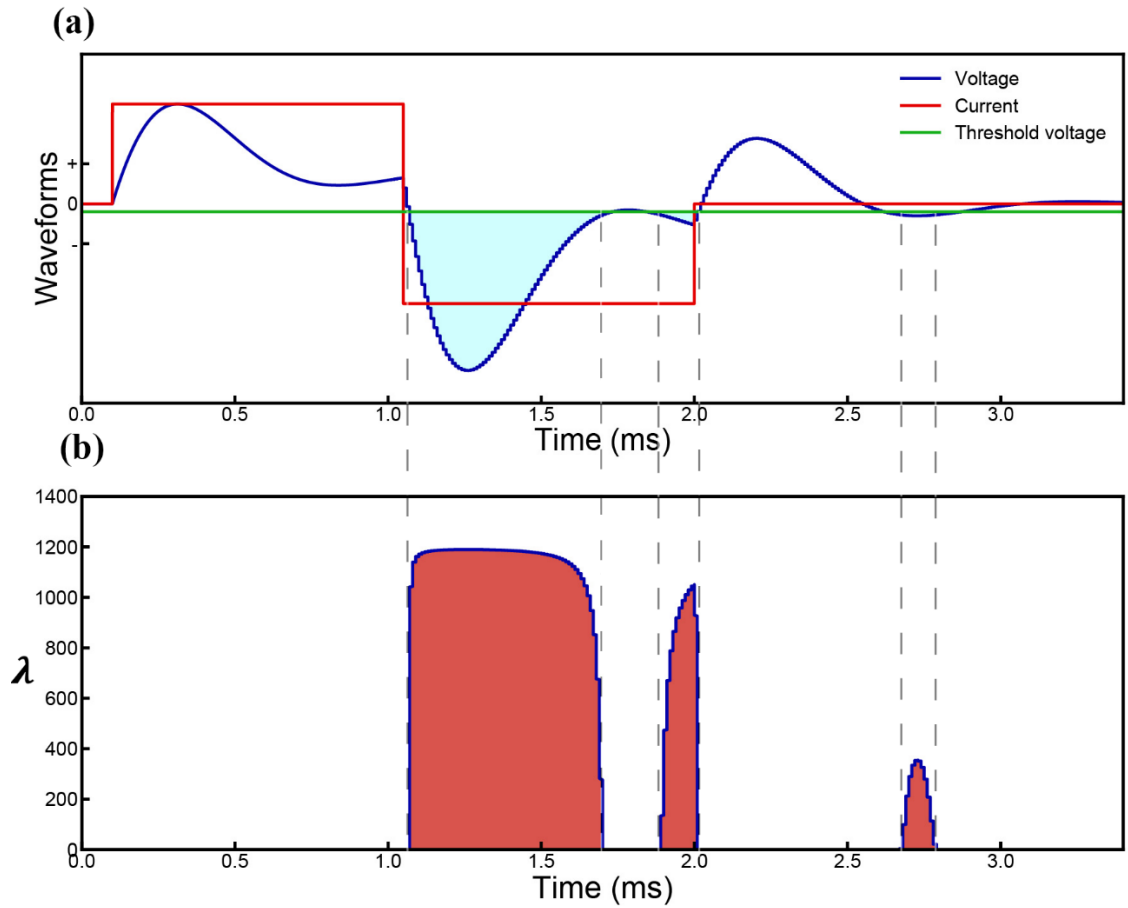

FIG. 1. Parameter illustration of the probability calculus. (a) An illustrative case with multiple effective voltage areas. Red line represents the input current (biphasic in this case), blue line represents the resultant voltage across the cell membrane, and the green line represents the threshold voltage for action potential generation. (b) Corresponding $\lambda$ curve converted from the voltage curve in (a). The probability, $P$, will change monotonically with the area of the $\lambda$ curve, $S_\lambda$.

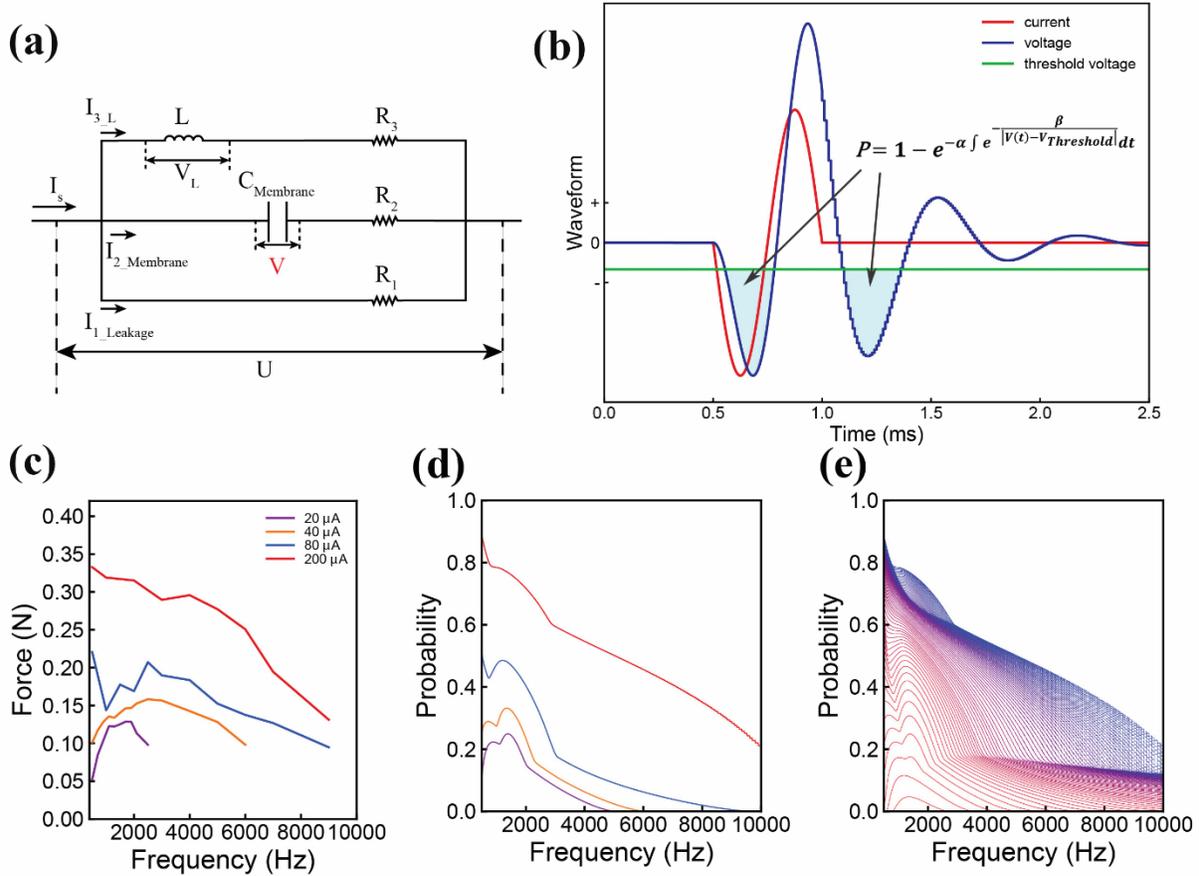

FIG. 2. Illustration of the Circuit-Probability (C-P) theory with experiment and modeling results from the Common Peroneal (CP) nerve stimulation with sine-wave current. (a) The equivalent parallel RLC circuit of the neural tissue; (b) A graph of the applied current (red line) and resulting voltage (blue line) waveforms produced across the capacitor of the circuit as shown in (a); This response is derived from the probability calculus in equation: $\boldsymbol{P = 1 - e^{-\alpha \int e^{-\frac{\beta}{|V(t) - V_{Threshold}|}} dt}}$; (c) the force mapping result recorded from the TA muscle by nerve stimulation. Four different current amplitudes were used at different frequencies, spanning from 500 Hz to 9000 Hz; (d) the corresponding modeling results showing the local minima and maxima predicted by the C-P theory; (e) a detailed probability mapping showing how the shape of the probability curve changes from low current, which exhibits the resonance effect, to high current, which has monotonically decreasing trend.

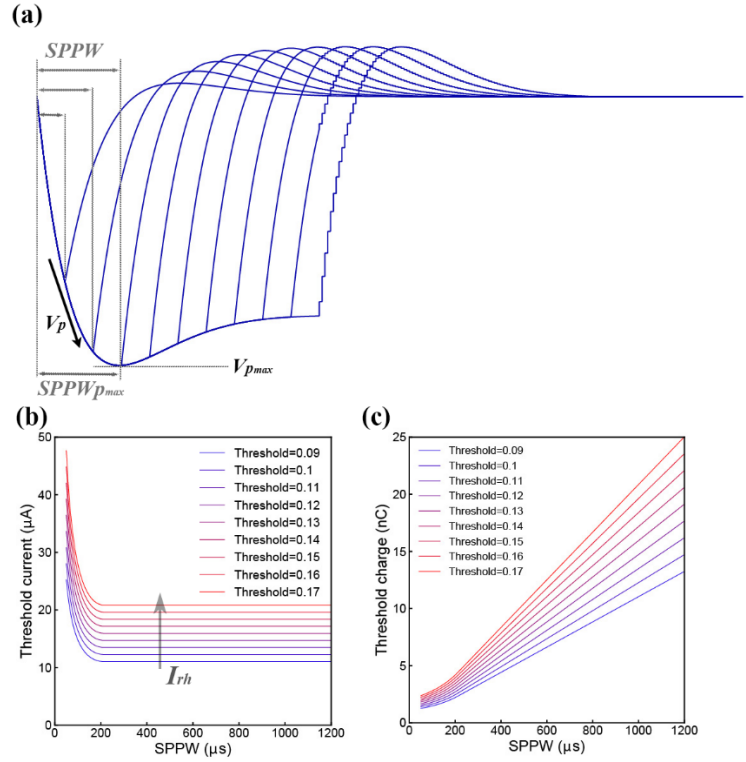

FIG. 3. Derivation of the Strength–duration relationship. (a) Illustrative voltage waveforms generated by negative monophasic square waveform current; (b) The threshold current amplitude ($I_{th}$) decreases as the $SPPW$ increases in a nonlinear fashion; (c) The relationship between threshold charge ($Q_{th}$) and $SPPW$ is linear.

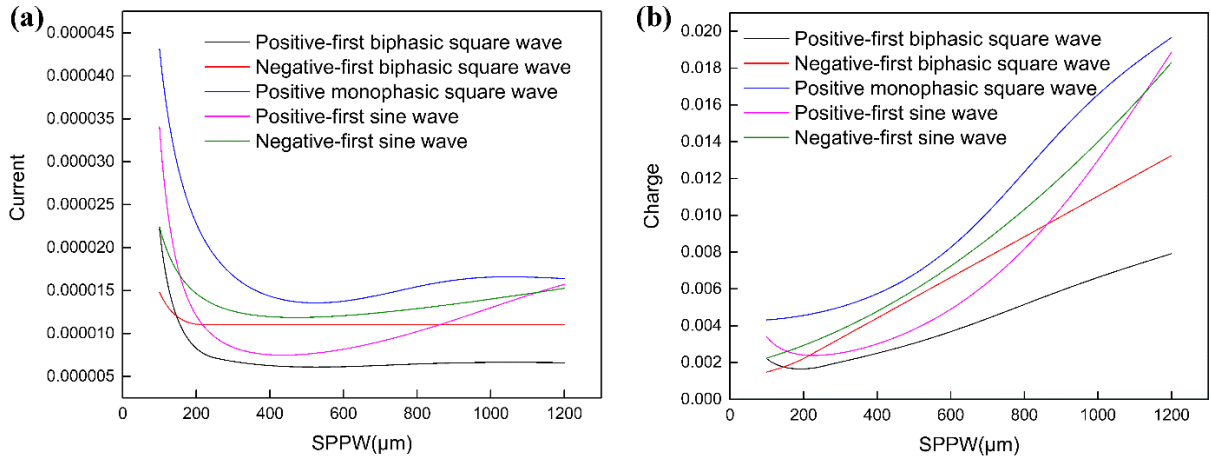

FIG. 4. (a) The relationship between the threshold current amplitude ($I_{th}$) and the *SPPW* for different current waveforms; (b) The relationship between threshold charge ($Q_{th}$) and *SPPW* for different current waveforms.

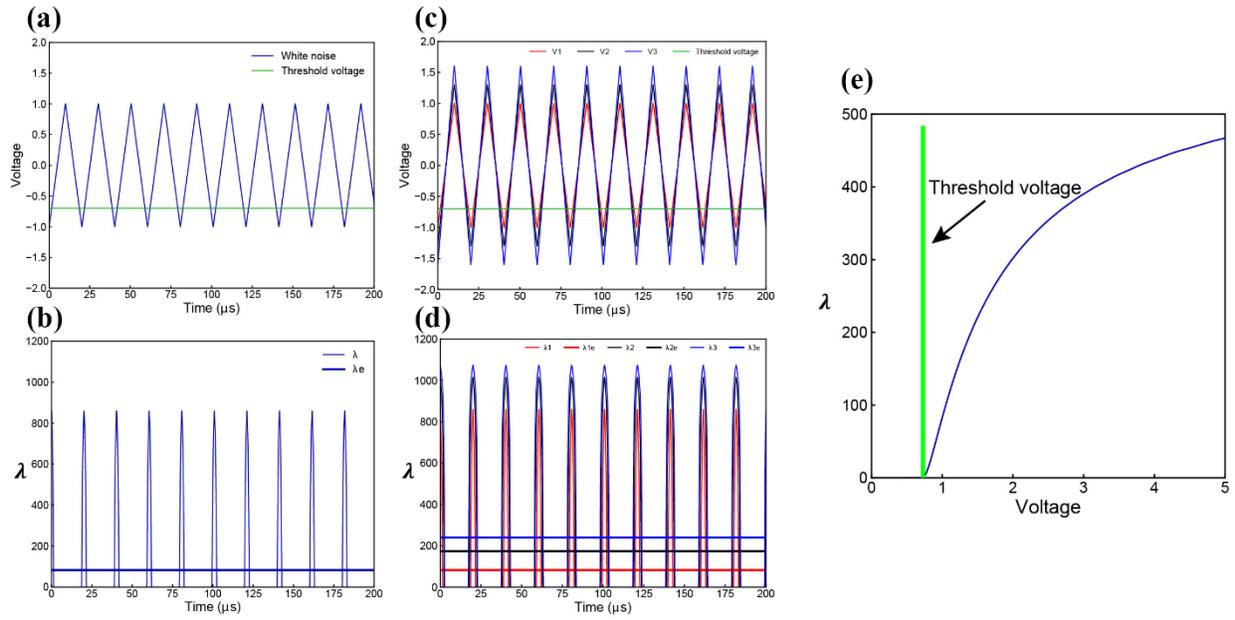

FIG. 5. Derivation of LNP model from C-P theory. (a) A simplified white noise voltage waveform; (b) The corresponding $\lambda$ curve of the voltage waveform in (a), this $\lambda$ curve can be averaged to a $\lambda_e$ curve; (c) Noise with increasing amplitude; (d) The corresponding $\lambda$ and $\lambda_e$ curves of the noise waveforms in (c); (e) The nonlinear curve of $\lambda_e$ versus the noise amplitude $V_w$.

Table 1. Modeling parameters

| No | $R_1(\Omega)$ | $R_2(\Omega)$ | $R_3(\Omega)$ | $C$(F) | $L$(H) | $\alpha$ | $\beta$ | $V_{Threshold}$(V) |
|---|---|---|---|---|---|---|---|---|
| 1 (d&e) | 345000 | 5000 | 10000 | 9n | 1.9545 | 2000 | 0.1 | -0.6 |
| 3 (b,c) | 16579 | 100 | 3000 | 12n | 2.1109 | NA | NA | from -0.09 to -0.17 |
| S4.1 (b) | 16579 | 100 | 3000 | 12n | 2.1109 | 1200 | 0.01 | -0.08 |
| S4.2 (a) | 11052 | 100 | 3000 | 12n | 0.5277 | 2000 | 0.04 | -0.048 |
| S5 (a-i) | 5181 | 100 | 200 | 12n | 0.1938 | 2000 | 0.1 | -0.1 |
| S5 (a-ii) | 5181 | 100 | 2000 | 12n | 0.1938 | 2000 | 0.1 | -0.1 |
| S5 (b-i) | 10362 | 100 | 200 | 12n | 0.1938 | 2000 | 0.1 | -0.1 |
| S5 (b-ii) | 10362 | 100 | 2000 | 12n | 0.1938 | 2000 | 0.1 | -0.1 |
| S5 (c-i) | 20723 | 100 | 200 | 12n | 0.1938 | 2000 | 0.1 | -0.1 |
| S5 (c-ii) | 20723 | 100 | 2000 | 12n | 0.1938 | 2000 | 0.1 | -0.1 |
| S6.2.1 (a) | 80000 | 300 | 1700 | 18n | 0.1086 | NA | NA | NA |
| S6.2.1 (b) | 2656 | 1800 | 800 | 18n | 0.0813 | NA | NA | NA |
| S6.2.1 (c) | 2656 | 1800 | 800 | 18n | 0.0813 | NA | NA | NA |
| S6.2.1 (d) | 2000 | 1350 | 500 | 10n | 0.1464 | NA | NA | NA |
| S6.2.1 (e) | 3701 | 350 | 500 | 10n | 0.1464 | NA | NA | NA |
| S6.2.1 (f) | 9000 | 1350 | 500 | 10n | 0.2326 | NA | NA | NA |
| S6.3.1.1 (b) | 60286 | 1800 | 2000 | 4n | 5.2335 | 600 | 0.8 | -0.7 |
| S6.3.1.2 (b) | 72343 | 4600 | 14400 | 4n | 5.2335 | 1500 | 0.06 | -9.69 |
| S6.3.1.3 (b) | 100 | 100 | 300 | 100n | 0.1621 | 45000 | 0.0075 | -0.006 |
| S6.4.1(b) | 12384 | 1200 | 18000 | 10n | 4.9687 | 13000 | 0.5 | -0.35 |
| S6.4.2(b) | 5000 | 30 | 200 | C1=400n; C2=5000n | 0.0702 | 2000 | 0.015 | -0.009 |
| S6.5.1(b) | 90000 | 100 | 600 | 12n | 0.1629 | 17000 | 0.58 | -0.22 |

**Acknowledgement**

We would like to thank for the experiment setup support from Han Wu, Shih Chiang Liu, Astrid, Shuhao Lu, Li Jing Ong and Dian Sheng Wong. We also would like to thank for the animal experiment support from Gammad Gil Gerald Lasam. We have our special acknowledgement to James T. Fulton for his pioneer research of neuroscience published on the Internet.

This work was supported by grants from the National Research Foundation Competitive research programme (NRF-CRP) 'Peripheral Nerve Prostheses: A Paradigm Shift in Restoring Dexterous Limb Function' (NRF-CRP10-2012-01), National Research Foundation Competitive research programme (NRF-CRP) 'Energy Harvesting Solutions for Biosensors' (R-263-000-A27-281), National Research Foundation Competitive research programme (NRF-CRP) 'Piezoelectric Photonics Using CMOS Compatible AlN Technology for Enabling The Next Generation Photonics ICs and Nanosensors' (R-263-000-C24-281), Faculty Research Committee (FRC) 'Thermoelectric Power Generator (TEG) Based Self-Powered ECG Plaster - System Integration (Part 3)' (R-263-000-B56-112) and HIFES Seed Funding 'Hybrid Integration of Flexible Power Source and Pressure Sensors' (R-263-501-012-133).

# Unveiling Stimulation Secrets of Electrical Excitation of Neural Tissue Using a Circuit Probability Theory

Hao Wang*[1,3,5], Jiahui Wang[1,2,3,5], Xin Yuan Thow[2], Sanghoon Lee[1,2,3,5], Wendy Yen Xian Peh[2], Kian Ann Ng[2], Tianyiyi He[1,3,5], Nitish V. Thakor[2], Chengkuo Lee*[1,2,3,4,5]

[1]Department of Electrical and Computer Engineering, National University of Singapore, Singapore 117583
[2]Singapore Institute for Neurotechnology (SINAPSE), National University of Singapore, Singapore 117456
[3]Center For Intelligent Sensor and MEMS, National University of Singapore, Singapore 117581
[4]NUS Graduate School for Integrative Sciences and Engineering, National University of Singapore, Singapore 117456
[5]Hybrid Integrated Flexible Electronic Systems, National University of Singapore, Singapore 117583

# Contents

# S1 Probability calculus

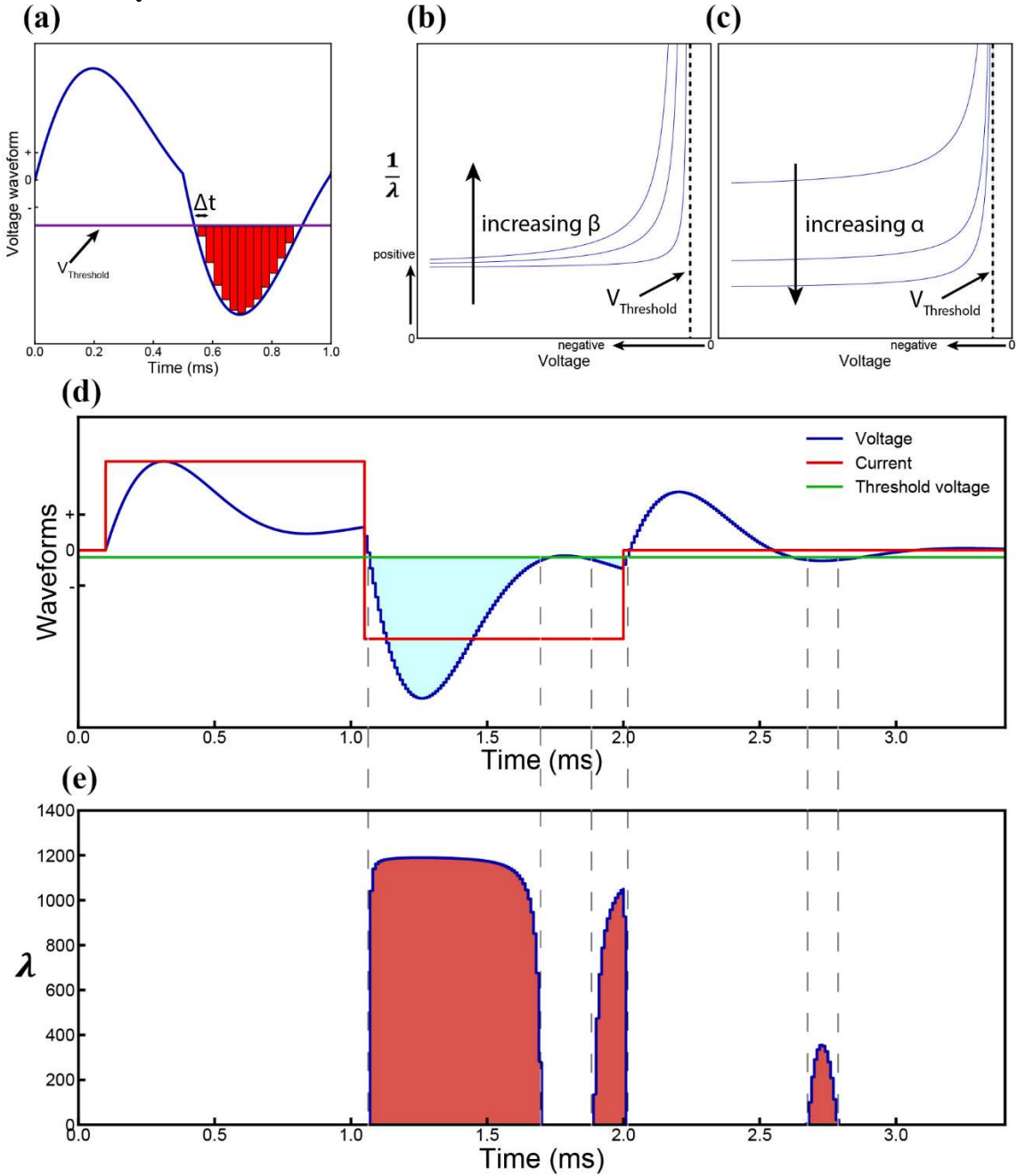

**Figure S1. Parameter illustration of the probability calculus.** (**a**) Derivation of the probability calculus with an effective voltage area; (**b**) The effect of $\beta$ upon the expectation $\frac{1}{\lambda}$, a lower $\beta$ will result in a higher slope; (**c**) The effect of $\alpha$ upon the expectation $\frac{1}{\lambda}$, a higher $\alpha$ will result in a lower saturation level; (**d**) An illustrative case with multiple effective voltage areas. Red line represents the input current (biphasic in this case), blue line represents the resultant voltage across the cell membrane, and the green line represents the threshold voltage for action potential generation. (**e**) Corresponding $\lambda$ curve converted from the voltage curve in (**d**). The probability, $P$, will change monotonically with the area of the $\lambda$ curve, $S_\lambda$.

Here, we describe the probability section of the C-P theory.

In the electrical stimulation of a neuron, we assume that electron transition causes the opening of sodium ion channel, which then generates the action potential (AP). Electron transition is a quantum phenomenon, which is random. Hence, the generation of APs can be described with a probability distribution, which is an exponential distribution as a Quantum event:

$$f(\lambda, t) = \begin{cases} 1 - e^{-\lambda t}, & t \geq 0 \\ 0, & t < 0 \end{cases} \quad \lambda > 0$$

Here $f(\lambda, t)$ represents the probability of AP to be generated within a time duration of $t$. $\frac{1}{\lambda}$ is the expected time until AP is generated. Since electron transitions need to overcome the energy barrier, which is affected by the applied voltage, a higher voltage should have a higher probability, and thus shorter expected time, of generating AP. Therefore, we can consider that the expected time, $\frac{1}{\lambda}$, to be a function of voltage:

$$\frac{1}{\lambda} = g(V)$$

We have three electrophysiological considerations when modeling $g(V)$:

**Consideration 1**: $\frac{1}{\lambda}$ to be infinitely large when the voltage, $V$, is more positive than the threshold voltage, $V_{Threshold}$. In this condition, the AP cannot be generated.

**Consideration 2**: $\frac{1}{\lambda}$ to be inversely proportional with the amplitude of $|V - V_{Threshold}|$, when $V$ is more negative than $V_{Threshold}$.

**Consideration 3**: $\frac{1}{\lambda}$ to approach a minimal level when $|V - V_{Threshold}|$ goes to infinite. So $\frac{1}{\lambda}$ should get saturated at a certain value.

With these three considerations, one possible form of $g(V)$ can be expressed as:

$$\frac{1}{\lambda} = g(V) = \frac{1}{\alpha} \times (e^{\frac{\beta}{|V-V_{Threshold}|^n}} - c)$$

The equation can be re-written as:

$$\lambda = \frac{1}{g(V)} = \alpha \times \frac{1}{e^{\frac{\beta}{(|V-V_{Threshold}|)^n}} - c}$$

Here, $\alpha, \beta, n$ and $c$ are adjustment parameters, where $\alpha > 0$, $\beta > 0$, $n > 1$ and $0 \leq c \leq 1$.

To simplify the modeling, here we assume that $n = 1$ and $c = 0$.

Then the complete expression of $\lambda$ is:

$$\lambda = \begin{cases} \alpha \times e^{-\frac{\beta}{|V-V_{Threshold}|}}, & V < V_{Threshold} \\ 0, & V \geq V_{Threshold} \end{cases}$$

Considering that voltage is a function of time, $V(t)$, then at time point $t_1$, within an infinitely small duration, $\Delta t$, if $V < V_{Threshold}$, the possibility of not generating AP is:

$$P'(V(t_1), \Delta t) = 1 - f(\lambda(V(t_1)), \Delta t) = 1 - (1 - e^{-\lambda(V(t_1)) \times \Delta t}) = e^{-\lambda(V(t_1)) \times \Delta t}$$

Here $V(t_1)$ is considered as a constant value within this $\Delta t$ duration.

For a duration of $T$, it can be divided into $N$ parts equally when each part is $\Delta t$ as shown in Figure S1(a). Thus, the total probability of not generating AP within the duration of $T$ is:

$$P'(V(t), T) = \prod_{n=1}^{n=N} P'(V(t_n), \Delta t) = \prod_{n=1}^{n=N} e^{-\lambda(V(t_n)) \times \Delta t} = e^{\sum_{n=1}^{n=N} -\lambda(V(t_n)) \times \Delta t}$$

The total probability of generating AP within the duration of $T$ is:

$$P(V(t), T) = 1 - P'(V(t), T) = 1 - e^{\sum_{n=1}^{n=N} -\lambda(V(t_n)) \times \Delta t} = 1 - e^{\sum_{n=1}^{n=N} -\alpha \times e^{-\frac{\beta}{|V(t_n) - V_{Threshold}|}} \times \Delta t}$$

Then replacing $\Delta t$ with $dt$, the above equation can be re-written as the continuous integral function:

$$P = 1 - e^{-\alpha \int e^{-\frac{\beta}{|V(t) - V_{Threshold}|}} dt}, \quad V(t) < V_{Threshold}$$

There are three parameters in the equation, $\alpha$, $\beta$ and $V_{Threshold}$. Figure S1(a-c) shows how these three parameters affect the stimulation results.

As shown in Figure S1(a), the probability of stimulation with a specific voltage waveform is affected by the part below the $V_{Threshold}$, which is the red area and called voltage area ($S_V$). When the amplitude of $V_{Threshold}$ becomes more negative, this $S_V$ decreases, lowering the stimulation strength.

$\beta$ is the parameter that determines how fast the expected time $\frac{1}{\lambda}$ decreases to the minimum value with increasing $V$. As shown in Figure S1(b), a lower $\beta$ value will increase the slope of the curve of $\frac{1}{\lambda}$ versus $V$. One extreme example is that when this $\beta$ is very small, the $\frac{1}{\lambda}$ will be a constant value, which is $\alpha^{-1}$.

$$\frac{1}{\lambda} = \alpha^{-1} \times e^{\frac{\beta}{|V - V_{Threshold}|}} \approx \alpha^{-1} \times e^0 = \alpha^{-1}$$

Then the stimulation result is only affected by the duration when $V(t) < V_{Threshold}$.

$$P = 1 - e^{-\alpha \int e^{-\frac{\beta}{|V(t) - V_{Threshold}|}} dt} \approx 1 - e^{-\alpha t}, \quad V(t) < V_{Threshold}$$

$\alpha$ is the parameter determines the minimum value of the $\frac{1}{\lambda}$ as shown in Figure S1(c). A higher $\alpha$ value can induce a higher probability of AP generation.

We can roughly estimate how $P$ will change with $S_V$. This is important in finding the correct circuit parameters for fitting the experiment data. For most of the situations, we need to estimate how $S_V$ changes with current pulse width based on the force mapping results.

However, since the probability calculus equation is nonlinear, a larger $S_V$ does not necessarily generate a higher $P$. Especially when two voltage curves are of different shapes and amplitudes, the sizes of $S_V$ cannot be used to accurately determine which one can achieve a higher $P$.

Thus, to make the relation between $S_V$ and $P$ clearer, we can convert the $V$ curve to a $\lambda$ curve. An illustrative case is shown in Figure S1(d) and (e). For a biphasic square wave current (red line shown in Figure S1(d)), a voltage waveform can be calculated (blue line shown in Figure S1(d)). Then, there are three parts of $S_V$, where the voltage waveform is below $V_{Threshold}$. Since $\lambda$ is a function of $V(t)$, the corresponding $\lambda$ curve plotted with time is shown in Figure S1(e) and the area of $\lambda$ curve is denoted as $S_\lambda$.

Here the probability calculus equation can be re-written as:

$$P = 1 - e^{-\alpha \int e^{-\frac{\beta}{|V(t)-V_{Threshold}|}} dt} = 1 - e^{-\int \lambda(t)dt} = 1 - e^{-S_\lambda}$$

In this simplified equation, $P$ changes with $S_\lambda$ monotonically. Every $V$ curve can be converted to a $\lambda$ curve. By observing the change of $S_\lambda$, we can reliably estimate and explain the change of $P$.

## S2 Circuit analysis
### S2.1 Equivalent Circuit for general tissue modeling

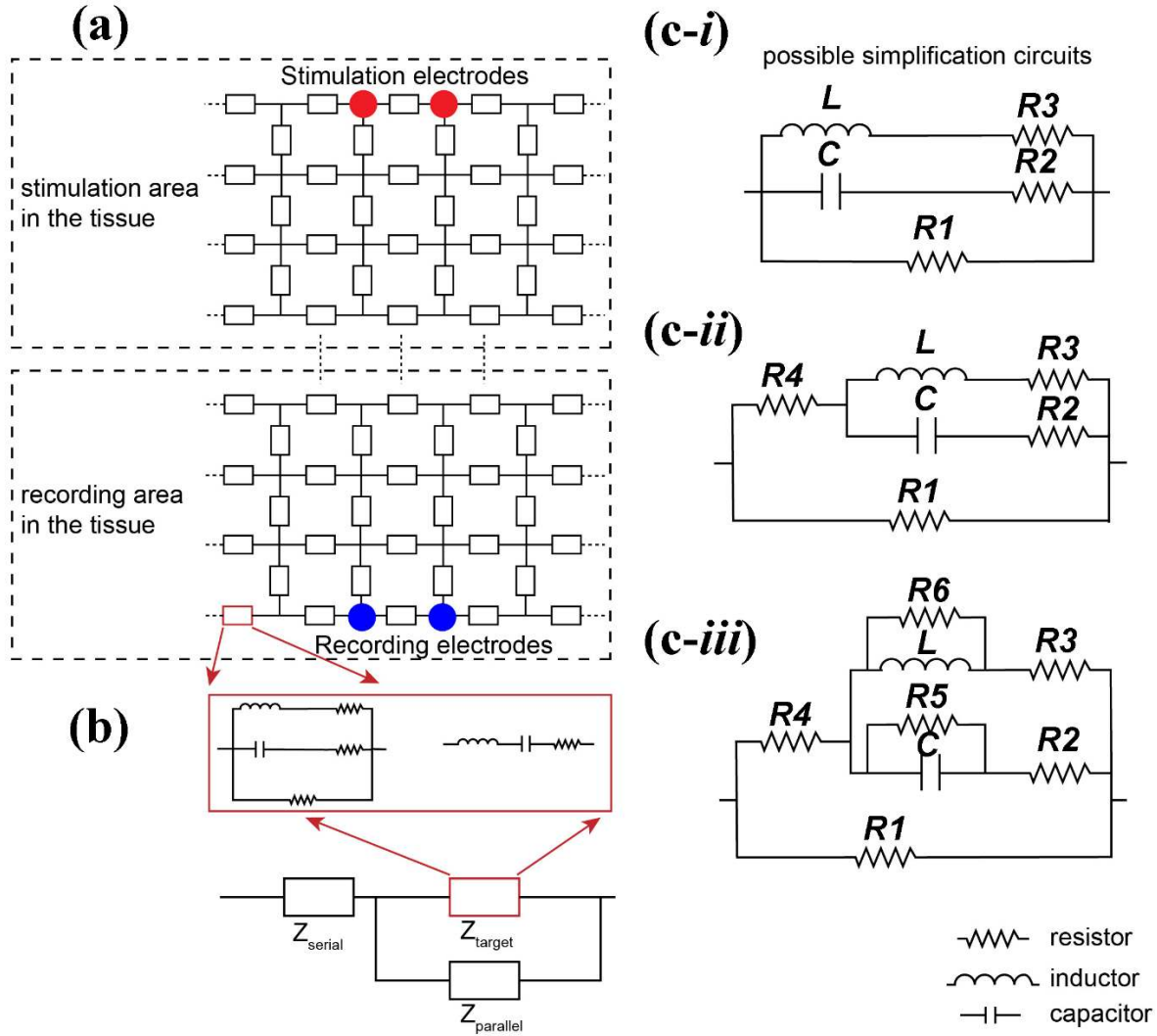

**Figure S2.1 Circuit model of neural tissue**: **(a)** An illustrative distributed parameter circuit to model the whole neural tissue with stimulating and recording electrode pairs which can be implanted at different locations (the two red dots represent one such stimulation pair and the blue dots represent one such recording pair), so the stimulus artifact recorded by the recording electrodes is only one part of the voltage delivered from the stimulating electrodes; **(b)** The external circuit, excluding the target block, can be simplified as a serial and a parallel impedance, while the impedance of the target block can be modeled as either a serial or a parallel RLC circuit; **(c)** Different configurations based on the basic parallel RLC circuit can be used to approximate the circuit in **(b)**.

Earlier, we used an individual RLC circuit to model the stimulated tissue and fit the force mapping curves. Ideally the whole tissue, including the neural and non-neural part, can be modeled as a distributed parameter circuit network as shown in Figure S2.1(a). Electrodes interface to this network to either introduce current (stimulating, shown in red) or measure voltage (recording, shown in blue). In this circuit network, a region of tissue is lumped as a block with certain impedance, which can be modeled as either a parallel RLC circuit or a serial RLC circuit as shown in Figure S2.1(b).

It is not feasible to get the accurate parameters and circuit structures of every block. Hence, simplification of the distributed parameter circuit into a lumped parameter circuit is necessary. Simplification causes distortion of circuit characteristic, resulting in distortions that can only be examined by comparing the experiment results to the predicted values.

In general, each block in the circuit network can be simplified, as shown in Figure S2.1(b). The impedance of the targeted neural tissue block is $Z_{target}$, which is connected to a serial impedance, $Z_{serial}$, and a parallel impedance, $Z_{parallel}$[1]. Because the current source is used in this study, $Z_{serial}$ can be neglected due to Kirchoff's circuit law, while $Z_{parallel}$ is simplified as a resistor, $R_1$. As shown earlier to explain the force mapping (Figure 2), the neural tissue, $Z_{target}$, responds similarly as a parallel RLC circuit. Several circuit parameters are proposed in Figure S2.1(c-i to iii). In Figure S2.1(c-i), only the five necessary parameters, $R_1$, $R_2$, $R_3$, $L$ and $C$, are presented. Figure S1(c-ii) adds further details by connecting a resistor, $R_4$, in series with the $L$ and $C$ branches, which produces a unique damping effect upon the resulting voltage curve in the probability calculus. Two additional resistors, $R_5$ and $R_6$, can be connected in parallel with $L$ and $C$ to further improve the precision of the circuit model (Figure S2.1(c-iii)). These two resistors can tune the voltage amplitude, damping and resonance frequency of the resulting probability curves. With these considerations, we recognize that while every additional circuit component significantly increases the complexity of fitting the circuit parameters, it also adds precision to represent the complexity and variability in tissue. Currently, the most simplified circuit (Figure S2.1(c-i)) is sufficiently accurate in reproducing the general trend of the testing results. In the future, circuits including more components, such as Figure S2.1(c-ii and c-iii), can be used for a better data fitting.

## S2.2 Simple RLC circuit

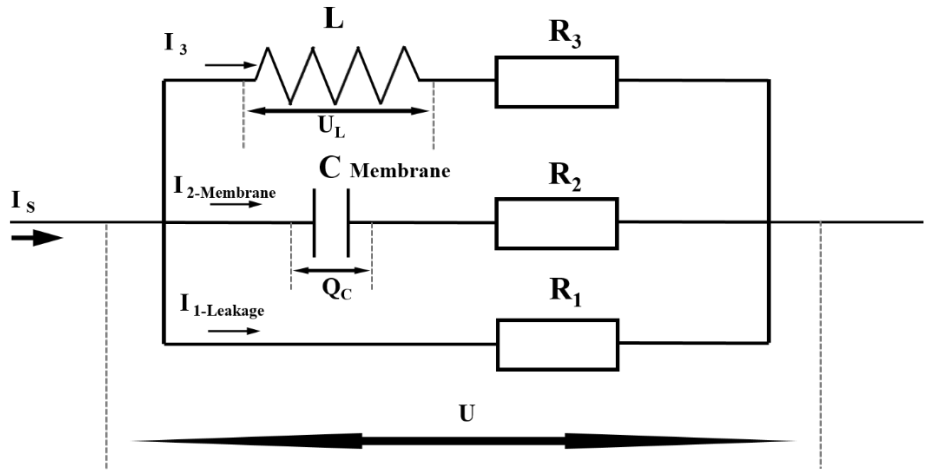

Figure S2.2.1. Simple parallel RLC circuit

In the equivalent parallel RLC circuit, there are 5 circuit parameters.

$L$: Inductance of the circuit;

$C$: Capacitance of the cell membrane;

$R_1$: Leakage resistance of the circuit;

$R_2$: Resistor connected in series with the membrane capacitor;

$R_3$: Resistor connected in series with the inductor.

Here we use a set of differential equations to calculate the voltage waveform upon the membrane capacitor with the current of arbitrary waveforms.

All variables for circuit analysis are as follows:

$\Delta t$: The time interval between each step of calculation;

$I_s$: The input current. This current can be of arbitrary waveform;

$\Delta I_s$: The change of $I_s$ between each time step $\Delta t$;

$I_1$: The current through leakage resistor;

$\Delta I_1$: The change of $I_1$ between each time step $\Delta t$;

$I_2$: The current through membrane capacitor;

$\Delta I_2$: The change of $I_2$ between each time step $\Delta t$;

$I_3$: The current through inductor;

$\Delta I_3$: The change of $I_3$ between each time step $\Delta t$;

$U$: The total voltage upon the whole circuit;

$\Delta U$: The change of $U$ between each time step $\Delta t$;

$Q_C$: The charge upon the membrane capacitor;

$\Delta Q_C$: The change of $Q_C$ between each time step $\Delta t$;

There are three branches of the whole circuit, all branches should share the same voltage $U$ and can be expressed as three equations:

$$I_1 \times R_1 = U \text{ (Leakage branch voltage)}$$

$$\frac{Q_c}{C} + I_2 \times R_2 = U \text{ (Membrane capacitor branch voltage)}$$

$$L \times \frac{dI_3}{dt} + I_3 \times R_3 = U \text{ (Inductor branch voltage)}$$

The above equations can be re-written as difference equations:

$$(I_1 + \Delta I_1) \times R_1 = U + \Delta U \qquad (1)$$

$$\frac{Q_c + \Delta Q_c}{C} + (I_2 + \Delta I_2) \times R_2 = U + \Delta U \qquad (2)$$

$$L \times \frac{\Delta I_3}{\Delta t} + (I_3 + \Delta I_3) \times R_3 = U + \Delta U \qquad (3)$$

The boundary condition is:

$$\Delta I_1 + \Delta I_2 + \Delta I_3 = \Delta I_s \qquad (4)$$

The expression of $\Delta Q_c$ is:

$$\Delta Q_C = I_2 \Delta t + \frac{\Delta I_2 \Delta t}{2} \qquad (5)$$

Based on the equation from (1) to (5), the $\Delta U$ can be solved as:

$$\Delta U = \frac{\Delta I_s + I_1 - \frac{U}{R_1} - \frac{1}{R_2 + \frac{\Delta t}{2C}} \times \left(U - \frac{Q_C}{C} - \frac{I_2 \times \Delta t}{C} - I_2 \times R_2\right) - \frac{1}{\frac{L}{\Delta t} + R_3} \times (U - I_3 \times R_3)}{\frac{1}{R_1} + \frac{1}{R_2 + \frac{\Delta t}{2C}} + \frac{1}{\frac{L}{\Delta t} + R_3}} \qquad (6)$$

In this equation, $I_1, I_2, I_3, Q_C$ and $U$ are known variables of the current state. $\Delta t$ is the time step we set. $\Delta I_s$ is the input to calculate the change of $U$, which is $\Delta U$.

Then all other differential variables can be solved as:

$$\Delta I_1 = \frac{U + \Delta U}{R_1} - I_1$$

$$\Delta I_2 = \frac{U + \Delta U - \frac{Q_C}{C} - \frac{I_2 \times \Delta t}{C} - I_2 \times R_2}{R_2 + \frac{\Delta t}{2C}}$$

$$\Delta I_3 = \frac{U + \Delta U - I_3 \times R_3}{\frac{L}{\Delta t} + R_3}$$

$$\Delta Q_C = I_2 \Delta t + \frac{\Delta I_2 \Delta t}{2}$$

Now based on the previous state, which is represented by $I_1, I_2, I_3, Q_C$ and $U$, we can calculate the next state, which is presented as:

$$I'_1 = I_1 + \Delta I_1$$

$$I'_2 = I_2 + \Delta I_2$$

$$I'_3 = I_3 + \Delta I_3$$

$$U' = U + \Delta U$$

$$Q'_C = Q_C + \Delta Q_C$$

Put the new state, $I'_1, I'_2, I'_3, Q_C'$ and $U'$ and the input $\Delta I_s$ into the equation (6) to calculate the next state. Repeat this procedure to calculate the waveform of each variable by the input of $I_s$ with a time step of $\Delta t$, as shown in Figure S2.2.2.

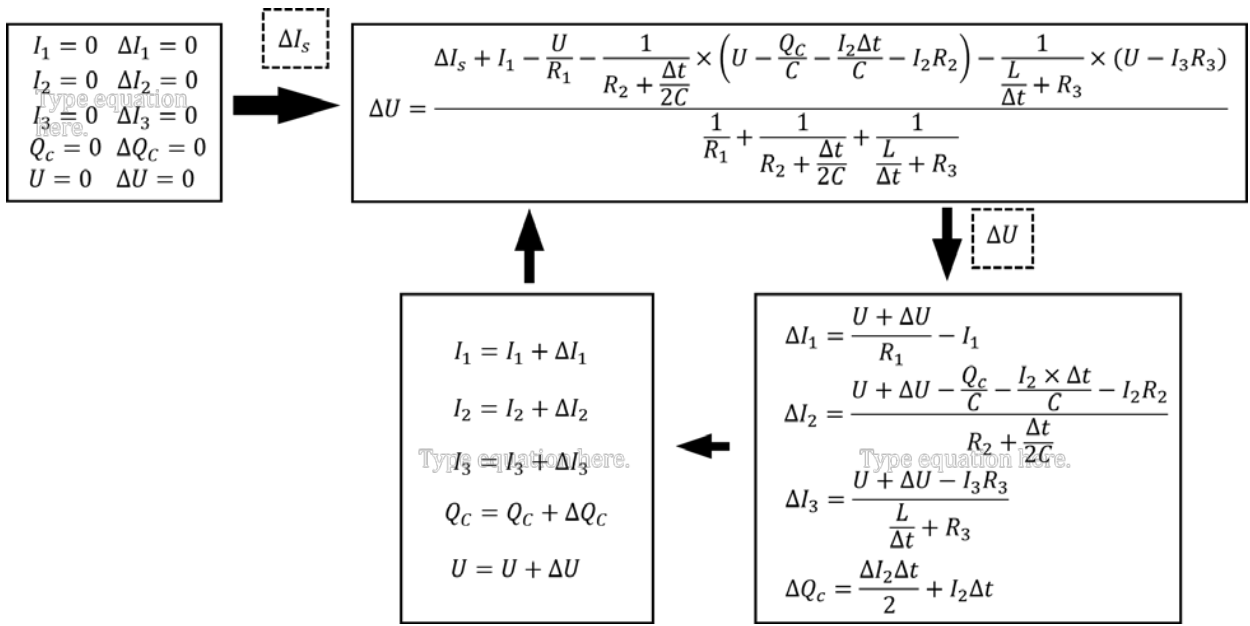

Figure S2.2.2. Calculation process for circuit analysis

Among all these variables, $Q_C$, which is the charge upon the membrane capacitor, is our target variable. With the waveform of $Q_C$, the voltage upon the membrane capacitor can be obtained by $V_C = \frac{Q_C}{C}$, here $V_C$ is the voltage of the membrane capacitor.

## S2.3 Revised RLC circuit

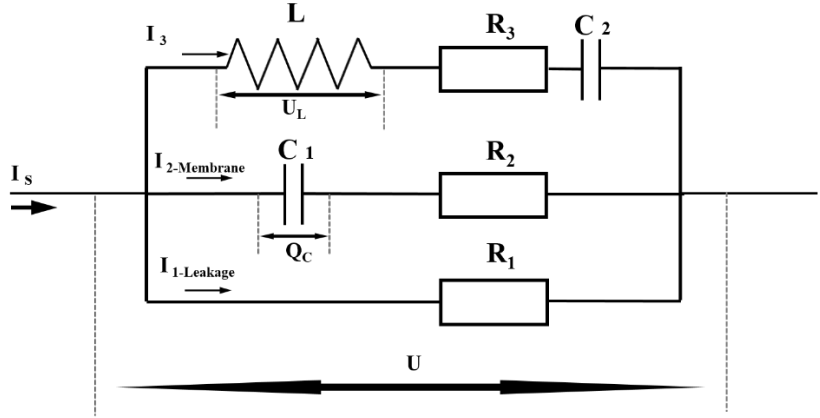

Figure S2.3. Revised RLC circuit with additional capacitor

To fit the force mapping curve in Figure S6.4.2, a revised parallel circuit is proposed. An additional capacitor is connected in series with the inductor branch, denoted as $C_2$, then the $C_{membrane}$ is denoted as $C_1$. The same method can be applied to solve this circuit. Here we just give all the revised differential equations as follows:

$$L \times \frac{\Delta I_3}{\Delta t} + (I_3 + \Delta I_3) \times R_3 + \frac{Q_{C_2} + \Delta Q_{C_2}}{C_2} = U + \Delta U \text{ (Inductor branch voltage)}$$

$$\frac{Q_{C_1} + \Delta Q_{C_1}}{C_1} + (I_2 + \Delta I_2) \times R_2 = U + \Delta U \text{ (Membrane capacitor branch voltage)}$$

$$(I_1 + \Delta I_1) \times R_1 = U + \Delta U \text{ (Leakage branch voltage)}$$

$$\Delta I_1 + \Delta I_2 + \Delta I_3 = \Delta I_s \text{ (Boundary condition)}$$

Then $\Delta U$ can be solved as:

$$\Delta U = \frac{\Delta I_s + I_1 - \frac{U}{R_1} - \frac{U - \frac{Q_{C_1}}{C_1} - \frac{I_2 \Delta t}{C_1} - I_2 R_2}{R_2 + \frac{\Delta t}{2C_1}} - \frac{U - I_3 R_3 - \frac{Q_{C_2}}{C_2} - \frac{I_3 \Delta t}{C_2}}{R_3 + \frac{\Delta t}{2C_2} + \frac{L}{\Delta t}}}{\frac{1}{R_1} + \frac{1}{R_2 + \frac{\Delta t}{2C_1}} + \frac{1}{\frac{L}{\Delta t} + R_3 + \frac{\Delta t}{2C_2}}}$$

Then all other differential variables can be solved as:

$$\Delta I_3 = \frac{U + \Delta U - I_3 R_3 - \frac{Q_{C_2}}{C_2} - \frac{I_3 \Delta t}{C_2}}{\frac{L}{\Delta t} + R_3 + \frac{\Delta t}{2C_2}}$$

$$\Delta I_2 = \frac{U + \Delta U - I_2 R_2 - \frac{Q_{C_1}}{C_1} - \frac{I_2 \Delta t}{C_1}}{R_2 + \frac{\Delta t}{2C_1}}$$

$$\Delta I_1 = \frac{U + \Delta U}{R_1} - I_1$$

$$\Delta Q_{C_1} = I_2 \Delta t + \frac{\Delta I_2 \Delta t}{2}$$

$$\Delta Q_{C_2} = I_3 \Delta t + \frac{\Delta I_3 \Delta t}{2}$$

## S3 Experiment preparation and testing setup
## S3.1 Common Peroneal (CP) nerve test electrode configuration, experiment procedure and testing setup
**Flexible Neural Clip (FNC)**

To demonstrate reliable and repetitive in vivo experiments, we used flexible neural clip (FNC) interface. This interface allows easy and conformal implantation on a variety of peripheral nerves in a manner analogous to clipping a paper clip. The FNC can be implanted onto a peripheral nerve easily by inserting the nerve between the clip-strip and clip-springs after slightly bending the clip-springs (Figure S3.1(a i-iii)). This interface provides not only conformal contact with the nerve, but also gentle pressure on the nerve to keep the clip interface in place. Figure S3.1(b) shows the clip dimensions (length, width, and thickness of the clip-springs, clip-strip, and clip-cavities). The size of the clip-cavity was 700 µm x 500 µm, and the clip strip was 650 µm x 900 µm. The width of clip-spring was the same as the length of cavity (700 µm) to maintain the spring elasticity of the polyimide. Two active electrodes (each 17672 µm$^2$) were located on the center of the clip-strip with the distance of 350 µm for reliable position and stimulation of the implanted nerve. This FNC can cover bigger (diameter: ~600 µm) or smaller sizes (diameter: ~300 µm) of nerves owing to the functionality of the clip design during acute *in vivo* test.

The flexible neural clip (FNC) consisted of a polyimide-Au-polyimide sandwiched structure fabricated by micro-electro-mechanical system (MEMS) technology. The detailed fabrication process is described in previous paper [2]. To enhance electrochemical characteristics of stimulation, the released electrodes were coated with electrodeposited iridium oxide film (EIROF), which show shows the largest CSC and lowest impedance [3]. The coated IrO$_2$ on Au sensing electrodes showed a good impedance value (1.9 ±0.09 kΩ at 1 kHz, n=10), and a cathodic charge storage capacity (56.4 ±2.42 mC/cm$^2$, n=10) for the stimulation. These values are comparable or even better to materials used previously in the literature for neural stimulation [2-4]. This result demonstrates that the IrO$_2$ coated electrodes can be used for *in vivo* stimulation experiments.

**Rat preparation for In vivo test**

Adult female Sprague-Dawley rats (200-300g) were used for acute in vivo experiments in this study. All procedures were performed in accordance with protocols approved by the Institutional Animal Care and Use Committee of the National University of Singapore. The methods were carried out in accordance with the 143/12 protocol. For each experiment, the rat was anesthetized with a mixture (0.2 ml/100 g) of ketamine (37.5 mg/ml) and xylazine (5 mg/ml) intraperitoneally (I.P.), and supplementary doses of 0.1 ml/100 g were injected for maintenance. For the sciatic nerve branch experiments, after an adequate depth of anesthesia was attained, the right sciatic nerves were exposed through a gluteal splitting incision for FNC interface. The FNCs were then implanted on common peroneal (CP) nerve (Figure S3.1(d)) and tibial nerve (Figure S3.1(e)), respectively for each experiment. Stimulations through the FNC were conducted for CP nerve and tibial nerve, respectively (Figure S3.1(c)). Force was measured with a dual-range force sensor (Hand dynamometer, Vernier, USA with NI-DAQ USB-6008, National instruments, USA) tied to the ankle of the animal.

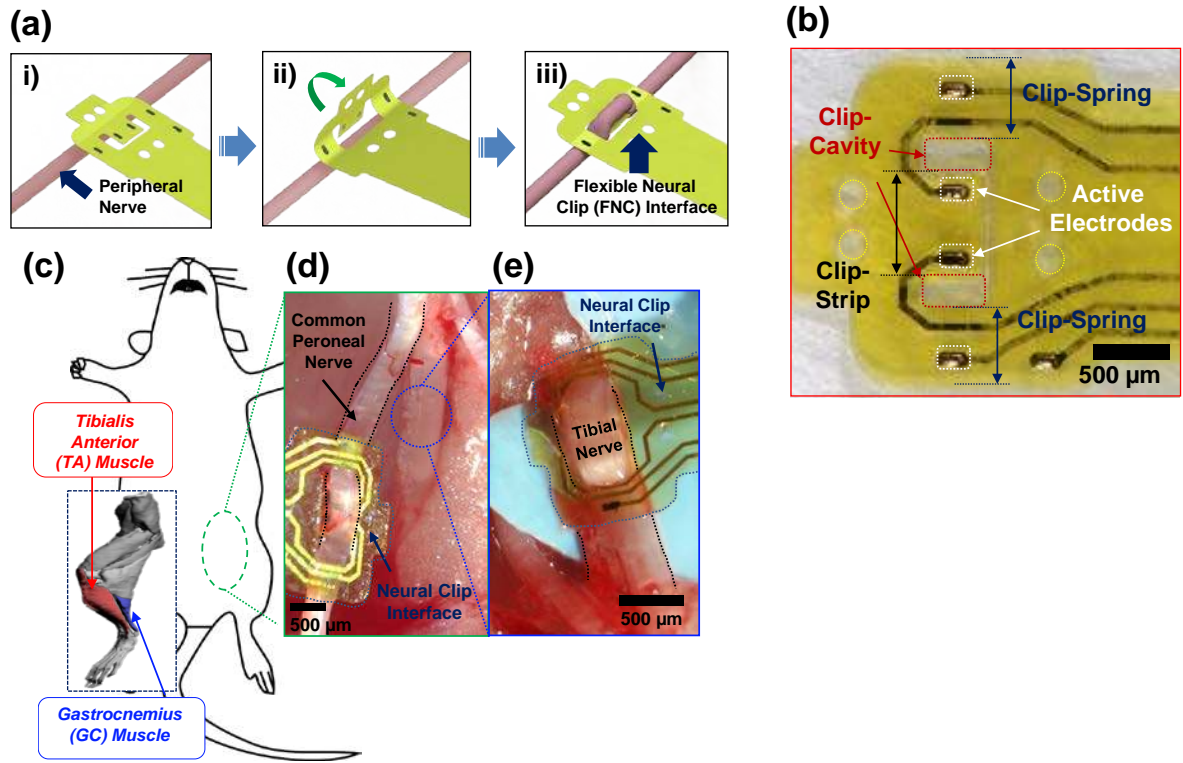

Figure S3.1. (a) Schematic diagram of the steps involved in implanting the FNC on a peripheral nerve (i-iii). (b) Photomicrographs of the fabricated FNC and (inset) clip-head. (c) Schematic diagram and photomicrographs of sciatic nerve branches; (d) common peroneal nerve and (e) tibial nerve.

**S3.2 Cortical stimulation with sciatic nerve recording test electrode configuration, testing setup, detailed procedure and testing parameters**

**Animals**

Female Sprague-Dawley rats (200-300g) were used in the acute experiments. Rats were housed and cared for in compliance with the guidelines of the National Advisory Committee for Laboratory Animal Research (NACLAR) and were humanely euthanized after the experiment. During the experiment, isoflurane was used to induce general anesthesia ( Aerrane®, Baxter Healthcare Corp., USA) prior to injection of ketamine/xylazine (37.5% ketamine, 5% xylazine, 0. 2ml / 100 g Comparative Medicine, NUS). Paw retraction reflex and breathing rate was used to assess the depth of anesthesia, and the core body temperature at 37ºC was maintained using a heating pad (Stryker T/pump, Kent Scientific Corp., USA). Rats were then placed in a stereotaxic frame (Kopf instruments, USA). Microsurgical techniques were used to expose the left sciatic nerve, the tibialis anterior and the skull, which was then subjected to a craniotomy to access the right motor cortex.

**Stimulation and Recording**

A tungsten single shank electrode (0.5MΩ, Microprobe Inc, USA) was implanted into the motor cortex in several locations at a depth of 2.0mm (near the layer V motor cortex pyramidal neurons) until the tibialis anterior was observed to move with a stimulation (100 μA). Once a movement was observed, the location was used for the subsequent experiment.

A thin-filament longitudinal intrafascicular electrode was used in the sciatic nerve to record stimulation artifacts and nerve activity during stimulation. Furthermore, using a string tied to the hind paw of the rat attached to a dynamometer (Hand dynamometer, Vernier, USA with NI-DAQ USB-6008, National instruments, USA), force was recorded simultaneously.

## S3.3 Pelvic nerve test electrode configuration, testing setup, detailed procedure and testing parameters

**Animal subject and surgery**

Female Sprague-Dawley rats (200-300 g) were used in the acute experiments. The animals were housed in pairs in individually ventilated cages, maintained in a 22-24°C room with a 12 h light–dark cycle, and given ad libitum access to food and water. All procedures were performed in accordance with protocols approved by the Institutional Animal Care and Use Committee of the National University of Singapore. For each experiment, the animal was anesthetized with a mixture (0.2 ml/100 g) of ketamine (37.5 mg/ml) and xylazine (5 mg/ml) intraperitoneally (I.P.) for induction, and a supplementary dose of 0.1 ml/100 g was injected I.P. for maintenance as required. The animal was placed in the supine position, and kept warm with a water-circulating heating pad. Laparotomy was performed, and underlying muscles were cut and adipose and connective tissues were gently teased apart to expose the pelvic nerve branches for electrical stimulation. To record muscle activity from the external urethral sphincter (EUS), the pubic bone overlying the urethra was cut and fat tissues were teased apart.

**Pelvic Nerve stimulation and EUS EMG recording**

Hook electrodes made from platinum iridium wires (A-M systems, 0.005" bare, 0.008" coated) were implanted onto the pelvic nerve branches unilaterally and silicone elastomer (Kwik-Sil, World Precision Instruments, FL, USA) was used to encase the electrode-nerve interface. A commercial isolated stimulator (A-M systems model 2100, WA, USA) was used to deliver either cathodic, anodic or biphasic rectangular pulse for pelvic nerve stimulation at a repetition frequency of 2 Hz and current amplitude of 70 µA and different pulse widths (20, 60, 100, 150, 200, 300, 400, 500 µs) used for each set of stimulation. A pair of fine stainless steel wires (304, California Fine Wire, CA, USA) with exposed tips was sutured to the top of the exposed urethra beneath the dissected pubic bone to record EUS EMG signals. EMG signals were amplified by using an Intan preamplifier 2216, and acquired at 20 kHz with the Intan RHD2000 data acquisition board (Intan Technologies), with a 50 Hz notch filter. Stimulation pulse markers were sent from the stimulator to the data acquisition board (DAQ) collecting EMG data simultaneously for data synchronization.

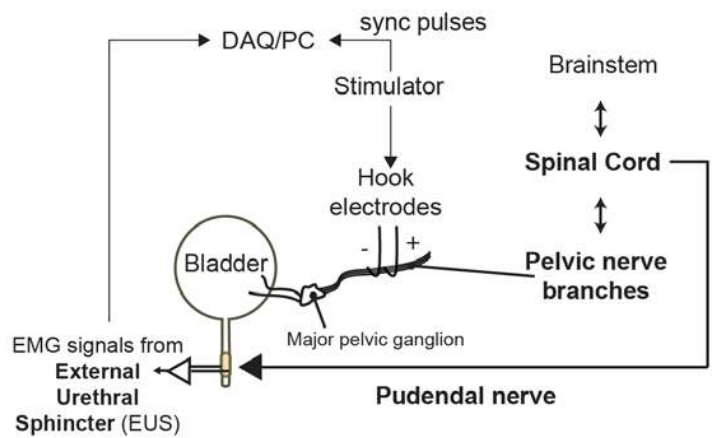

Figure S3.3 Pelvic Nerve stimulation and EUS EMG recording

## S3.4 Tibialis Anterior (TA) muscle test electrode configuration, experiment procedure and testing setup

Sprague-Dawley rats (around 450g) were used in the acute experiments. Rats were housed and cared for in compliance with the guidelines of the National Advisory Committee for Laboratory Animal Research (NACLAR) and were humanely euthanized after the experiment. During the experiment, isoflurane was used to induce and maintain general anesthesia (Aerrane®, Baxter Healthcare Corp., USA). The Tibialis Anterior (TA) muscle was exposed for electrode implantation. Our home-made double-side polyimide electrode (Figure S3.4) was sutured into the muscle belly, transversal to muscle fibers in TA muscle. Current stimulation was delivered from A-M SYSTEMS model 4100 isolated high-power stimulator. Every one second, a train of pulses of 60 Hz was applied. In the experiment of comparing four waveforms, positive-negative biphasic pulses, negative-positive biphasic pulses, positive monophasic pulses, and negative monophasic pulses were applied. Force was measured with a dual-range force sensor (Hand dynamometer, Vernier, USA with NI-DAQ USB-6008, National instruments, USA) tied to the ankle of the animal.

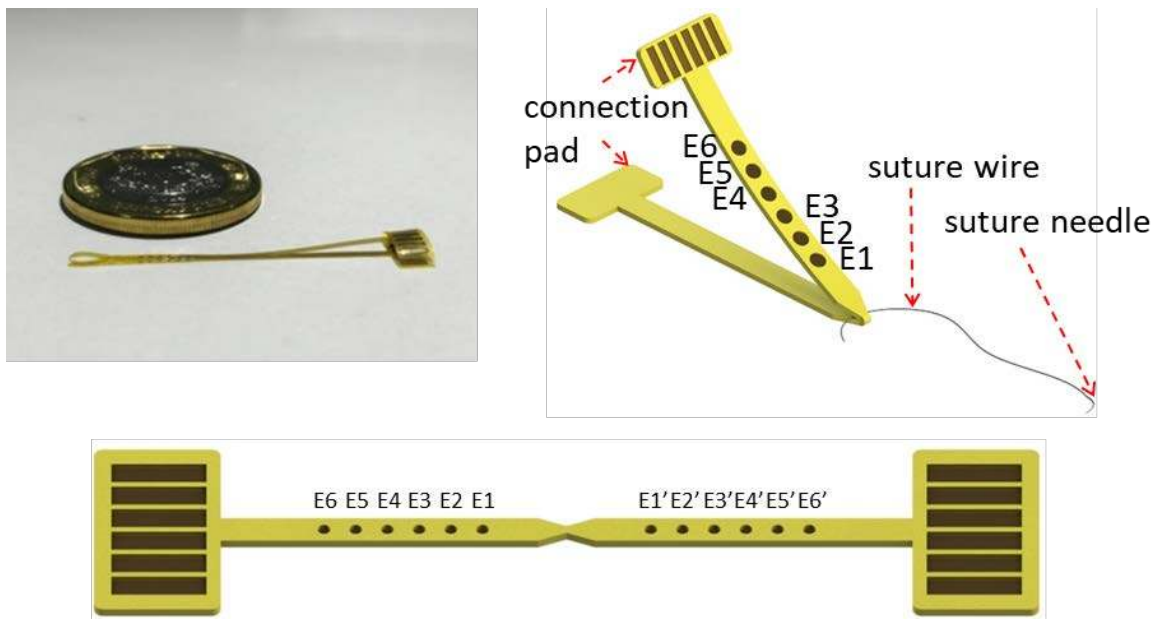

Figure S3.4 Detailed configuration of the home-made electrode for TA muscle stimulation

# S4 A case study demonstration and an illustration of force mapping patterns using different parameters

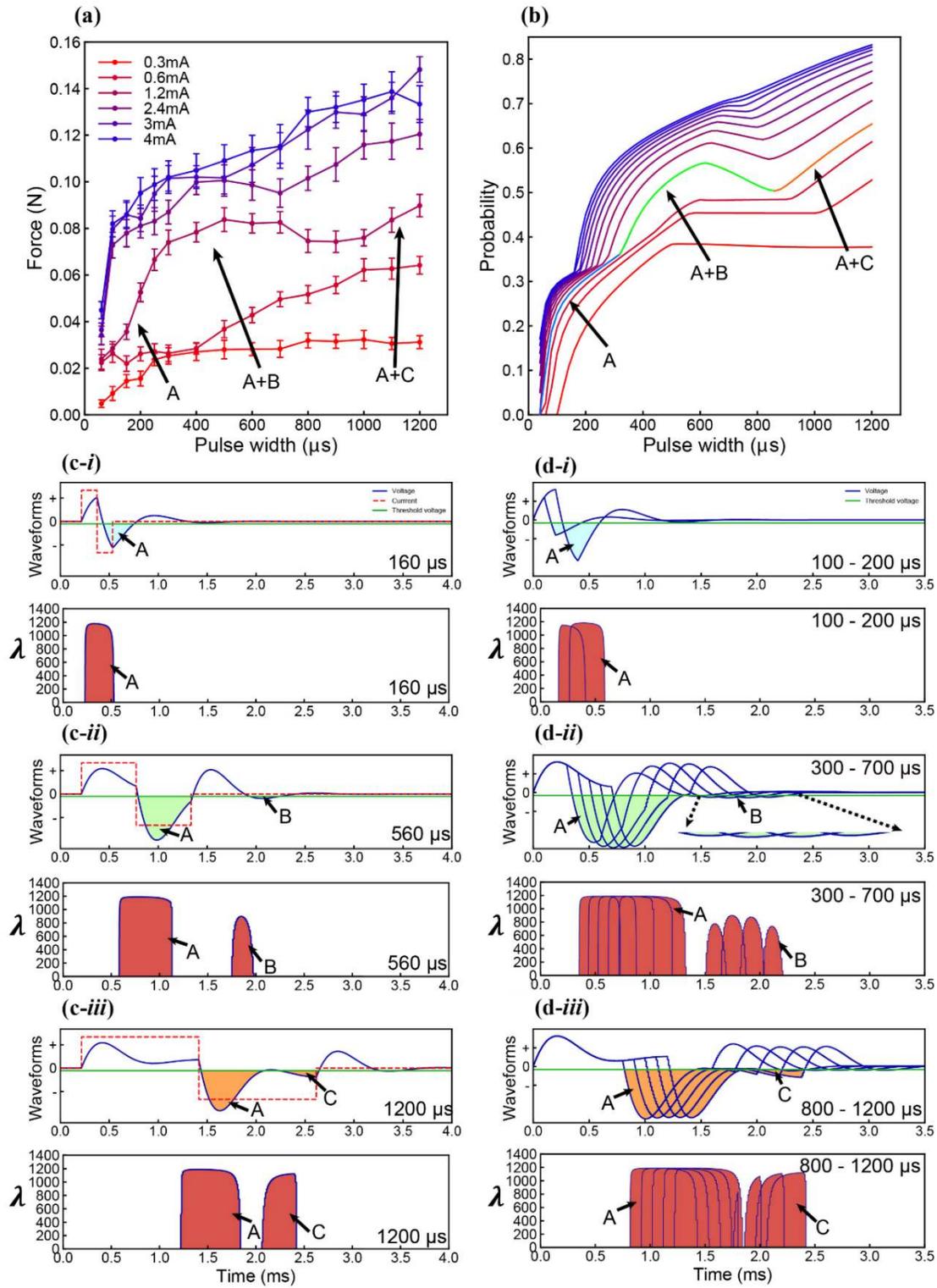

**Figure S4.1 Measurement and modeling results of the tibialis anterior (TA) muscle stimulation by positive-first biphasic square waveform current. TA muscle contraction force was measured by attaching a rat hindlimb to a force dynamometer. (a)** Measured force of the TA muscle stimulation against pulse width, with different stimulating current amplitudes (force mapping). The pulse width refers to the SPPW; **(b)** Probability calculation across pulse width when using different current amplitudes (probability mapping); the colored curve, showing distinctive frequency dependent effect, is divided into three regions: region A, region A+B and region A+C; **(c-i)** to **(c-iii)** Compares the change in the resultant voltage waveform shape and $\lambda$ curve with increasing pulse duration. When a constant-current positive-first biphasic square waveform of different pulse widths was applied (red dashed line), the resultant Voltage waveform, $V$ (blue solid line), is the tissue/circuit response to the current. Pulse duration periods result in different voltage response shapes within an RLC circuit; **(d-i)** to **(d-iii)** shows the changing trend of $V$ and corresponding $\lambda$ at different pulse width.

In this section, the C-P theory is used to explain the force mapping results from Tibialis Anterior (TA) muscle stimulation. TA muscles were stimulated with positive-first biphasic waves by varying amplitudes and SPPW (single phase pulse width) using our electrode. The resultant force was recorded and compared to the modeling results of the C-P theory (Figure S4.1(a) to (b)). The detailed testing setup and configuration of the electrode used can be found in **Supplement S3.4**.

In Figure S4.1 (a), the nonlinear shape of the force mapping curve changes with current amplitude (similar to Figure 1(c)). The 0.3 mA curve shows force saturation above 300μs SPPW. The 0.6 mA curve shows no force increase until 400 μs. However, the 1.2 mA curve shows a pulse width dependent response between 150 μm and 800 μm (Region A+B). This pulse width dependent response gradually fades away with increasing current amplitude (2.4 mA, 3 mA and 4 mA curves).

From the curves in Figure S4.1 (a), a set of parameters for both the circuit and the probability calculus were iteratively determined to generate a probability mapping (Figure S4.1 (b)). The details of these parameters can be found in Table 1- S4.1(b). These parameters were found by retroactively matching the resultant probability mapping of different input amplitudes to the measured force curves with different current amplitudes (0.3-4 mA). The probability mapping estimated from the C-P theory fits well with the force mapping, predicting a pulse width dependent response in Region A+B.

However, for the curves with low current (0.3 mA and 0.6 mA in Figure S4.1(a)), the probability mapping curves (the lowest three curves in Figure S4.1(b)) do not correlate well with the force mapping curve. This suggests that muscle fiber groups stimulated at low and high current amplitudes may not be the same, requiring different sets of circuit parameters for modeling. A combined probability mapping and parameter determination of these two groups of muscle fibers can also be found in Figure S4.2. Nevertheless, in this work we mainly focus on the force curves when the current is higher than 1.2 mA to explain how the pulse width dependent response occurs and disappears by increasing the current amplitude.

Assuming a fixed set of parameters in the C-P theory, it is estimated that distinct voltage response waveforms will be formed at a specific SPPW. In Figure S4.1 (c), we show three voltage waveforms ($V$, blue solid lines) synchronized with their input current (dashed red line). Based on the circuit parameters detailed in Table 1-S4.1(b), we estimate the voltage across the membrane capacitor ($C_{Membrane}$) to reproduce these curves. These parameters predict three distinct voltage responses (Figure S4.1 (c(i)-(iii))) at three different pulse widths. When the pulse width is 100 μs (Figure S4.1 (c-i)), there is only one region of $V$, which is lower than the threshold voltage ($V_{Threshold}$), indicated as region A. By increasing the SPPW to 500 μs, the waveform of $V$ will change as shown in Figure S4.1 (c-ii). Firstly, the area of region A becomes larger. Furthermore, the RLC circuit of the C-P theory predicts that a second region exceeding $V_{Threshold}$ will occur, which is indicated as region B. This region B occurs after the input current waveform ceases due to the voltage oscillations, which is a result of the inductance in the C-P theory. For the SPPW of 1200 μs, region B will disappear as shown in Figure S4.1 (c-iii). However, within the duration of the stimulating current waveform, a new region exceeding $V_{Threshold}$ will appear, indicated as Region C. It has to be emphasized here that region A and region C are not connected in this case.

Figure 4.1(d) shows the corresponding modeled voltage response curves as shown in Figure S4.1 (c), but with smaller pulse width steps to better show the distinct changing trends in the activating regions A, B and C. These three regions have their own different trends (Figure S4.1 (d-i) to Figure S4.1 (d-iii)). Region A increases to saturation with increasing SPPW from 100 µs to 200 µs. Region B exists only within the pulse width range of 300 µs to 700 µs, as shown in Figure S4.1 (d-ii). Its area first increases and then decreases, in a pulse dependent manner. Thus, the corresponding probability curve also shows a pulse width dependent response from 300 µs to 700 µs. The region C only appears when the pulse width is higher than 800 µs, as shown in Figure S4.1 (d-iii). This region will continue to increase with increasing pulse width. As explained in Figure 1, a voltage waveform can be replotted as a $\lambda$ figure. So, here for a better illustration, the corresponding $\lambda$ figures are also added below the voltage figures. The changing trend of each region is shown more clearly.

With a fixed $V_{Threshold}$, not every mapping curve produces all three regions with increasing pulse width. In the force mapping data, all three regions can only be observed in the curve of 1.2 mA in Figure S4.1 (a). The corresponding probability mapping curve (Figure S4.1 (b)) is represented by the multi-colored curve, with blue denoting region A, green denoting region A+B and orange denoting region C. This curve represents a predicted stimulation amplitude that would have a pulse width dependent response, resulting in the three different regions shown in Figure S4.1 (b). For the blue section, only region A contributes to the stimulation, so the probability has a monotonically increasing trend (Figure S4.1 (d-i)). For the range from 300 µs to 700 µs, regions A and B contribute to the stimulation. Region A is almost constant while region B shows a pulse width dependent response trend (Figure S4.1 (d-ii)). Hence the probability mapping curve also shows a significant pulse width dependent change, which is the green section. When the pulse width is greater than 700 µs, region B is almost zero, but region C begins to take effect (Figure S4.1 (d-iii)). So the probability mapping curve can further increase after 700 µs, which is the orange section. This colored curve shows the individual effects induced by region B and region C because these two regions do not appear simultaneously. The sections of A+B and A+C are not overlapped. However, with an increasing current amplitude, the range of pulse width affected by region B and region C both broaden and partially overlap with each other. Therefore, the pulse width dependent response induced by region B will gradually fade away and the whole curve only shows a monotonically increasing trend at the high current amplitude.

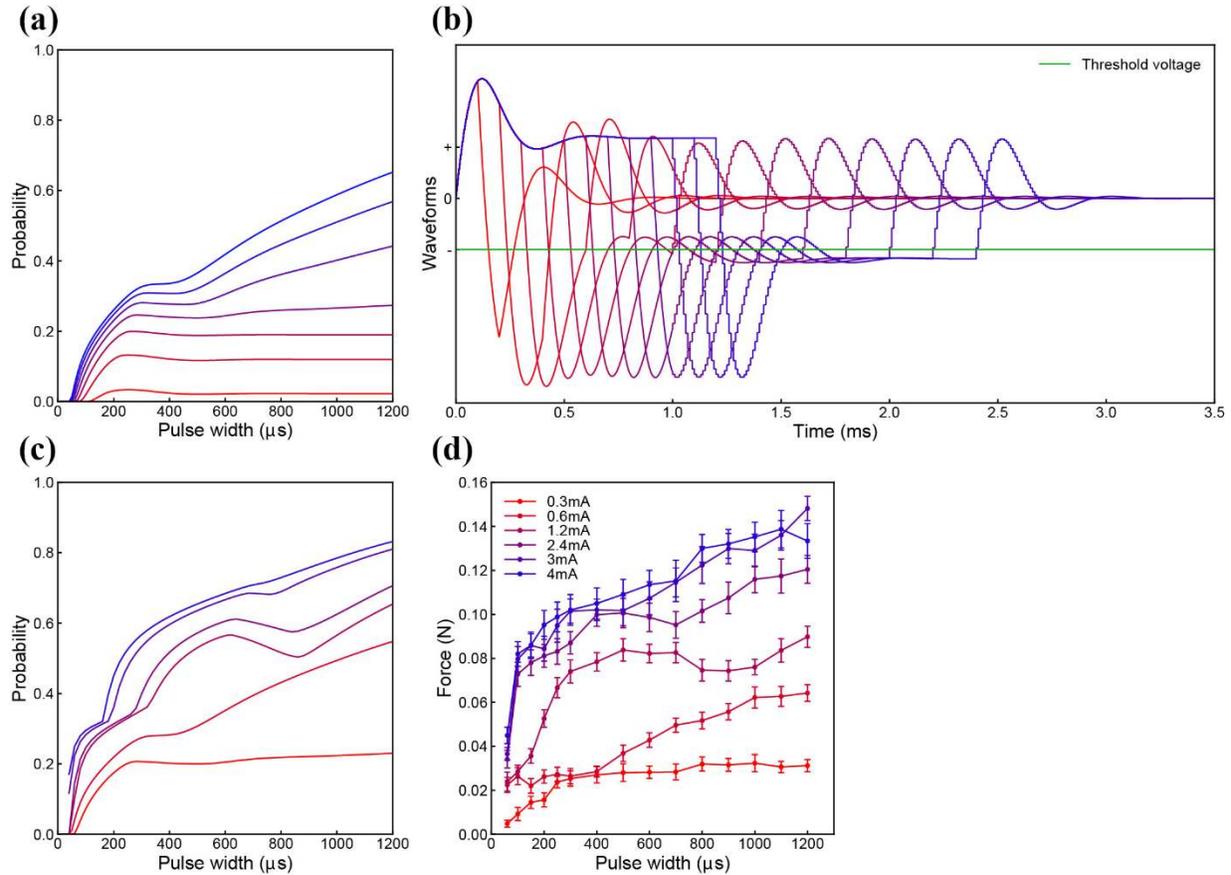

Figure S4.2. Combined modeling result of TA muscle stimulation. (a) the probability mapping of the second group of muslce; (b) The corresponding voltage waveform of (a); (c) The combined probability mapping of two group of muscles; (d) The force mapping result of TA muscle stimulation.

Since the two lowest curves in Figure S4.1(b) has a relatively large mismatch with the counterparts in the force mapping data, another set of the parameters is required to fit these two curves. It means that with different current amplitude, different number of muscle group can be stimulated, which is already reported by previous study [4]. Based on the pattern shown by these two curves, a detailed probability mapping result with the same pattern is shown in Figure S4.2 (a). The corresponding voltage waveform is shown in Figure S4.2 (b). The detailed modeling parameters can be found in Table 1- S4.2 (a). A combined probability mapping result is shown in Figure S4.2 (c). The lowest two curves are from the second muscle group in Figure S4.2 (a) and other curves are from Figure S4.1 (b). The force mapping data is shown in Figure S4.2 (d) for comparison with the probability mapping result. The whole changing trend of the force mapping curves can be well reproduced by the modeled probability mapping.

## S5 Various probability mapping pattern by biphasic square current waveform

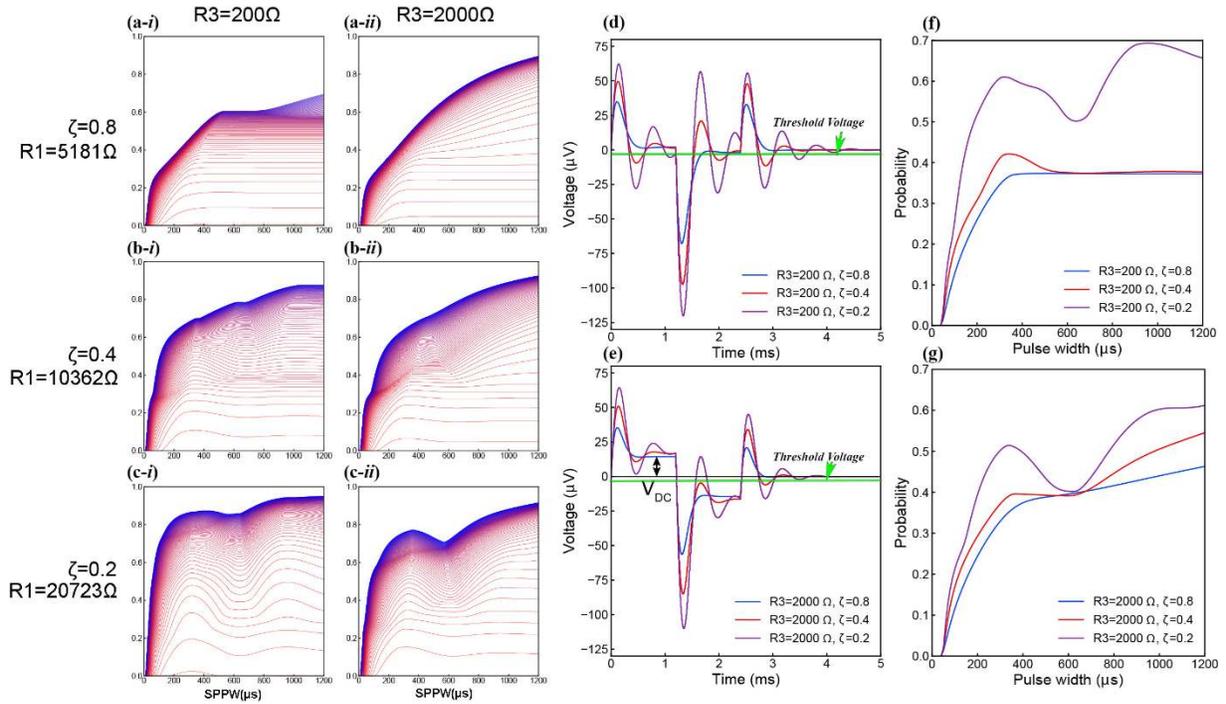

**Figure S5. A general illustration of the probability mapping of positive-first biphasic square waveform current stimulation generated by different circuit parameters.** **(a)** to **(c)** show the probability mapping when varying the damping factor $\zeta$ from 0.8 to 0.2, showing a stronger resonance effect with a lower $\zeta$, with (*i*) the left figures modeled when $R_3 = 200\,\Omega$, and (*ii*) the right figures modeled when $R_3 = 2000\,\Omega$; **(d&e)** Illustrative voltage waveforms comparison of SPPW=1200 µs when the damping factor $\zeta$ from 0.8 to 0.2. **(d)** shows three voltage waveforms when $R_3 = 200\,\Omega$; (e) shows three voltage waveforms when $R_3 = 2000\,\Omega$. A lower $\zeta$ results in a higher voltage amplitude with a stronger oscillation. A higher $R_3$ results in a higher $V_{DC}$; **(f&g)** Illustrative probability curves with the same current selected from **(a)** to **(c)**. A lower $\zeta$ will result in a higher probability curve with stronger resonance effect. A higher $R_3$ will result in an increasing trend at high SPPW range.

As biphasic square wave is the most commonly used current waveform in neural tissue stimulations, it will be used in this illustration.

Firstly, we modify the circuit by changing parameters $R_1$ and $R_3$ from the lumped parameter circuit (Figure 2(c)) for comparison (Figure S5 (a-c)). The probability mapping pattern tends to show a more distinctive resonance effect by increasing $R_1$ (Figure S5 (a-i) to S5 (c-i)). Meanwhile, at the high SPPW range, the probability mapping curves tend to have an increasing trend with a higher $R_3$, and keep constant with a lower $R_3$ (from Figure S5 (i) to (ii)).

Before explaining the details of the illustration in Figure S5, two key index parameters need to be introduced first.

The first one is the damping factor of a parallel RLC circuit:

$$\zeta \approx \frac{1}{2R_1}\sqrt{\frac{L}{C}}$$

This is an approximation form of $\zeta$ when $R_2$ and $R_3$ are negligible. The effect of $\zeta$ on the voltage upon the capacitor $C$ is shown in Figure 6(d). When $\zeta$ is low, the voltage amplitude and oscillation are both higher, and vice versa. Meanwhile, the quality factor of the parallel RLC circuit is

$$Q \approx R_1 \sqrt{\frac{C}{L}} = \frac{1}{2\zeta}$$

This $Q$ will determine the resonance effect in the probability mapping. When $\zeta$ is low, which means $Q$ is high, the resonance effect will be more distinctive, and vice versa.

In these equations, $L$ and $C$ are mainly determined by the resonance frequency, which typically can be roughly estimated from the force mapping data. Thus, $\zeta$ can be directly tuned by changing $R_1$ in the model to fit the force mapping curve. However, when $R_2$ and $R_3$ cannot be neglected, both $Q$ and $\zeta$ cannot be written as analytical formulae. But these two approximation equations still can be used to estimate the scale of $R_1$.

Another index parameter is the DC voltage, $V_{DC}$:

$$V_{DC} = I_A \times (R_3 \,/\!/\, R_1) = I_A \times \frac{R_3 \times R_1}{R_3 + R_1}$$

$I_A$ is the amplitude of the square wave current. Typically, $R_3 \ll R_1$, so this equation can be simplified as:

$$V_{DC} \approx I_A \times R_3$$

This is the voltage upon the capacitor when a DC current is applied. In Figure S5 (f), all three voltage curves will finally approach this $V_{DC}$. This $V_{DC}$ is small in Figure S5 (d) because $R_3$ is low. Since $\zeta$ is mainly determined by $R_1$, here $V_{DC}$ is mainly determined by $R_3$.

Assuming $L$ and $C$ are already obtained from the estimated resonance frequency observed in the force mapping results, then these two index parameters, $\zeta$ and $V_{DC}$, are mainly determined by $R_1$ and $R_3$. In this comparison as shown in Figure S5 (a)-(c), only $R_1$ and $R_3$ are changed, the rest circuit parameters and probability calculus parameters are all kept constant. The detailed modeling parameters of each case can be found in Table 1-S5(a-c).

The probability mapping patterns of different $\zeta$ are compared first (from Figure S5 (a-i) to S5 (c-i)), assuming $R_3$=200 Ω. The resonance effect becomes more significant with the decreasing $\zeta$. This is because a lower $\zeta$ can induce a higher $Q$, resulting in stronger resonance of the voltage waveform. Also, in Figure S5 (c-i), there are two resonance peaks, at the SPPW set as $SPPW_{P_1}$ and $SPPW_{P_2}$, calculated as:

$$SPPW_{P_1} \approx \frac{1}{2} \times \frac{1}{f}$$

$$SPPW_{P_2} \approx \frac{3}{2} \times \frac{1}{f}$$

$f$ is the resonance frequency set in the modeling. In this comparison, $f$ is set as 1600Hz, so $SPPW_{P_1} \approx$ 312.5 μm and $SPPW_{P_2} \approx$ 937.5 μm (Figure S5 (c-i)). If the SPPW is long enough or $f$ is high enough, there can be more resonance peaks. A general equation for the $n$th resonance peak is:

$$SPPW_{P_n} \approx \frac{2 \times n - 1}{2} \times \frac{1}{f}$$

When the force mapping curve has multiple resonance peaks, the above equation can better estimate the resonance frequency. This multiple resonance peak effect can be seen in the cortical stimulation in **Supplementary S6.5**. In that case, three resonance peaks can be observed, indicating a small $\zeta$, which means a large $R_1$ and a higher $f$.

Next, $R_3$ is changed from 200 Ω to 2000 Ω, causing the pattern changes as seen from (i) to (ii) in Figure S5 (a-c).

There are two major differences between (i) and (ii). Firstly, the resonance effect is less distinctive when $R_3$ is higher because of a higher $\zeta$. In fact, $\zeta$ is affected by every resistor in the RLC circuit. It increases with any increase in the value of parallel resistors, which are $R_2$ and $R_3$, and decreases with any increase in the value of the serial resistor, which is $R_1$.

Secondly, a higher $R_3$ results in a higher $V_{DC}$ (Figure S5 (d) and (e)), lowering the current amplitude required to make $V_{DC}$ exceed $V_{Threshold}$. Once $V_{DC}$ exceeds $V_{Threshold}$, the effective voltage area can increase with SPPW, resulting in a monotonically increasing trend with SPPW in probability mapping curves. In the comparison between (i) and (ii), at high SPPW range, the curves tend to keep constant in (i) but tend to further increase in (ii).

The probability mapping curves of the same current in Figure S5 (a-c) are plotted in Figure S5 (f) and (g) for a more detailed comparison. In both Figure S5 (f) and (g), a lower damping, which means a higher $R_1$, not only induces a more distinctive resonance effect, but also leads to a higher probability of action potential firing. For example, in sciatic nerve stimulation, where the neural probe is wrapped around the nerve, some current will flow through the outer surface of the nerve. When the nerve surface is wet due to saline solution or tissue fluid, $R_1$ will decrease, resulting in an immediate drop in evoked activity for the same current.

It is easy to calculate the probability mapping with a specific set of parameters. But the inverse process, deriving a specific set of parameters from the mapping curves, cannot be easily realized. In real modeling, we need to derive both the circuit and the probability calculus parameters just from the force mapping curves. For the curves shown in Figure S5 (c-i), there are distinctive resonance peaks and clear curve shapes that change with low current to high current stimulation; this observation provides clues to capturing the modeling parameters. Fitting the circuit model is challenging for the cases of Figure S5 (a) and (b). Furthermore, when very high or low stimulus currents are applied, the force mapping curves generated are quite similar. Only specific currents can generate curves that can provide information about the resonance frequency, damping factor and DC voltage. When such curves are missing from testing, which frequently happens in our measurement experiments, then we can only apply exhaustive method to find the correct modeling parameters.

**S6 Experiment data**
**S6.1 Detailed force mapping curve and voltage waveform of sinewave test on CP nerve**

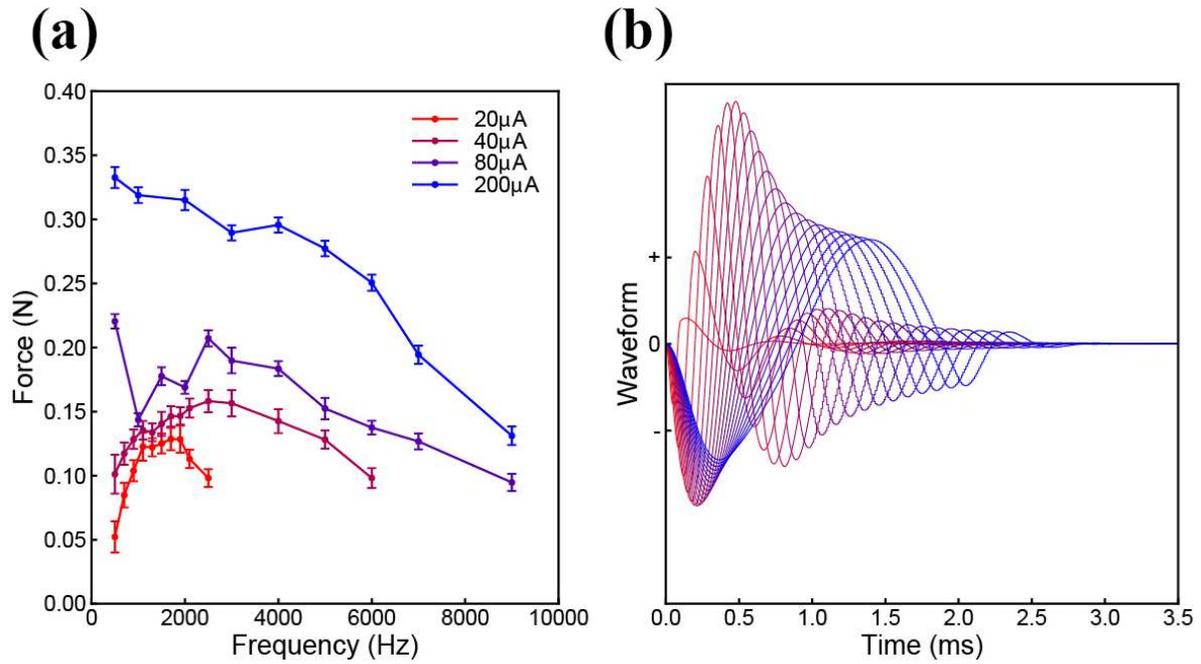

Figure S6.1 (a) Force mapping result of sinewave test on CP nerve with error bar; (b) The corresponding voltage waveform of sinewave current with same amplitude but different pulse width.

## S6.2 Stimulus artifact data

### S6.2.1 Experimental measurement and modeling of the stimulus artifact of pelvic nerve and cortical stimulation with different current waveforms and SPPW

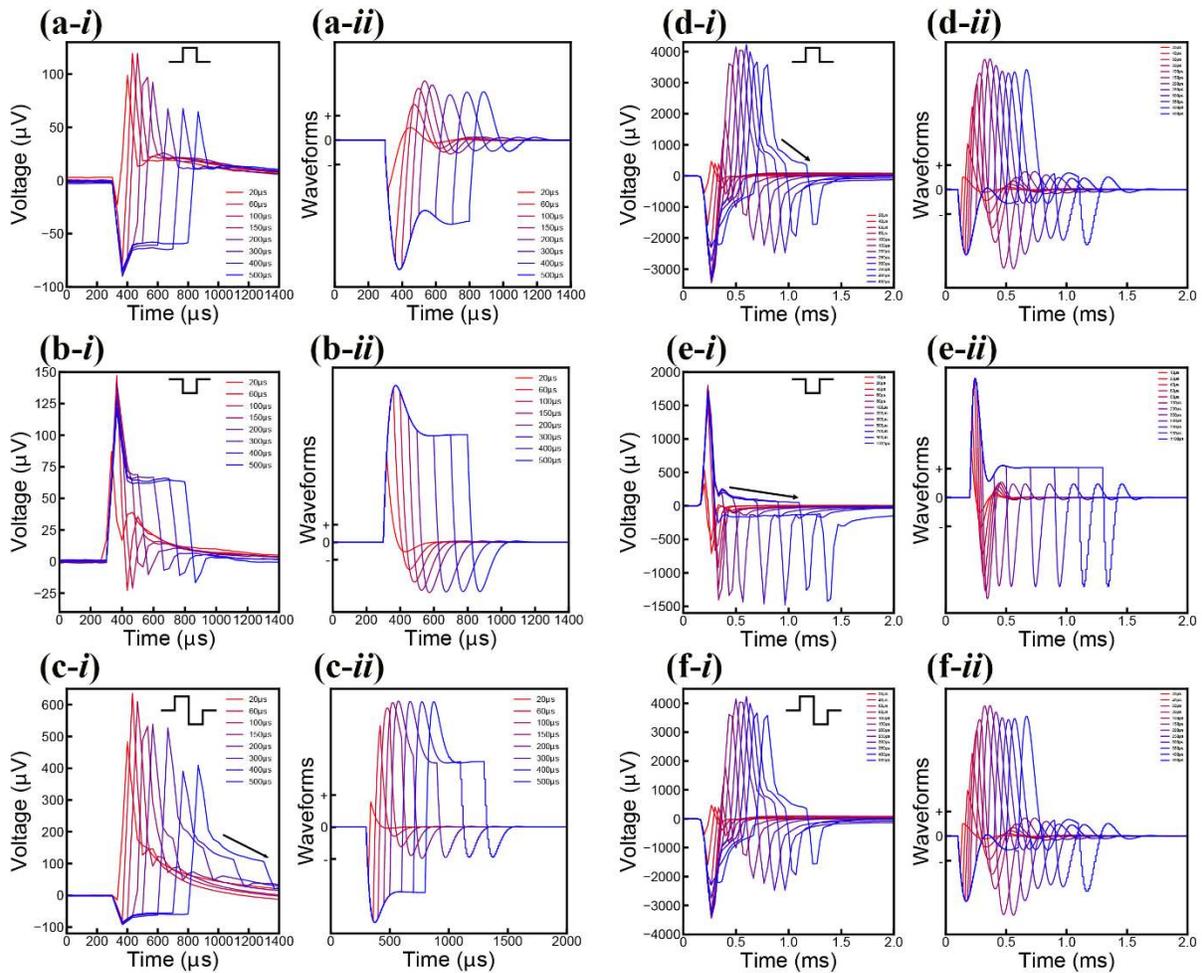

**Figure S6.2.1 Experimental measurement and modeling (notation *-i* and *–ii*, respectively) of the stimulus artifact of pelvic nerve (a to c) and cortical stimulation (d to f) with different current waveforms and SPPW. (a-c)** Pelvic nerve stimulation and modeling results: **(a)** positive monophasic square wave, **(b)** negative monophasic square wave and **(c)** positive-first biphasic square wave; **(d-f)** Cortical stimulation and modeling results: **(d)** positive monophasic square wave, **(e)** negative monophasic square wave and **(f)** positive-first biphasic square wave; **(i)** left figures refer to the measured data, **(ii)** right figures refer to the modeling results. The modeling results match well with the measurement data, validating the parallel RLC circuit used in this study.

The neural tissue is modeled as a circuit shown in Figure S2.1(a), and the resulting voltage response, which is normally considered as a stimulus artifact, can be measured from the recording electrodes. The profile of the voltage response recorded can be accounted for with the choice of the correct circuit, where it represents a fraction of the voltage delivered from the stimulator. Here we will use the most simplified parallel RLC circuit as shown in Figure S2.2.1 for modeling and use the voltage upon $C$ to represent the voltage response of the tissue.

Experiments used positive monophasic square waveforms, negative monophasic square waveforms and positive-first biphasic square waveforms as commonly used in neural stimulations, to generate a voltage response. The voltage responses were measured in two additional kinds of tissue models, in pelvic nerve stimulation/recording and cortical tissue stimulation/sciatic nerve recording. The detailed testing procedure, setup and current amplitude can be found in the **Supplement S3.2** and **S3.3**.

All pulse widths refer to the single phase pulse width (SPPW) in both the monophasic and biphasic waveforms. Figure S6.2.1 (a)-(f) shows the stimulus artifact recorded for each type of applied square waveform. For each current waveform, graphs show multiple overlapping curves corresponding to increasing pulse width but constant amplitude (starting range: 10-20 µs; maximum range of 500 µs or 1100 µs; range depends on the condition of the subject and the depth of anesthesia) to clearly show how the waveform changes (all (*i*) in Figure S6.2.1). Corresponding modeling results are shown in (*ii*) figures. All the modeling parameters can be found in Table 1- S6.2.1 (a-f).

As shown in Figure 3, the modeling from the circuit of Figure S2.2.1 can generally reproduce the shapes of the voltage response. However, in Figure S6.2.1 (c), (d) and (e), the voltage waveforms show some mismatch with the testing results. All curves appear similar to an RC discharging curve overlapped with a parallel RLC circuit response. The voltage curve gradually drops at the position (indicated by black arrows), whereas the voltage is predicted to keep constant in the modeling. This indicates that an additional RC discharge effect should be considered to improve the precision of the circuit. Detailed data of stimulus artifact in other experiments can be found in **Supplement S6.2.2 and S6.2.3**. In summary, these data further validates the existence of a parallel RLC circuit response, and shows that the circuit simplification used in our C-P theory reproduces experimental results with sufficient accuracy.

## S6.2.2 Stimulus artifact of cortical stimulation

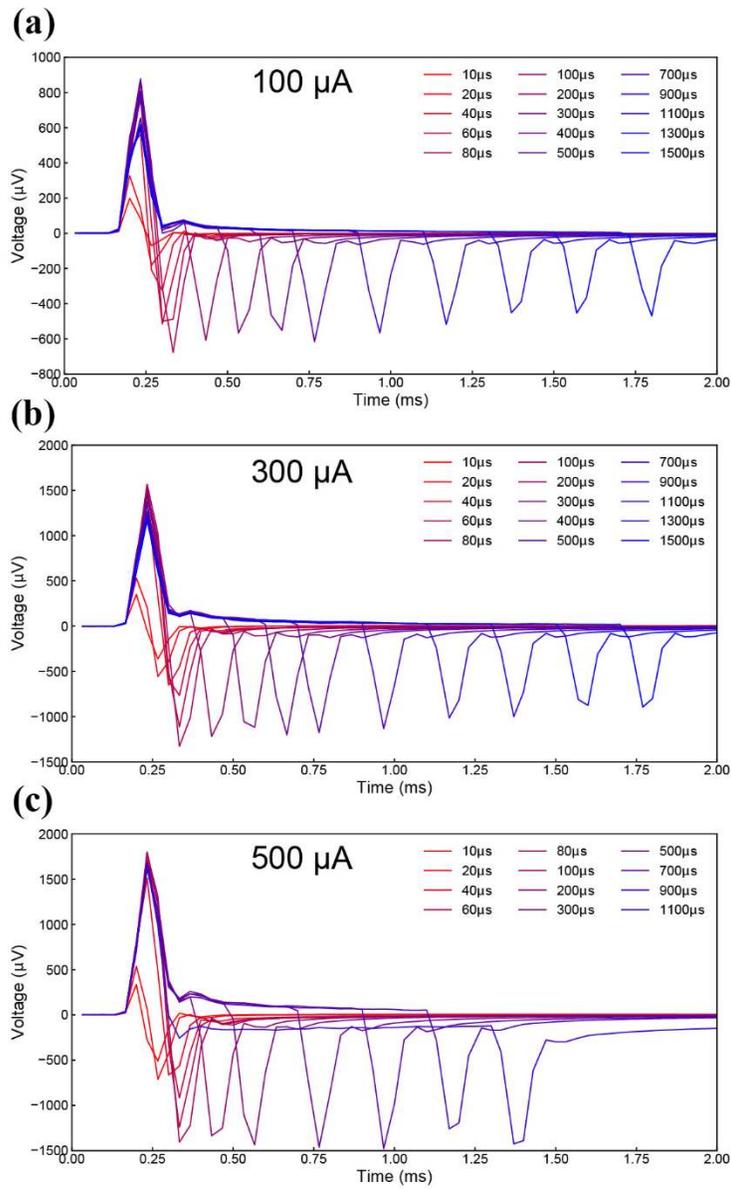

Figure S6.2.2.1 Stimulus artifact of cortical stimulation with different current amplitude of negative monophasic square waveform. The SPPW range is from 10 µs to 1500 µs.

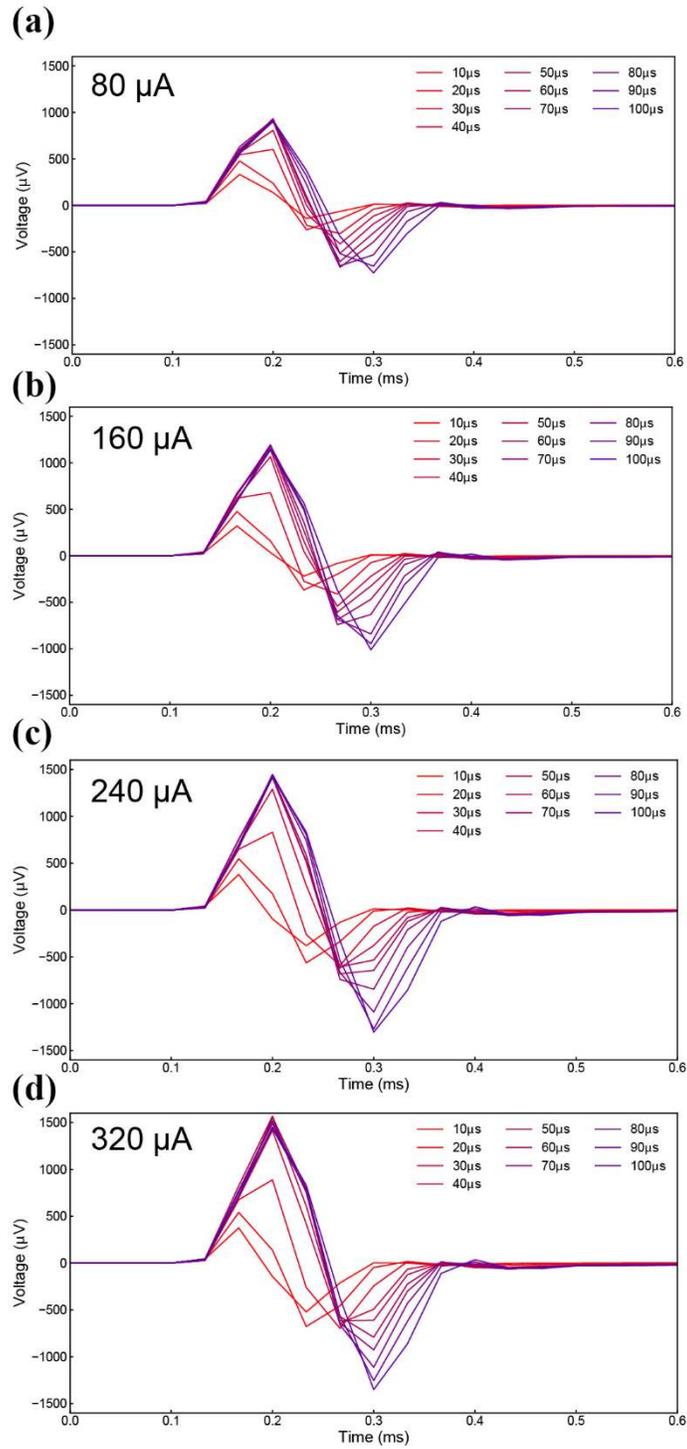

Figure S6.2.2.2 Stimulus artifact of cortical stimulation with different current amplitude of negative monophasic waveform. The SPPW range is from 10 µs to 100 µs.

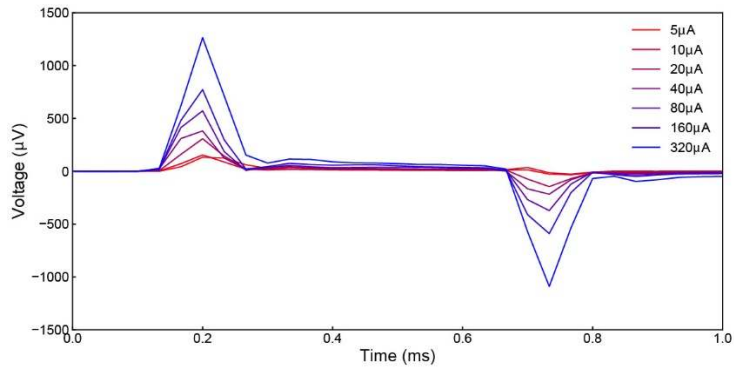

Figure S6.2.2.3 Stimulus artifact of cortical stimulation with negative monophasic waveform current of 500 µs SPPW. The current is from 5 µA to 320 µA.

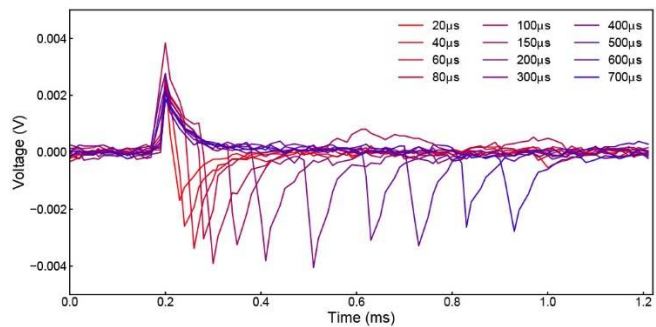

Figure S6.2.2.4. High sampling rate system (70 kHz) recording of stimulus artifact of cortical stimulation with negative monophasic waveform current of 100 µA. The SPPW is from 20 µs to 700 µs.

Figure S6.2.2.1 to Figure 6.2.2.4 show other testing data of stimulus artifacts for cortical stimulation.

In Figure S6.2.2.1, different amplitudes of negative monophasic square waveform current were applied. The shape and the changing trend of the artifacts follow the same pattern as shown in Figure 6.2.1(e). The current amplitude has no effect upon the waveform of the stimulus artifact. The similar test with a lower SPPW range, from 10 µs to 100 µs, is shown in Figure S6.2.2.2. A more detailed test with different amplitudes but constant SPPW is shown in Figure S6.2.2.3. The current amplitude has no effect upon the shape of the stimulus artifacts.

Since the sampling rate of our recoding system is 30 kHz, which may cause large distortion of the recording of high frequency signal. Thus the same stimulus artifact is recorded with another recording system with higher sampling rate, which is 70 kHz, as shown in Figure S6.2.2.4. The recording result shows the same pattern as the data of 30 kHz sampling rate system, only with more noise.

These results affirm the repeatability of the stimulus artifact recording in our tests.

## S6.2.3 Stimulus artifact of pelvic stimulation

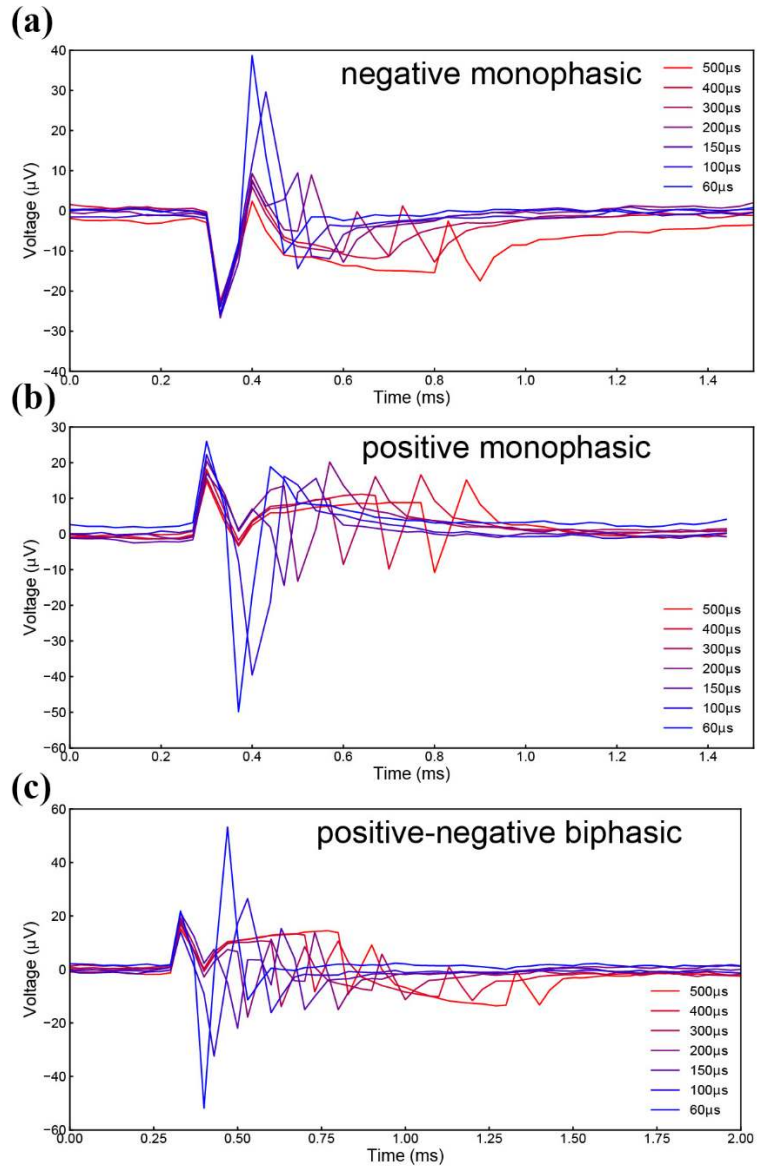

Figure S6.2.3. Stimulus artifact of pelvic stimulation with different current waveform of 70 µA; (a) Positive monophasic square waveform; (b) Negative monophasic square waveform; (c) Positive-first biphasic square waveform.

Another set of stimulus artifact recording data of pelvic nerve stimulation is shown in Figure S6.2.3. This set of data shows a very high resonance frequency, indicating that the recording system may not be able to capturing the waveform of the signal well. So we do not try fitting it by modeling. But it still shows a clear voltage response of the RLC circuit. The voltage amplitude reaches maximum at a certain pulse width, which is 60 µs in this case.

## S6.3 Experiments of TA muscle stimulation
### S6.3.1 Four waveforms comparison of TA muscle stimulation

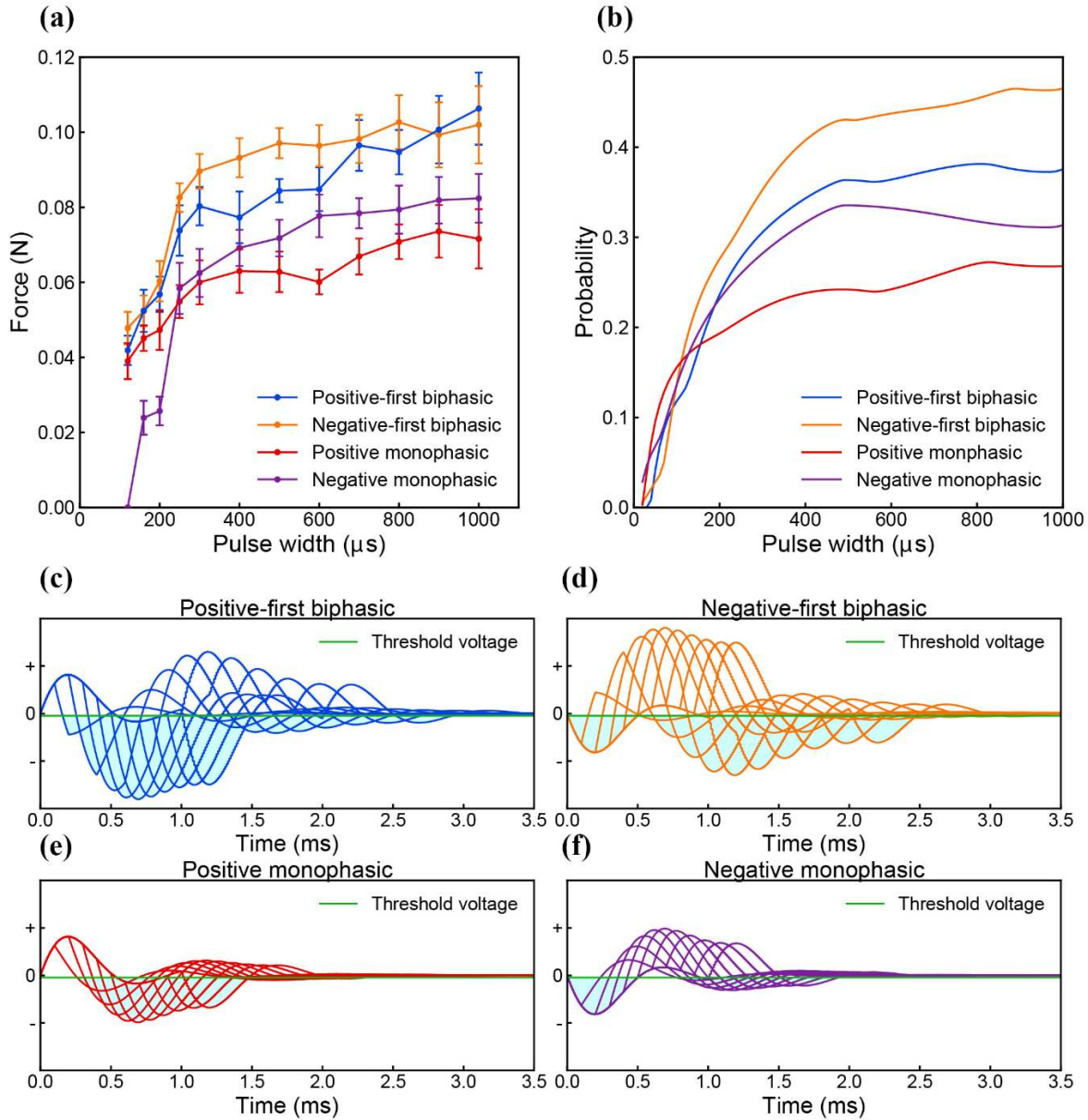

Figure S6.3.1.1 Four waveforms comparison of TA muscle stimulation with the same current amplitude. (a) Force mapping results; (b) Probability mapping results; (c)-(f) Voltage waveforms of four different current waveforms: (c) Positive-first biphasic square waveform; (d) Negative-first biphasic square waveform; (e) Positive monophasic square waveform; (f) Negative monophasic square waveform.

It is widely reported that negative current will be more effective for nerve stimulation because the depolarization of the ion channel is gated by negative voltage. However, the difference of the stimulation result between biphasic and monophasic square waveform cannot be well explained. Since in the C-P theory the probability mapping curve is directly determined by the voltage waveform, current of different square waveforms will generate different probability mapping curves. With the correct circuit and probability calculus parameters, the probability mapping curves of different current waveforms of the same amplitude can be modeled.

To validate this prediction, a force mapping comparison of different current waveforms upon TA muscle were conducted. Four typical current waveforms of the same amplitude, which is 180 µA, were used. The force mapping results are shown in Figure S6.3.1.1 (a). These four current waveforms show completely different efficiency for muscle stimulation even though they share the same current amplitude. Since the shape of these four curves do not reveal very clear circuit parameters such as resonance frequency and damping factor, just as the pattern as shown in Figure S5(b-i), exhaustive method was applied to capture the modeling parameters (Table 1- S6.3.1.1 (b)). The modeling results are shown in Figure S6.3.1.1 (b). The corresponding voltage waveforms of these four types of square wave current are shown in Figure S6.3.1.1 (c) to (f). It should be noted that the voltage waveforms of Figure S6.3.1.1 (c) and (d) are just of opposite polarity since they share the same current waveform with opposite polarity. Figure S6.3.1.1 (e) and (f) are of the same situation. However, the positive and negative phase of the whole voltage waveform are asymmetrical, making the effective voltage area for probability calculus form different shapes, amplitudes and changing trends. Apparently, these different effective voltage areas will result in different probability mapping curves as shown in Figure S6.3.1.1 (b). The same test was conducted for three times. The testing and modeling results of another two tests are shown in Figure S6.3.1.2 and Figure S6.3.1.3. All these three tests reveal that these four types of square wave current generate force mapping curves with different shape and stimulation efficiency.

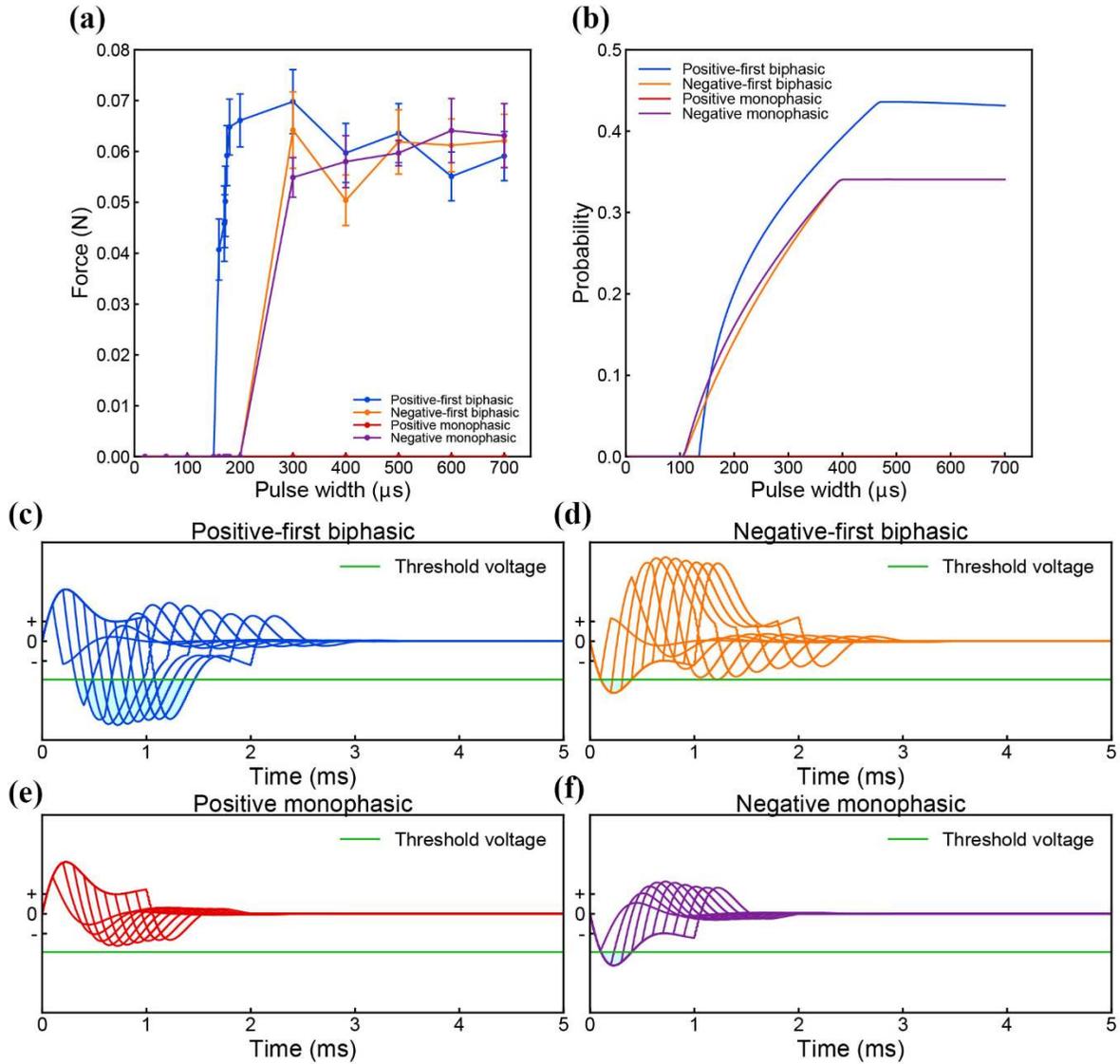

Figure S6.3.1.2 Four waveform comparison test of TA muscle stimulation with 520 µA current amplitude. (a) Force mapping data; (b) Probability mapping result; (c) Voltage waveform of positive-first biphasic square waveform current; (d) Voltage waveform of negative-first biphasic square waveform current; (e) Voltage waveform of positive monophasic square waveform current; (f) Voltage waveform of negative monophasic square waveform current;

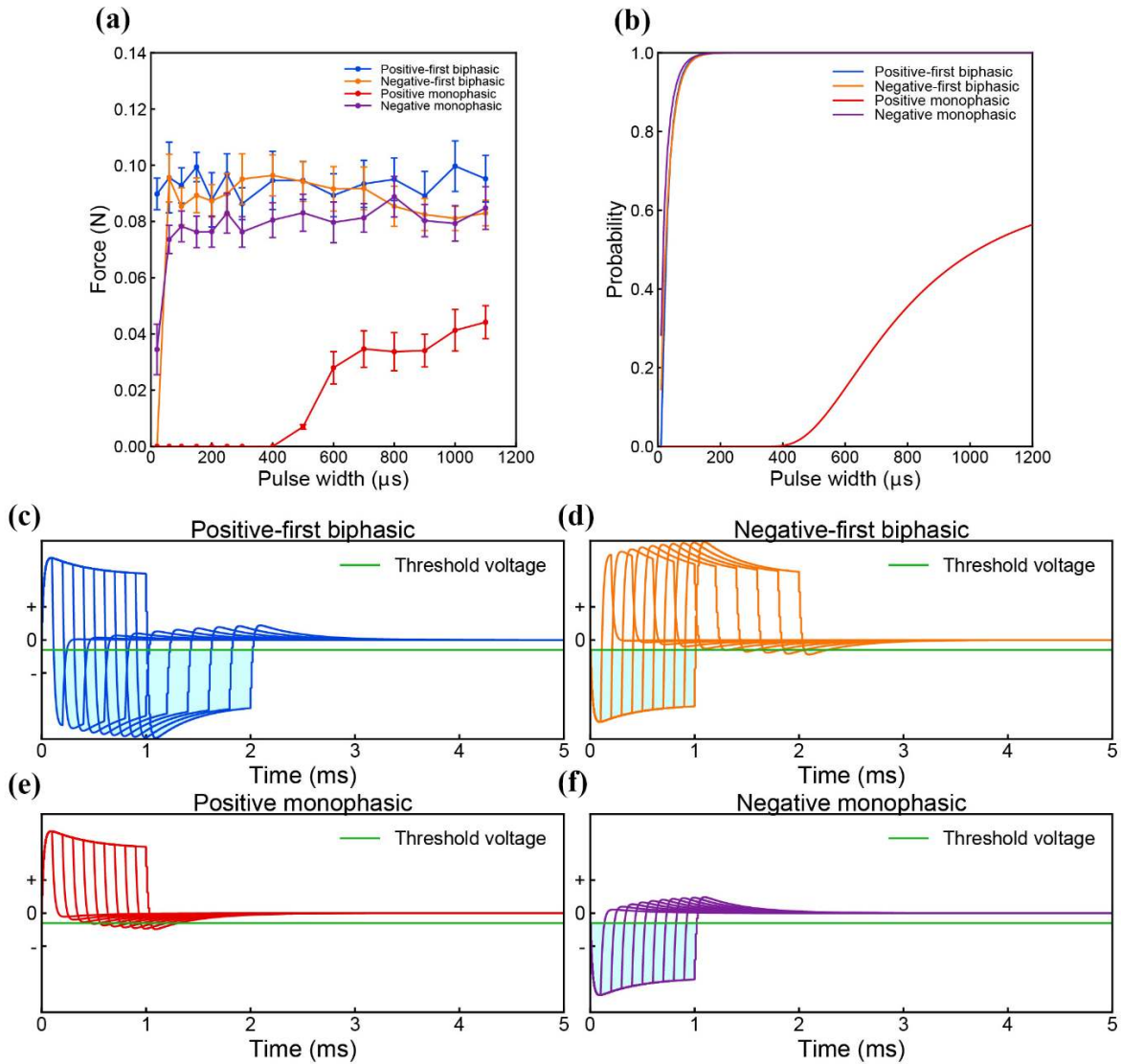

Figure S6.3.1.3 Four waveform comparison test of TA muscle stimulation with 500 µA current amplitude. (a) Force mapping data; (b) Probability mapping result; (c) Voltage waveform of positive-first biphasic square waveform current; (d) Voltage waveform of negative-first biphasic square waveform current; (e) Voltage waveform of positive monophasic square waveform current; (f) Voltage waveform of negative monophasic square waveform current;

Figure S6.3.1.2 and S6.3.1.3 show the four waveform comparison tests with 520 µA and 500 µA current amplitude in another two experiments, respectively. In Figure S6.3.1.3, apart from the positive square wave, other three waveforms show almost the same force mapping curve, so in the probability mapping result, these three curves are overlapped with each other. The detailed modeling parameter is shown in Table 1-S6.3.1.2 (b) and S6.3.1.2 (b).

## S6.4 Experiments of CP nerve stimulation
### S6.4.1 CP nerve stimulation results by negative monophasic square wave current pulse

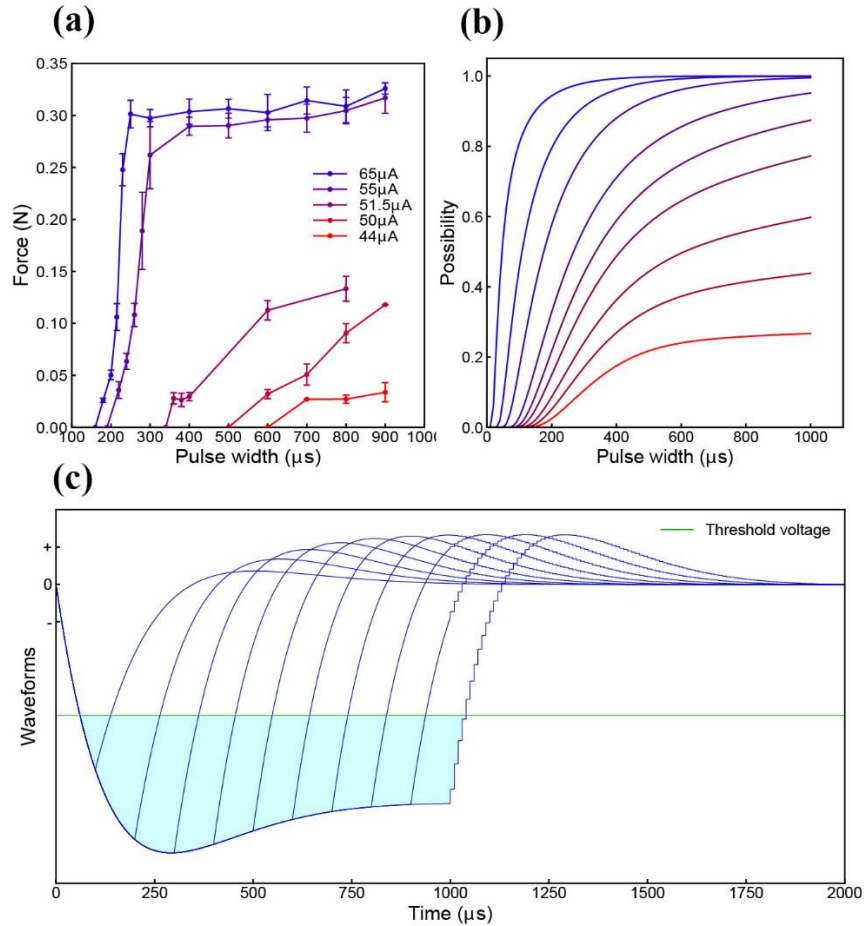

Figure S6.4.1 Measurement and modeling results of the CP nerve stimulation by negative monophasic square waveform current. (a) Force mapping results; (b) Probability mapping results; (c) Corresponding voltage waveforms.

Since we consider C-P theory as a general theory for describing the electrical activation of different neural and non-neural tissues, we also did similar tests on the CP nerve, the cortex and the pelvic nerve.

The force mapping curve of CP nerve stimulation with negative monophasic square wave current is shown in Figure S6.4.1 (a). The current is varied from 44 µA to 65 µA and the SPPW mapping range is from 100 µs to 900 µs. The force mapping pattern is quite similar as the one shown in Figure S5(a). A set of parameters (Table 1- S6.4.1 (b)) is captured for the modeling result as shown in Figure S6.4.1 (b). The resonance frequency used in this modeling is 714 Hz. All the curves show a monotonically increasing trend with SPPW, just as the force mapping curve. The corresponding voltage waveform is plotted in Figure S6.4.1 (c). In this case, $V_{DC}$ is much higher than $V_{Threshold}$, making the effective voltage area increase monotonically with SPPW.

**S6.4.2 Four waveforms comparison of CP nerve stimulation**

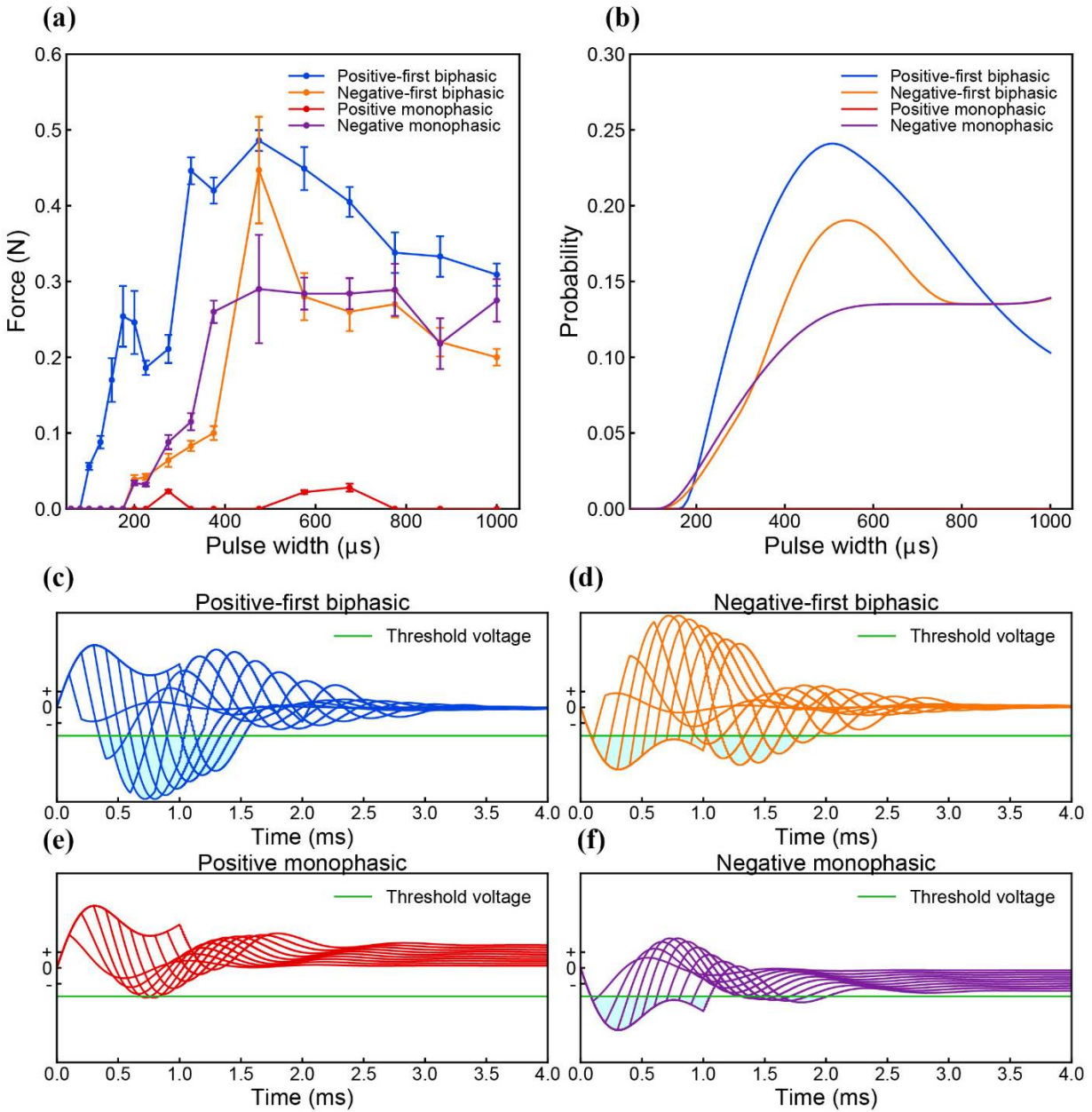

Figure S6.4.2 Four waveforms comparison of CP nerve stimulation of the same current amplitude. (a) Force mapping results; (b) Probability mapping results; (c)-(f) Voltage waveforms of four different current waveforms: (c) Positive-first biphasic square waveform; (d) Negative-first biphasic square waveform; (e) Positive monophasic square waveform; (f) Negative monophasic square waveform.

The force mapping comparison of four different square waveforms of CP nerve stimulation is shown in Figure S6.4.2 (a). The curves of positive-first biphasic and negative-first biphasic waveform all show a distinctive resonance effect at SPPW of 475 µs, indicating that the resonance frequency is 1052 Hz. Compared with the resonance frequency used in Figure S6.4.1(b), which is 714 Hz, the frequency in this test is higher. This also indicates that this resonance frequency will vary with different individuals.

The curve of monophasic negative waveform does not show a clear resonance peak and get close to or ever higher than the curve of negative-first biphasic waveform at high SPPW range. This trend is very weird because normally the biphasic waveforms should have a higher stimulation efficiency than monophasic waveforms. In Figure S6.3.1.1 (c)-(f), it is very clear that the voltage curve of biphasic waveforms should give a larger effective voltage area than that of monophasic waveforms. Even by exhaustive method, we are not able to capture a set of proper parameters to fit these curves in Figure S6.4.2 (b). So the circuit used for modeling in this test is revised based on the circuit shown in Figure S2.1(c). An additional capacitor is connected in series with the inductor. The detailed circuit structure and analysis method can be found in the **Supplementary S2.3**. The corresponding modeling result is shown in Figure S6.4.2 (b) and the detailed parameters can be found in Table 1-S6.4.2(b). The voltage waveforms of these four waveforms are shown in Figure S6.4.2 (c)-(f).

## S6.5 Experiments of cortical stimulation
### S6.5.1 Cortical stimulation results by negative monophasic square wave current pulse

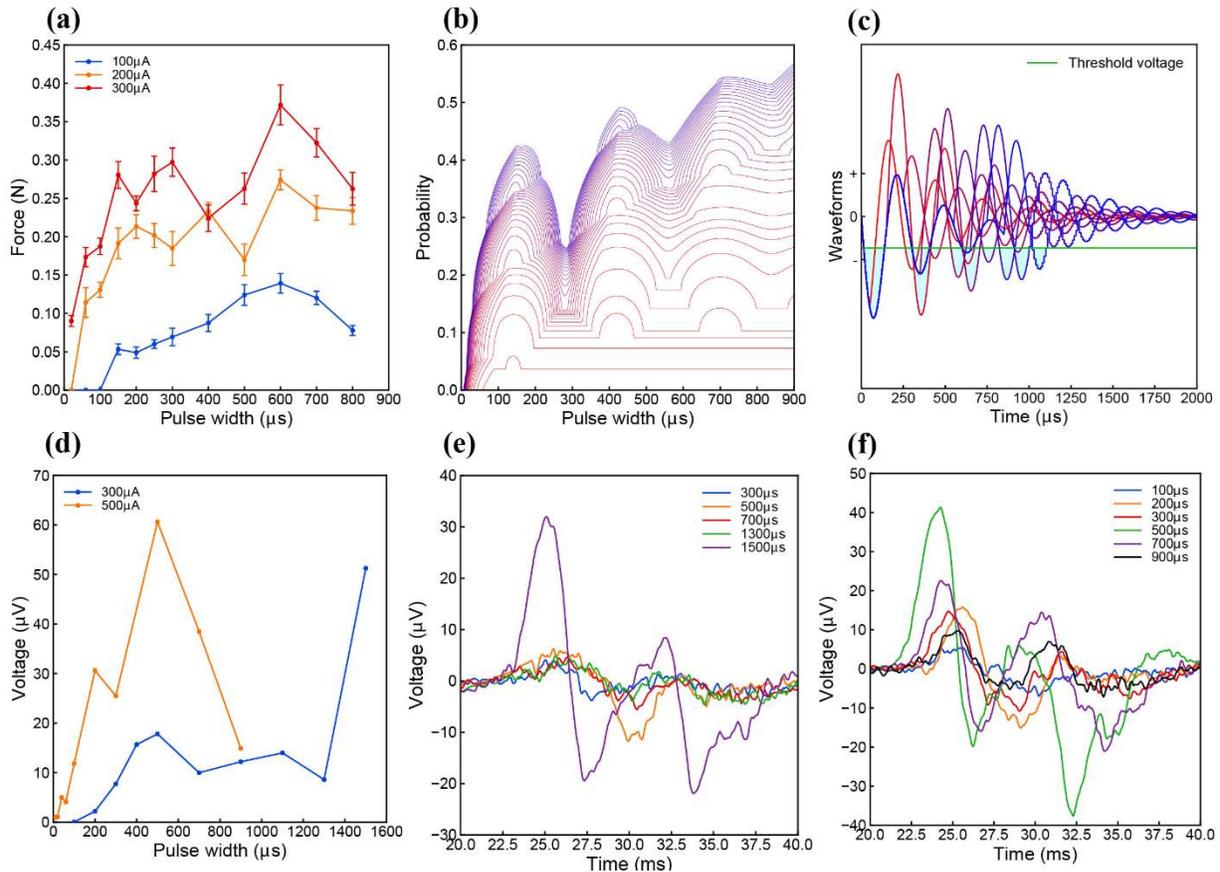

Figure S6.5.1 Measurement and modeling results of the cortical stimulation by negative monophasic square waveform current. (a) Force mapping results; (b) Probability mapping results; (c) Corresponding voltage waveforms; (d) Amplitude of EMG signal with two different current amplitudes; (e) Selected EMG waveforms of 300 µA stimulation; (f) Selected EMG waveforms of 500 µA stimulation;

This C-P theory can also be applied on cortical stimulations. The detailed testing setup and procedure can be found in the **Supplementary S3.2**. In the cortical stimulation, only negative monophasic square wave current was used for the force mapping test as shown in Figure S6.5.1 (a). The force mapping curves show three resonance peaks, which is quite similar to the situation to be discussed in Figure S5(c-i). The multiple resonance peaks indicate that the resonance frequency is higher than that of the TA muscle and the CP nerve, and the damping factor is very low. The probability modeling result is shown in Figure S6.5.1(b) and the detailed modeling parameters can be found in Table 1-S6.5.1(b). The resonance frequency captured is 3600 Hz. The corresponding voltage waveform is shown in Figure S6.5.1(c). Due to the low damping factor, the voltage waveform have a strong oscillation, showing a very complex changing trend. Other similar testing data can be found in the Figure S6.5.2. All these data will follow the similar probability mapping pattern shown in Figure S6.5.1(b).

To correlate the force mapping results, we also measured the EMG signal from the sciatic nerve by cortical stimulation with negative monophasic square wave current of 300 µA and 500 µA. Some selected EMG waveforms are shown in Figure S6.5.1(e) and (f). The complete testing data including the stimulus artifacts can be found in the Figure S6.5.3. The peak-to-peak amplitude of the EMG signal is measured and shown in Figure S6.5.1(d). The EMG signal also shows multi-peak resonance, which is similar as the force mapping results in Figure S6.5.1(a). It seems that the second resonance peak of the force mapping data is between the SPPW from 300 µs to 400 µs. However, the second resonance peak of the EMG recoding is at around 500 µs. This resonance peak shift should be induced by the individual difference.

One point to be emphasized here is that the amplitude of the EMG signal cannot fully represent the number of action potentials generated. Since in our theory the action potential is evoked based on probability, all the action potentials are not generated at the same time but with a time distribution. Especially when there are more than one effective voltage areas, there will be more than one group of action potentials. The actual measured EMG signal is the result of all individually action potentials with a complex phase combination. Even with the same quantity of action potentials generated, meaning the same force generated, the different phase combination will still affect both the amplitude and the pulse duration of the EMG signal measured in the real tests. In both Figure S6.5.1(e) and (f), the EMG waveforms of different SPPW not only differ in amplitude but also in pulse with and shape. So currently the EMG results shown in Figure S6.5.1(d) cannot be fitted with the probability mapping. But we still can get some important information from EMG results such as the resonance frequency.

## S6.5.2 Other force mapping data of cortical stimualtions

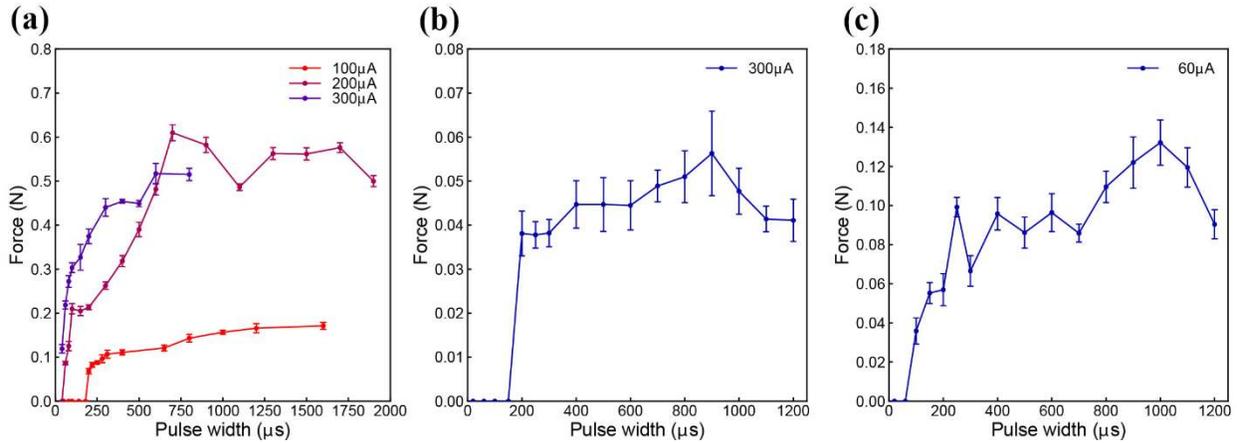

Figure S6.5.2. Cortical stimulations in three experiments using negative monophasic square waveform current.

Figure S6.5.2 shows the force mapping data cortical stimulations in another three experiments. The current waveforms are all negative monophasic square waveform. Because the cortex is too fragile, not all the tests can achieve perfect force mapping curve with several current amplitudes. A current amplitude that is too high or a SPPW that is too long will cause brain damage and further affect the testing result. In Figure S6.5.2(a), the curve of 300 µA is even lower than that of 200 µA when the SPPW is higher than 500 µs. The curve of 300 µA is not completed because the brain is damaged afterward and no further stimulation can be detected. The similar situation happened for Figure S6.5.2(b) and (c). In these two tests, only one force mapping curve can be completed. The brain was damaged afterward.

Nervertheless, the shape of the force mapping curves of all these three tests show an obvious multiple resonance peak effect, which agrees with the probability mapping shown in Figure 6.5.2(b). Since all these three tests are either with incompleted data or only with one curve, we did not try fitting these curves by modeling. However, all these curves resemble some probability mapping curves in Figure 6.5.2(b).

### S6.5.3 Complete EMG data of cortical stimulation

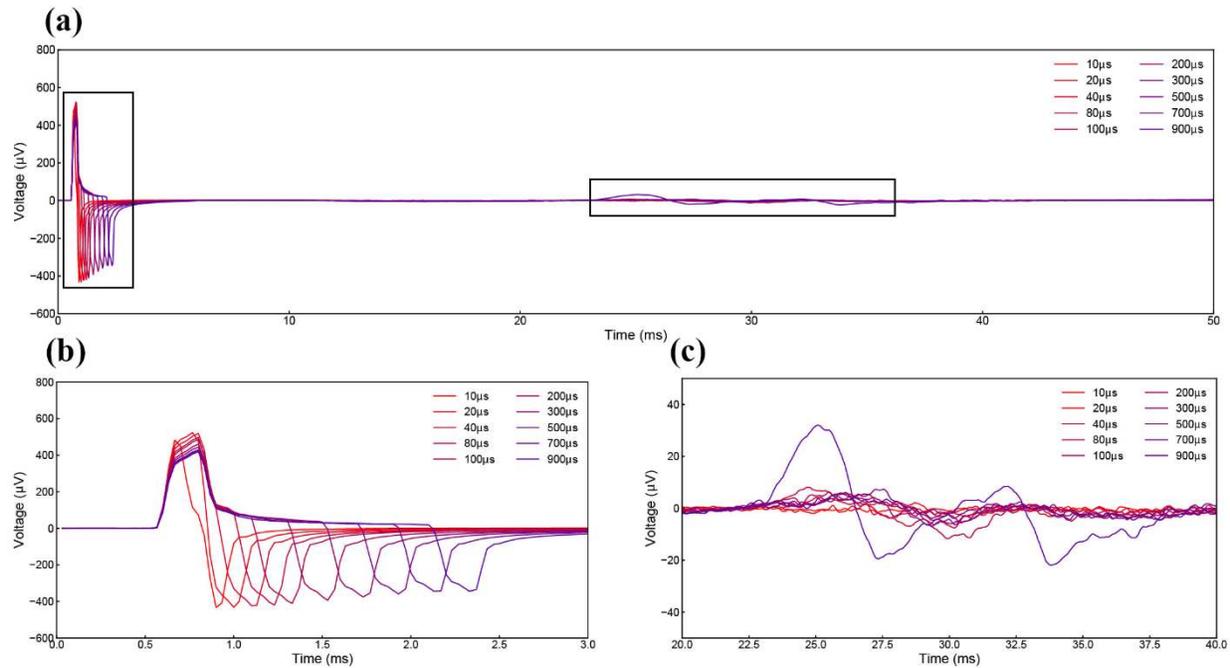

Figure S6.5.3.1 Complete EMG data of cortical stimulations with negative monophasic current waveform of 300 µA; (a) The complete raw EMG data; (b) The stimulus artifacts; (c) the EMG signal.

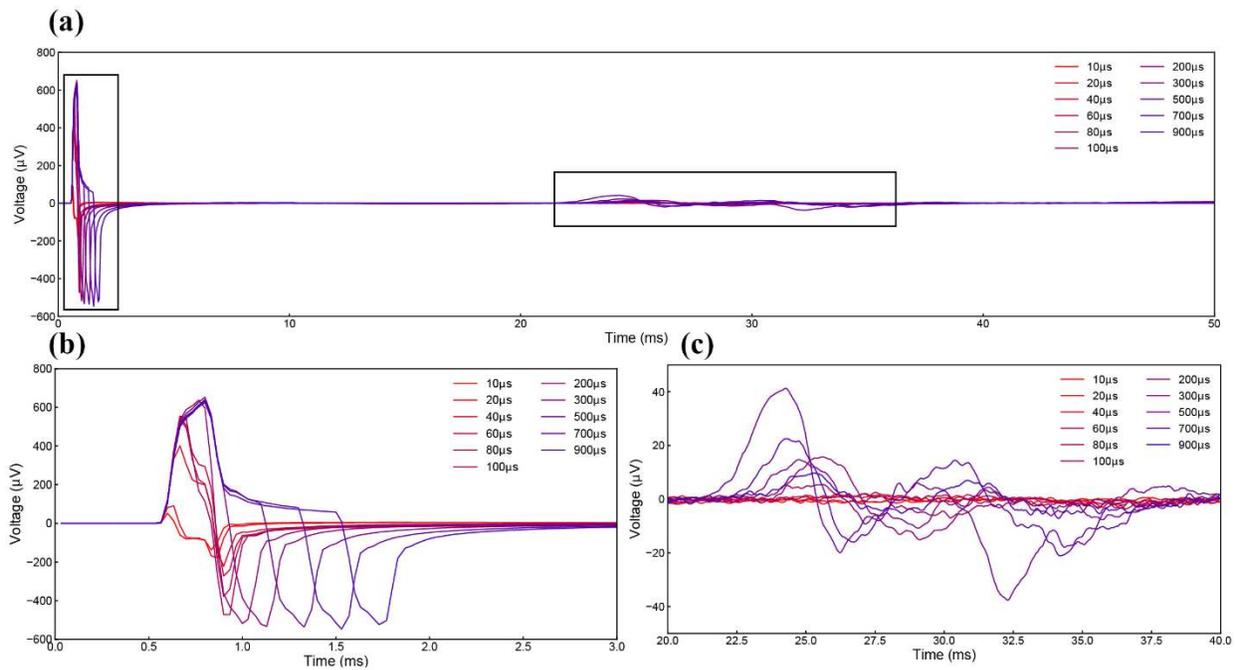

Figure S6.5.3.2 Complete EMG data of cortical stimulations with negative monophasic current waveform of 500 µA; (a) The complete raw EMG data; (b) The stimulus artifacts; (c) the EMG signal.

The detailed and complete EMG recording results of cortical stimulations with 300 µA and 500 µA are shown in Figure S6.5.3.1 nad S6.5.3.2, respectively. Each curve is an average result of 60 trials and no filter is further applied for signal processing. The artifact follows the same pattern as shown in Figure S6.2.1(e).

In Figure S6.5.3.2, the EMG signal peaks at SPPW of 500 µs and then significantly decreases with increasing SPPW, showing a distinctive resonance effect.

## S6.6 Experiments of pelvic nerve fibers stimulation
### S6.6.1 Pelvic nerve fibers stimulation results by three different current waveforms

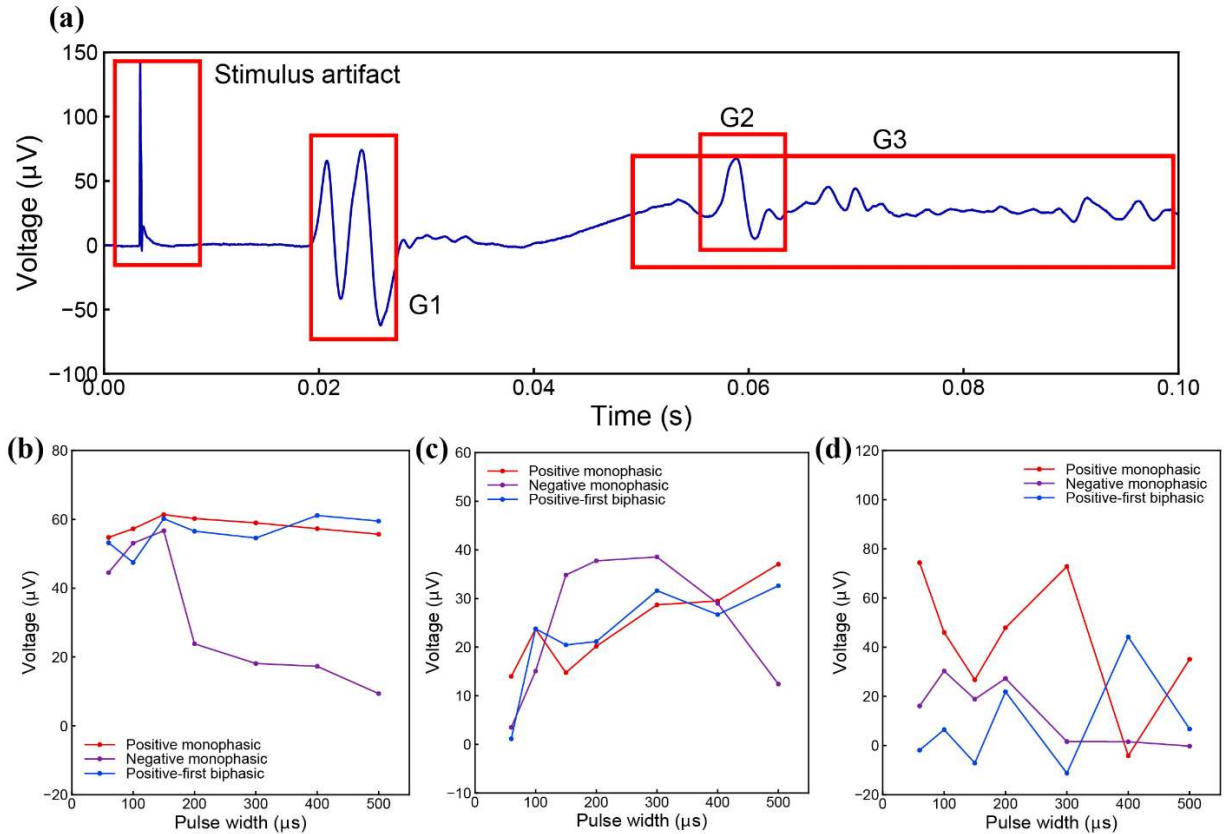

Figure S6.6.1 EMG measurement of the pelvic nerve stimulation by using three different current waveforms. (a) EMG sample of the pelvic nerve stimulation; (b) Amplitude of G1 with three different current waveforms; (c) Amplitude of G2 with three different current waveforms; (d) Amplitude of G3 with three different current waveforms.

The force mapping is a quite reliable method to represent the quantity of action potential because the force is proportional to the motor units recruited by the stimulation. However, this method is not always applicable. For some neural tissues, only ENG (Electroneurography) or EMG measurement is available. But as explained in the previous section, the amplitude of the EMG cannot fully represent the quantity of action potential due to the complex phase combination. The ENG signal will also have the same issue.

However, based on our theory, we still have chance to measure the resonance frequency if the resonance frequency really exists in this neural tissue. Figure S6.5.1(d) shows one successful case on cortical stimulation. As explained in Figure S5, a distinctive curve with clear resonance peak can only be achieved with a proper current amplitude. Meanwhile, as shown in Figure S6.4.2, not every kind of current waveform can show a clear resonance peak. It means the resonance peak can be obtained only with a proper current waveform of a proper amplitude.

Here we used the pelvic nerve as an example to show how to find the resonance frequency from the EMG signal. The detailed experiment setup, procedure and parameter can be found in the **Supplementary S3.3**. Figure S6.6.1(a) shows an EMG signal sample recorded by square wave current stimulation. Normally there will be four groups of signal in the raw data including the stimulus artifact. The real EMG signal can be generally classified as three groups of different time latency, labelled as G1, G2 and G3 as shown in Figure S6.6.1(a). The different time latency indicates that these three groups of the signal should transmit through different pathways. Three different types of current waveforms, positive monophasic square wave and negative monophasic square wave and positive-first biphasic square wave, with the same amplitude of 70 µA, are used for stimulation. All the detailed EMG data can be found in the **Supplementary S6.6.2**. Here only the amplitude of peak-to-peak voltage of each group is shown from Figure S6.6.1(b) to (d).

For group G1 in Figure S6.6.1(b), all three curves peak at SPPW of 150 µs. For the case of negative monophasic square wave, the curve has a significant drop when the SPPW is higher than 150 µs. This phenomenon cannot be explained by empirical models based on the calculation of charge or energy since more charge and energy induces a lower stimulation. However, in our new theory, this resonance effect is a direct prediction. The peak point at SPPW of 150 µs indicates that the resonance frequency of G1 is around 3333 Hz. The result of G2 also shows a similar situation (Figure S6.6.1(c)). The curve of negative monophasic square wave shows a resonance peak between SPPW of 200 µs and 300 µs, indicating a resonance frequency between 1600 Hz and 2500 Hz. And the result of G3 in Figure S6.6.1(d) shows a random behavior. Considering the long signal duration, this G3 should be movement artifact which is a side effect induced by electrical stimulation. The movement artifacts should have no strong correlation with the SPPW of the input current pulses. It also explains the random behavior of the signal amplitude shown in Figure S6.6.1(d). The different resonance frequencies of G1 and G2 indicate that nerve branches with different circuit parameters are involved in the pelvic nerve stimulation. Even without knowing more detailed circuit parameters, this EMG signal analysis still provide quite a lot of valuable information of the pelvic nerve. But as explained before, such resonance frequency can only be measured with some specific current amplitude and waveform.

## S6.6.2. Complete EMG data of pelvic stimulation

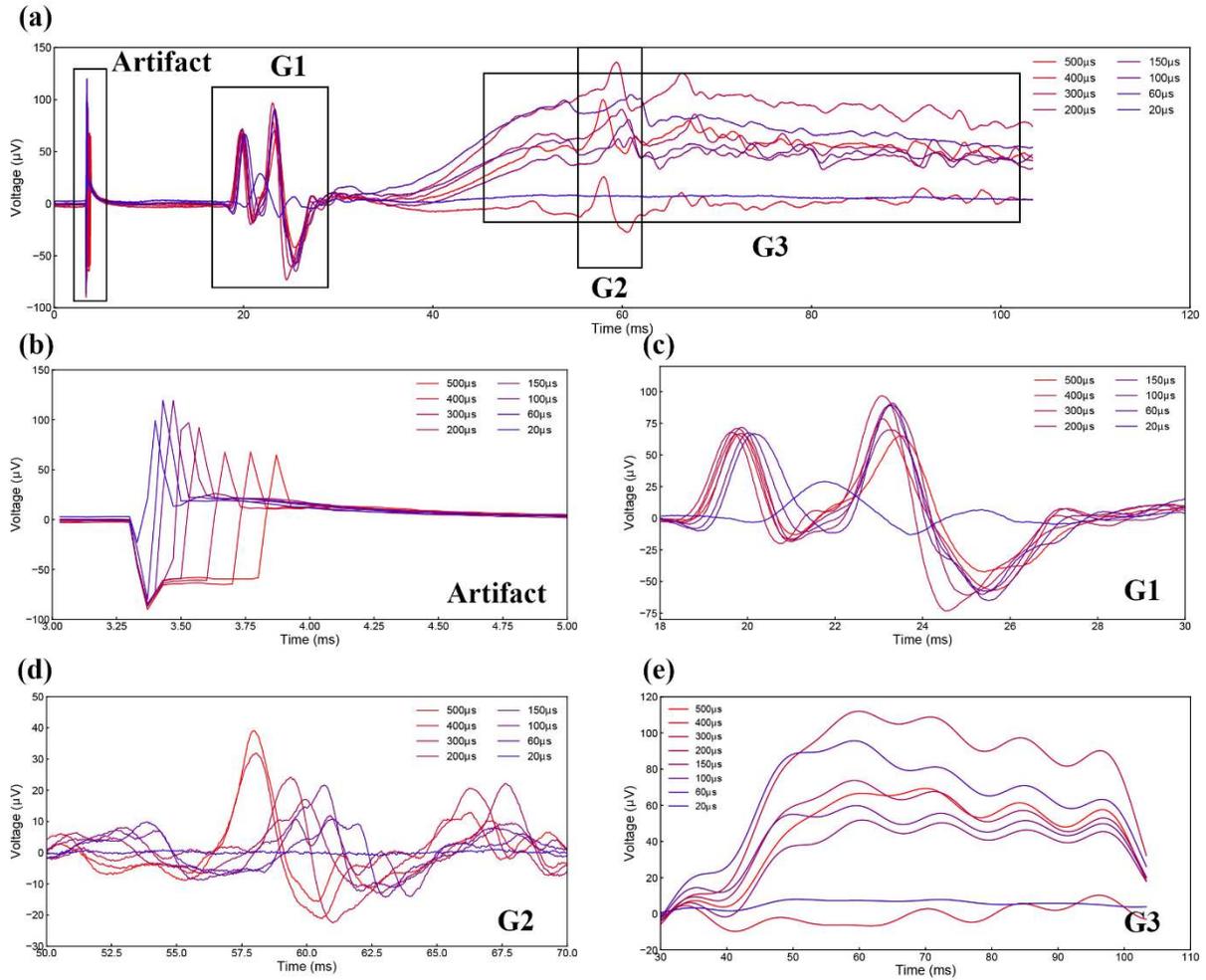

Figure S6.6.2.1 (a) Complete EMG recording result of pelvic nerve stimulation by using positive monophasic square waveform current. The current amplitude is 70 µA; (b) Stimulus artifact; (c) EMG signal of group G1; (d) EMG signal of group G2 by applying 80 Hz high pass filter; (e) EMG signal of group G3 by applying 80 Hz low pass filter.

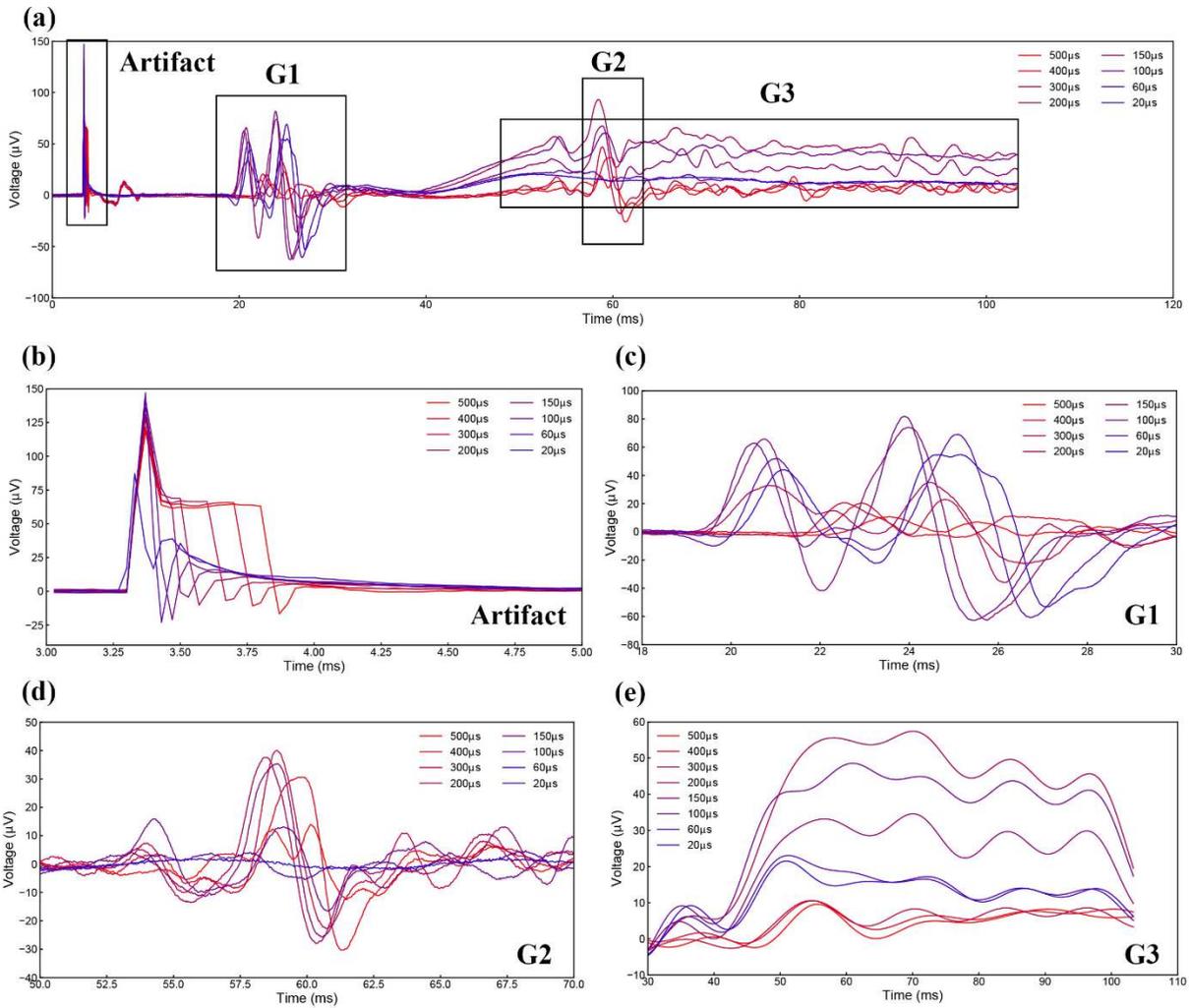

Figure S6.6.2.2 (a) Complete EMG recording result of pelvic nerve stimuation by negative monophasic square waveform current. The Current amplitude is 70 µA; (b) Stimulus artifact; (c) EMG signal of group G1; (d) EMG signal of group G2 by applying 80 Hz high pass filter; (e) EMG signal of group G3 by applying 80 Hz low pass filter.

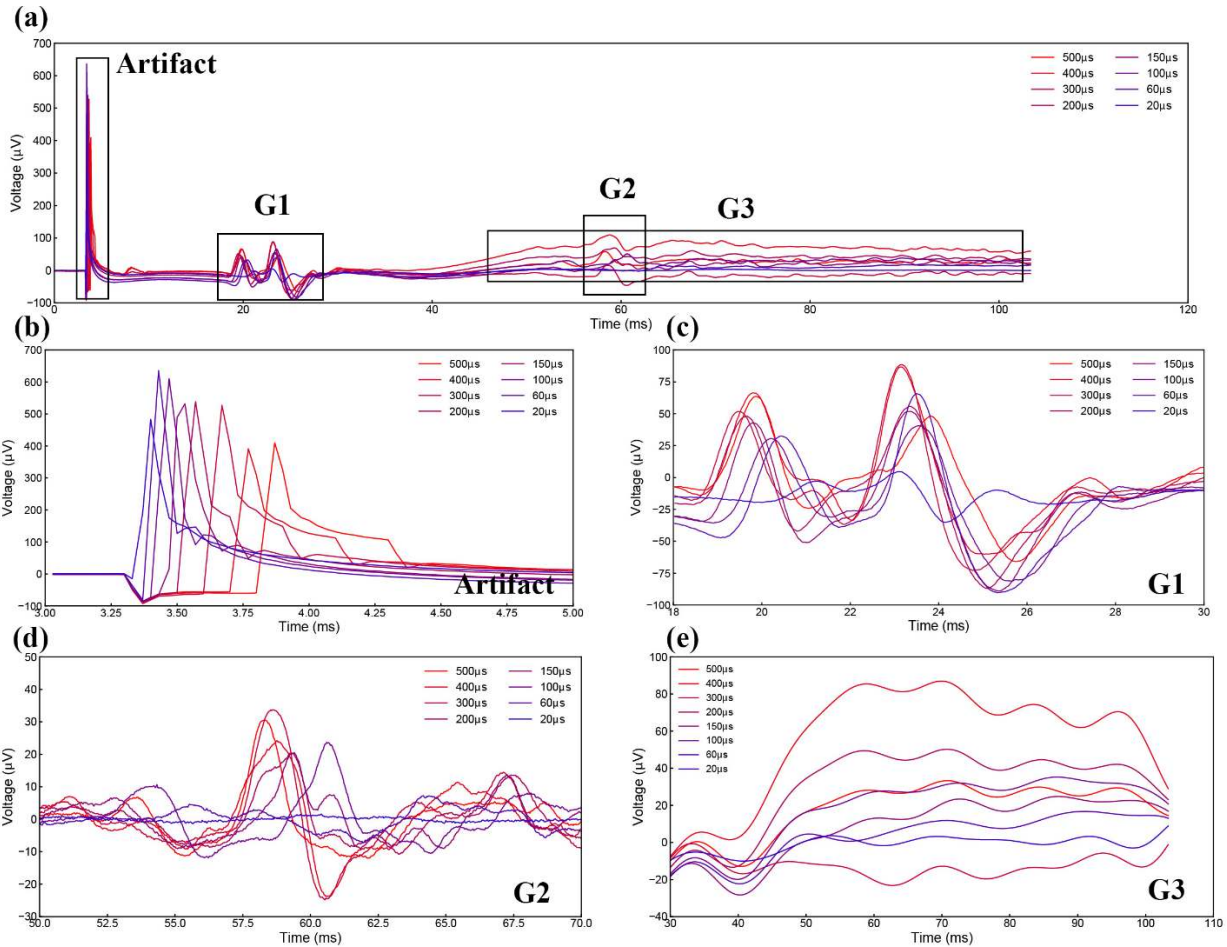

Figure S6.6.2.3 (a) Complete EMG recording result of pelvic nerve stimulation by positive-first biphasic square waveform current. The Current amplitude is 70 µA; (b) Stimulus artifact; (c) EMG signal of group G1; (d) EMG signal of group G2 by applying 80 Hz high pass filter; (e) EMG signal of group G3 by applying 80 Hz low pass filter.

Figure S6.6.2.1 to Figure S6.6.2.3 show the detailed EMG recording results by applying current with different waveforms. The summary of the peak-to-peak voltage of each EMG group, G1, G2 and G3, is shown in Figure S6.6.1(b)-(d). G2 and G3 are superpositioned with each other. To check the amplitude of these two EMG group independently, a 80 Hz high pass filter and 80 Hz low pass filter are applied for acquiring G2 and G3, respectively. In Figure S6.6.2.2, groups of G1 and G2 show distinctive resonance effect. Instead of the highest SPPW, G1and G2 achieve maximum peak-to-peak voltage at SPPW of 150 µs and 300 µs, respectively. This phenomenon can be well explained by C-P theory although we did not fit the curve with modeling.

### S7 Some more discussion
### S7.1 Equation for probability calculus

As explained in the theoretical part, the function of $\lambda$ is based on three empirical hypotheses. The general form of the function of $\lambda$ is

$$\lambda = \alpha \times \frac{1}{e^{\frac{\beta}{(|V-V_{Threshold}|)^n}} - c}$$

In this study, we directly set $n = 1$ and $c = 0$ to make the equation simpler. However, when $n \neq 1$ and $0 \leq c \leq 1$, other sets of parameters could still be captured to fit the force mapping curves. Although we derive the expression of $\lambda$ from three empirical considerations, the exact expression of $\lambda$ is only determined by its physical meaning. Before knowing the real physical meaning, which is out of the scope of this study, we are not able to determine the exact form of the function of $\lambda$. But we still have a conjecture which may help us find the correct form of the function of $\lambda$.

In the previous discussion section about LNP model, it is proved that the exponential distribution can be rewritten as Poisson distribution with the same $\lambda$. In Poisson distribution, there is a very clear physical meaning of the $\lambda$, which is the number of occurrences per interval of time. So the $\lambda$ can be considered as some kind of frequency, representing some sort of energy density. Since the implementation of the exponential distribution is from the hypothesis of quantum effect as explained in the theoretical part, we may also get some clues from quantum mechanics. The physical meaning of the $\lambda$ in Poisson distribution can let us associate with the Planck's law, which is given by

$$B_\nu(\nu, T) = \frac{2h\nu^3}{c^2} \frac{1}{e^{\frac{h\nu}{k_B T}} - 1}$$

Where $\nu$ is the frequency in the spectrum, $T$ is the absolute temperature, $k_B$ is the Boltzmann constant, $h$ is the Planck constant, $c$ is the speed of light in the medium and $B_\nu(\nu, T)$ is the spectral radiance of a body. Here $B_\nu$ can also be considered as a kind of energy density, which is the same as the $\lambda$. Meanwhile, $B_\nu$ is of the same form of $\lambda$ when $n = 1$ and $c = 1$, as shown below:

$$\alpha = \frac{2h\nu^3}{c^2}$$

$$\beta = \frac{h\nu}{k_B}$$

$$B_\nu(\nu, T) = \frac{2h\nu^3}{c^2} \frac{1}{e^{\frac{h\nu}{k_B T}} - 1} = \alpha \times \frac{1}{e^{\frac{\beta}{T}} - 1}$$

$$\lambda = \alpha \times \frac{1}{e^{\frac{\beta}{|V-V_{Threshold}|}} - 1} = \alpha \times \frac{1}{e^{\frac{\beta}{\Delta V}} - 1}$$

The resemblance between the Plank's law and the function of $\lambda$ may not be just a coincidence. It may reveal a deeper connection between the ion channel gating and quantum mechanics. In other words, the ion channel gating is induced by electron transition. In optogenetics, it is well known that photon can open the ion channel. One possible physical explanation is photoelectric effect. It is highly possible that electrical nerve stimulation, and even the propagation of neural signals, follows the similar mechanism.

The function of $\lambda$ with the same form of Plank's law can be called as Plank form $\lambda$. The function of $\lambda$ used in the modeling of this study, in which $n = 1$ and $c = 0$, can be called as the C-P form $\lambda$. Currently we have no substantial evidence to confirm which form is correct. The difference between these two forms are not very distinctive in modeling results. To simplify the parameter capturing, we still use the C-P form $\lambda$ in the modeling of this study. In the future, we may be able to prove this conjecture with a more detailed study of the ion channel.

**S7.2 Inductor in neural tissue**

The inductor involved in the C-P theory may be the most controversial part. Not only in this C-P theory but also in some other studies, the similar inductance effect of the neurons has been reported [5-6]. The myelinated neurons have much higher inductance than the unmyelinated neurons [6]. Nevertheless, in all previous neural models such as H-H model and the cable model, there is no such an inductor involved. Thus, the inductor may seem to be an unnecessary hypothesis for previous neural models.

However, in the C-P theory, the inductor is an inevitable component to make the whole theory established. C-P theory is a basic physical description instead of merely a computational model, and the existence of the inductor has already been confirmed in our testing results. Thus, the physical form of this inductor is a critical issue. Although there is no decisive evidence, quite a lot of clues imply that the myelin should be the component to provide the inductance in the neural tissue.

Firstly, the myelin wrapped around the nerve like a coil, which is a common and effective structure to generate inductance. Then the cell membrane and the myelin can form a RLC circuit and the whole axon can be considered a RLC cascade. However, in the cable theory, the myelin is just modeled as a resistor and the whole axon is treated as a RC cascade. As well known, failure to consider the inductance associated with any alternating electric signal passing along a coaxial cable, such as the first RC based undersea cable in human history, will lead to a disaster in practice. Only with a more sophisticated RLC design based on Maxwell's Equations for a coaxial structure, the new undersea cable was laid with great success. It is difficult to believe that as a result of long evolutionary process, the neuron is still working with a design which is already proved to be a failure in human engineering.

Moreover, modelling the myelin as an inductor can explain why magnetic field can also induce nerve stimulation. As been reported [7-8], the magnetic field can stimulate the nerve only when the magnetic field lines is along the axon, in other word, is exactly perpendicular to the sectional area of the myelin. It is a common sense that a changing magnetic field across the sectional area of a coil can generate an inductive voltage. This inductive voltage upon the myelin can be further coupled by the cell membrane, which is a capacitor, to induce the action potential.

Thirdly, if the myelin can be modeled as an inductor, the inductance value can be controlled by the profile of the myelin, either the length or the thickness. Let's just consider the length here. The length of the myelin will not only change the inductance of itself, but also affect the length of the Ranvier node, which will further determine the value of the capacitance induced by the cell membrane. A changing length of the myelin will change both the *L* and *C* in the RLC circuit, resulting in a frequency modulation. Such distinct myelin profile distribution is already discovered in pyramidal neurons in the neocortex [9]. Such phenomenon definitely cannot be explained if the myelin is still modeled as merely a resistor. However, it is quite easy to understand this phenomenon when the myelin is an inductor. Different axons in the brain will have different intrinsic resonance frequency, which is modulated by the myelin sheath, and this is also how the brain is differentiated into groups with different functions.

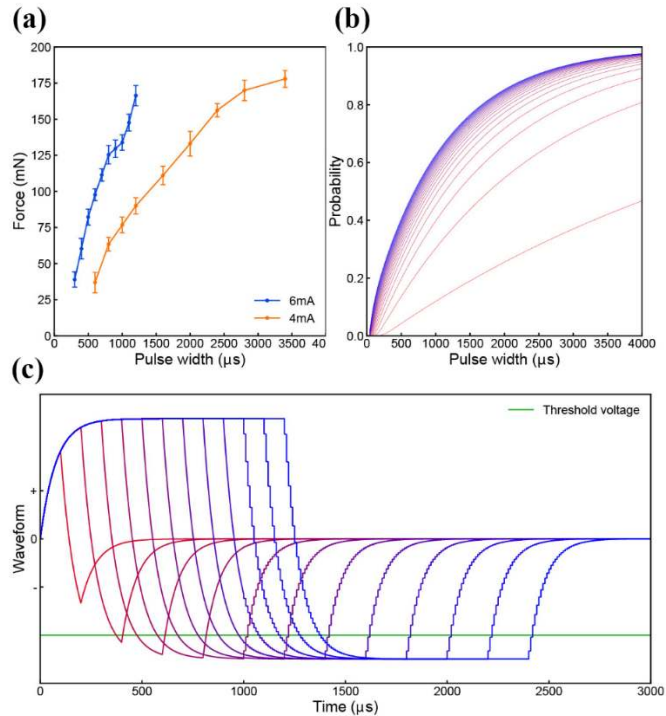

Figure S7.2 Measurement and modeling results of the atrophy TA muscle stimulation by positive-first biphasic square waveform current. (a) Force mapping results; (b) Probability mapping results; (c) Corresponding voltage waveforms.

Finally, as a direct prediction, the unmyelinated nerves should have no inductance or a very low inductance, hence no resonance effect can be detected. To confirm this, we also did a force mapping test on an atrophy muscle, in which the sciatic nerves were transected. Since the muscle fiber is unmyelinated, the force mapping curve should follow the pattern of a RC circuit as shown in Figure S7.2(b). The corresponding voltage waveform is shown in Figure S7.2(c). The two force mapping curves shown in Figure S7.2(a) show monotonically increasing trend as the probability mapping curves in Figure S7.2(b). Since the atrophy muscle is not as healthy as the normal muscle, it cannot stand a long term stimulation to obtain many force mapping curves. But we still consider this RC pattern from atrophy muscle as a circumstantial evidence implying that myelin is an inductor.

**S7.3 The stimulation efficiency of different current waveforms**

In previous empirical models for electrical nerve stimulation, the efficiency of different waveforms is also a key point to be discussed. Some ambiguous conclusions, such as negative pulse will be more effective for stimulation, will be drawn from these empirical models. And there is even an argument about whether exponential wave is the most energy-efficient waveform for nerve stimulation [10-11]. Two research groups got the opposite conclusion about nerve stimulation by exponential current wave.

Apparently, based on our C-P theory, it is unscientific to say one current waveform is more effective than another. The comparison result of different current waveforms is not only affected by their own waveforms, but also affected by the current amplitude, circuitry parameters and probability calculus parameters. The probability mapping curves with the same current amplitude but different waveform can even have a cross point, which can be seen in both testing and modeling results shown in Figure S6.3.1.1(a)&(b) and Figure S6.4.2(a)&(b). In the four waveforms comparison measured in three experiments, the negative-first biphasic is more effective than positive-first biphasic in one test and less effective in another two tests.

In the C-P theory, any situations can happen, such as the cross point of probability mapping curves of different waveforms and the efficiency switch between positive-first biphasic and negative-first biphasic (happen in three times of muscle tests of four waveforms comparison), and can be validated by testing results. Such phenomenon definitely cannot be explained by any previous theories or models that calculate either charge, current, voltage or energy.

Meanwhile, in the C-P theory, it is meaningless to investigate which current waveform is the most effective one for nerve stimulation. Other than the waveform, frequency will be a more important parameter to be considered. Normally the input current with the resonance frequency of the neural tissue will be more effective. And the stimulation result is also affected by the testing condition, such as $R_1$, which is affected by the humidity on the nerve surface. However, despite these complex situations, some general conclusions still can still be drawn from the C-P theory. For example, normally biphasic current waveform will be more effective than monophasic waveform because of a larger effective voltage area, and negative monophasic current pulses will be more effective than positive monophasic current pulses.

**S7.4 The polarity of the voltage upon the capacitor**

In this C-P theory, the voltage waveform upon the capacitor is the most critical factor affecting the stimulation results. The negative voltage, which should be the potential outside of the membrane versus the potential inside the membrane, is effective for stimulation since the sodium ion channel is gated by negative voltage. However, in the lumped parameter circuit used in this study, we are not able to define which terminal refers to either the outside part or inside part of the membrane. This is because this is a simplified circuit, the physical meaning of each terminal of the capacitor, which can be defined in a complete distributed parameter circuit, disappears. All capacitors referring to each membrane segment are simplified as one capacitor. So here we have to manually introduce a method to determine which terminal of the capacitor should be defined as positive. Currently, in all the stimulation tests, the downstream electrode, which is more close to the recording side, was defined as positive. The voltage in the modeling adopted the same polarity as the current waveform from the downstream electrode. For example, if the downstream electrode is connected with the positive terminal of the stimulator and the current is the positive first biphasic square waveform, the corresponding voltage waveform upon the capacitor also has a positive first phase. The validity of this method is proved in all modeling results.

## S7.5 About the sine waveform tests

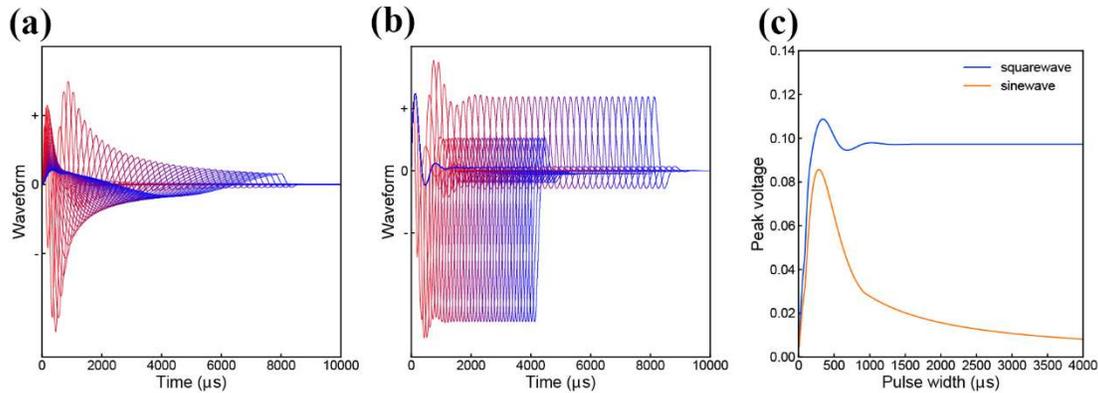

Figure S7.5 Waveform comparison of sinewave and square wave current. (a) An illustrative sinewave current with different SPPW; (b) An illustrative positive-first square waveform current with different SPPW; (c) Peak value comparison of the voltage waveform of sinewave and positive-first square wave with different SPPW.

Because of the parallel RLC circuit, theoretically sine wave should be the best waveform to capture the resonance frequency of a neural tissue. However, in actual test, the sinewave test was only successful on the CP nerve. In both muscle and cortical test, the sinewave can easily damage the tissue by only a few stimulation trials. Afterwards, no neural response can be further detected no matter what current waveform was applied.

One possible reason for this phenomenon is that the voltage response of the sinewave is with a too high Q factor while the voltage response of the square wave has a very low Q factor. An illustrative voltage waveform by sinewave and positive-first square wave current is as shown in Figure S7.5(a) and (b). And the Figure S7.5(c) shows how the peak value of the voltage waveform changes with SPPW. For the square waveform, the peak value will keep almost constant at high SPPW while the peak value of sinewave can drop to a very low value. So the voltage difference of square wave at different SPPW will not change a lot, meaning a low Q. Therefore, a gradual change of the force generated by different SPPW can be well recorded. However, the situation of sinewave is quite different. If a proper current amplitude at a very high or a very low SPPW is selected to generate a medium level force, the amplitude of the voltage waveform will increase significantly when the SPPW is tuned to be close to the resonance frequency, and then exceeds the threshold voltage to induce tissue damage. This phenomenon repeatedly happened in the sinewave test. Initially some data points can be recorded but suddenly from one data point, no further stimulation, either the force or the EMG signal, can be detected. Even if the current waveform was switched back to square waveform, there was still no response at any current amplitude. Apparently the neuron has been damaged. This result also indicate that sinewave may be very dangerous for stimulation of skeletal muscle and cortex.

## S7.6 A complete version of the thought experiment

To help readers comprehend the C-P theory, a thought experiment is proposed. Three basic questions are asked and each question has an exclusive answer. The C-P theory is the direct reasoning result of this experiment.

To facilitate understanding of the C-P theory, several key points are emphasized. Firstly, this thought experiment is not based on any previous theories or models, such as H-H model, cable theory and other empirical models. Secondly, the answers of these three questions are based on pure logic and some basic physical knowledge. Not much biological knowledge is involved. Thus, as a reasoning result of this thought experiment, the C-P theory is a physical theory other than a biological theory. As a physical theory, it is a priori rather than a posteriori which is the case for most of the previous biological theories and models. Thirdly, we derived the C-P theory from this thought experiment because we believe that the answer for each question is exclusive. Only when alternative answers exist, the correctness of this priori theory can be jeopardized.

The thought experiment is as follows:

For the electrical nerve stimulation, an electric input is applied and action potentials can be generated. This action potential is not the direct conduction of the electric input. It is the response of the ion channel to electric input.

### Q1: What is the real factor inducing the ion channel gating?

We know that the ion channel response is induced by the electric input. However, an electric input can generate changes in many different physical quantities such as current, voltage, charge, energy and electric field. In the H-H model, electric field and voltage is used to interpret the ion channel response. In other empirical models, people also try developing the relationship between the ion channel response and current, voltage, charge and energy. To find out the real factor to induce the ion channel, let's imagine two situations.

Firstly, imagine a nerve connected with a pair of electrodes. When we apply a current pulse, how will the current flow through the nerve? Some current will just flow through the outer surface of the nerve, some will flow through the tissue and some will flow through the ion channel. Apparently, not all the current can truly interact with the ion channel. If there is some liquid, such as blood, on the outer surface of the nerve, most of the current will just flow through the liquid instead of flowing through the tissue due to the short circuit between two electrodes induced by liquid. Thus the current truly interacting with ion channels only takes a small ratio of the injected current and this ratio can be significantly affected by external environment. The situation is the same with voltage, charge and energy. Although we still can summarize some vague relationship between the stimulation result and electric input, such as the strength-duration relationship, a precise and accurate mathematical relation between the stimulation result and electric input from the stimulating electrode in unavailable.

Secondly, imagine an individual ion channel embedded upon the cell membrane and surrounded by different molecules and all kinds of ions. In such a microscopic world, all interactions are actually induced by field. Normally the cell membrane can be considered as a capacitor, which means no ion can directly go through the membrane unless the ion channel is open. So all the positive and negative ions are now accumulated at the inside and outside surfaces of the cell membrane. However, since all these ions have no physical contact with the ion channel, the ion channel can only sense the electric field emitted by all these ions. So apparently, electric field is the key factor to interact with the ion channel.

### Q2: How to obtain this electric field?

Before answering this question, we need to think of another question first: **What kind of features should this electric field have?**

As well known, the cell membrane is impermeable to ions. All the ions are accumulated at both sides of the cell membrane. So the direction of major component of the electric field emitted by these ions should be perpendicular to the cell membrane surface, in other words, along the longitudinal dimension of the ion channel. Meanwhile, the strength of the electric field is proportional to the quantity of the ions. Since ions can only move with a certain speed, the accumulation of ions induced by external applied electric input requires a certain duration. This is also why a capacitor requires some time to get charged and discharged. It means no matter what kind of electric input is applied, for example, a certain current, a certain voltage or a certain quantity of charge, the electric field is always a time-varying field with gradual change. We need to know how this electric field changes with time. In summary, the two key features of the electric field, direction and how it change with time, are inevitable.

**Now we can consider how to obtain this electric field.**

One easy answer is that we can get the electric field distribution within a tissue by finite element modeling. In this method, the tissue can be simplified as a kind of medium with a set of parameters such as conductivity, permittivity and permeability. In fact, many research groups have adopted this method for decades. However, in this method, the direction of the electric field is only determined by the position of the electrodes, the boundary conditions and the medium parameters, but not the orientation of the cell membrane and the ion channels. Moreover, the information of how the electric field and the tissue condition changes with time is also lost, because the electric field distribution obtained from the modeling only reveals how the electric field spread from the electrodes, not how the ions moves within the tissue. The localized electric field around the ion channel is mainly determined by the movement of ions on both sides of the cell membrane. But the movement of ions is not only determined by the external applied electric field, but also constrained by the physical boundary of the cell membrane, the conditional permeability of ion channels and channel refractory periods. Electric field modeling can tell us how the electric field emitted from the electrode spreads in the tissue, but never tell us how this electric field drives the ions. In summary, the electric field obtained from previous modeling method does not meet the requirement mentioned above: with a correct direction and provide the information of how the electric field changes with time.

Let's think about the direction issue first. As explained, the electric field we need is perpendicular to the cell membrane. If we model the cell membrane as a capacitor, based on the ion quantity upon the surface of the cell membrane, we can directly calculate the voltage with the equation $V = Q/C$, where $Q$ is the charge quantity and $C$ is the equivalent capacitance. Meanwhile, the electric field across the capacitor can be calculated as $E = V/d$, where $d$ is the thickness of the cell membrane. It should be noted that the electric field calculated here is perpendicular to the cell membrane, which is exactly what we desire. It means that if we model the cell membrane as a capacitor, then the voltage calculated upon this capacitor will be proportional to the magnitude of the electric field we need and with the correct direction. Although the voltage $V$ is a scalar, we still can use it to characterize the electric field we need, which is a vector, because the information of the direction has already been included.

Then let's think about the time issue. Since the cell membrane is now modelled as a capacitor, we need to know how the voltage changes with time, in other words, the voltage waveform induced by an electric input. Apparently, this voltage waveform upon the capacitor is not only determined by the capacitor itself, but also by the peripheral circuit and the waveform of the electric input. Here we can take the parallel RC circuit as a simple case. The charging and discharging rate of the capacitor is determined by the time constant $\tau = RC$. The resistor $R$ connected in parallel with the capacitor $C$ can also change the charging and discharging rate then affect the voltage waveform. On the other hand, the voltage waveforms generated by a square wave current pulse and a sine wave current pulse are definitely different. It also can be expected that voltage waveform difference, induced by either different peripheral circuits or different current waveforms, should result in different nerve stimulation results. In summary, to obtain the exact voltage waveform, we need to construct a complete circuit which is equivalent to the neural tissue, including the cell membrane and all other peripheral circuit components. Meanwhile, we need to specify the detailed waveform of the electric input.

**In summary, we need to know the correct equivalent circuit. With this correct circuit, the correct voltage waveform upon the cell membrane can be calculated from the electric input.**

Up to now, we know the circuit is important. Let's pretend that we already know the exact circuit equivalent to a specific neural tissue. Then a current pulse is applied and an exact voltage waveform is generated upon the cell membrane.

**Q3: How an exact voltage waveform determines the gating state of a specific ion channel?**

Firstly, let's imagine an individual ion channel. If no voltage is applied, no action potential will be generated no matter how long time it takes. It means a 0V voltage cannot introduce any effect upon the ion channel. If a negative voltage with a certain amplitude and duration is applied, definitely action potential can be generated. So an obvious conclusion can be drawn here: the threshold voltage exists. At least, the voltage required to induce an action potential should exceed 0V.

Then let's pretend that we already know the threshold voltage, denoted as $V_T$. Now we can guess how the ion channel is affected by the voltage.

One simple guess is that the ion channel is open when the voltage, $V$, exceeds $V_T$. It means at the time point, $t_T$, when the $V \geq V_T$, the ion channel will open and an action potential will be generated. (Notice that voltage $V$ is a negative value) Then we can measure this $V_T$. No matter what kind of voltage waveform is applied, such as a sine wave or a square wave, the action potential is always generated at the time point $t_T$ when $V \geq V_T$. However, all the previous researches show that the working mechanism of ion channel is not so simple.

Then another guess is that $V$ should be kept for some duration, denoted as $T$, after exceeding $V_T$ to open the ion channel. Then we can apply a DC voltage $V_{DC}$ upon the ion channel and measure this $T$. When $V_{DC} < V_T$, no action potential can be generated. When $V_{DC} \geq V_T$, after a duration $T$, action potential can be generated. If we kept $V_{DC}$ as the same value, a constant $T$ can be measured. This $T$ may be a function of $V_{DC}$ ($T = f(V_{DC})$). A higher $V_{DC}$ can induce a shorter $T$. However, this DC voltage test has already be done before and the results show that ion channel does not work like this. Such a function $T = f(V_{DC})$ cannot be achieved.

We know there is a threshold voltage. We also know ion channel does not open at the time point when $V \geq V_T$. So what happen in this duration $T$. It looks like that something is accumulated within this $T$ to open the ion channel. And the result of ion channel opening is discrete. It suddenly switched from close state to open state. Although the state change of an individual ion channel is discrete, the number of excited nerve fibers in a nerve branch can have a continuous change by changing either the current amplitude or the pulse duration. There definitely is something can build the bridge between the microscopic discrete and macroscopic continuity.

To solve the issue above, let's treat the ion channel as a black box first and see whether we can find some clues. From the above discussion, we find that even we give the same input each time, the output from this black box is different. Is the thing hidden in the black box deterministic? If it is, then the same input should result in the same output. If the output is different, it means the input is definitely different. Surely we are not able to control everything identical in reality. There is always something not identical for each test. But as long as the system is deterministic, we can always know the system better with a more precisely controlled experiment condition. Unfortunately, previous researches don't show such an optimistic perspective. So what if the thing hidden in the black box is not deterministic? What if the ion channel is probabilistic? Just like the situation happened in quantum mechanism, everything about elementary particles is discrete and can only be described by probability, but everything built by these elementary particles in macroscopic is still continuous.

So if the ion channel is non deterministic, probability is the exclusive option to describe its working mechanism. Then the above question can also be answered: within the duration $T$, the thing getting accumulated is probability.

**Probability is the exclusive option to describe the gating of the ion channel.**

Up to here, we get a basic framework, which is called C-P (Circuit-Probability), for the analysis of electrical nerve stimulation.